\documentstyle[12pt]{article}

\def\hybrid{\topmargin -20pt  \oddsidemargin 0pt
      \headheight 0pt   \headsep 0pt
      \textwidth 6.25in 
      \textheight 9.5in 
      \marginparwidth .875in
      \parskip 5pt plus 1pt   \jot = 1.5ex}

\hybrid

\def\x{\times}

\def\o+{\oplus}
\def\ra{\rightarrow}
\def\beqa{\begin{eqnarray}}
\def\eeqa{\end{eqnarray}}

\newcommand{\ti}{\tilde}

\newcommand{\resetcounter}{\setcounter{equation}{0}}     

\begin{document}
\thispagestyle{empty}
\rightline{IFP-9908-UNC}
\rightline{IASSNS-HEP-99-80}
\rightline{hep-th/9908193}
\vspace{2truecm}
\centerline{\bf \LARGE On discrete Twist and Four-Flux}
\vspace{.5truecm}
\centerline{\bf \LARGE in $N=1$ heterotic/$F$-theory compactifications}

\vspace{1.5truecm}
\centerline{Bj\"orn Andreas$^{\dagger}$
\footnote{bandreas@physics.unc.edu, 
supported by U.S. DOE grant DE-FG05-85ER40219/Task A.}
and Gottfried Curio$^*$\footnote{curio@ias.edu, 
supported by NSF grant DMS9627351.\\ 
E-mail address after 1.July: curio@physik.hu-berlin.de}} 

\vspace{.6truecm}

{\em
\centerline{$^{\dagger}$Department of Physics and Astronomy}
\centerline{University of North Carolina at Chapel Hill, 
NC 27599-3255, USA}
\vspace{.4truecm}
\centerline{and}
\vspace{.4truecm}
\centerline{$^*$School of Natural Sciences, Institute for Advanced Study}
\centerline{Olden Lane, Princeton, NJ 08540, USA}}

\vspace{1.0truecm}
\begin{abstract}
We give an indirect argument for the matching $G^2=-\pi_* \gamma^2$
of four-flux and discrete twist
in the duality between $N=1$ heterotic string and $F$-theory.
This treats in detail the Euler number computation for the 
physically relevant case of a Calabi-Yau fourfold with singularities.
\end{abstract}

\newpage
\section{Introduction and Summary}

Compactification on an 
elliptic Calabi-Yau three-fold $Z$ with vector bundle 
$V$ embedded in $E_8\times E_8$ gives a four-dimensional 
heterotic string model of $N=1$ supersymmetry. 
Originally the case of $V$ the tangent bundle was considered
which lead to an unbroken gauge group $E_6$ times a hidden $E_8$.
The generalisation to an $SU(n)$ bundle $V_1$ gives unbroken GUT
groups like $SO(10)$ and $SU(5)$ (we will in the following focus on
the visible sector and therefore assume an $E_8$ bundle $V_2$ embedded
in the second $E_8$). 

Especially interesting is the case where $Z$ 
admits an elliptic fibration $\pi: Z\rightarrow B_2$ which has 
a section $\sigma$.
This makes possible an explicit description of the bundle by using the
spectral cover\footnote{The details of the spectral cover method
and the corresponding $F$-theory description are reviewed in the 
appendix.} $C$ of $B_2$. In this description the $SU(n)$ bundle is encoded
in two data: a class\footnote{unspecified cohomology classes
refer to $B_2$; below $B_2$ will often simply denoted by $B$} 
$\eta_1=6c_1-t$ in $H^{1,1}(B_2)$ and a class
$\gamma$ in $H^{1,1}(C)$ (the latter is connected to the possible
existence of chiral matter in these models [\ref{Cc3}], [\ref{Dc3}]).
In this case it is also possible to give a dual
description by $F$-theory on a Calabi-Yau four-fold $X^4$ which
is $K3$ fibered over $B_2$ and elliptically fibered over $B_3$ which
in turn is a ${\bf P}^1$ fibration described by the class $t$
over $B_2$. Having an unbroken gauge group
$G$ on the heterotic side corresponds then to having a section of $G$
singularities along $B_2$ in $X^4$. 

It was shown \cite{FMW} that an anomaly mismatch in the heterotic model
causes the occurence of a
number $n_5$ of five-branes wrapping the elliptic fiber\footnote{We
will stick in the following to the ansatz $\eta_1+\eta_2=12\sigma c_1$
which leads only to five-branes wrapping the elliptic fiber. In general
other curves would be wrapped as well \cite{donburt}, [\ref{DR}].
Note that in the cases of $A_5$ and $D_6$ we take
for $V_1$ a product bundle $SU(n_1^{(1)})\times SU(n_1^{(2)})$
with $\eta_1^{(1)}=2c_1$ and $\eta_1^{(2)}=4c_1-t$.} $F$
\beqa
c_2(Z)=c_2(V_1)+c_2(V_2)+n_5F
\eeqa
where the Chern classes were given by (we assume $B_2$ to be rational)
\beqa
c_2(Z)&=& 12\sigma c_1+12+10c_1^2\nonumber\\
c_2(V_1)&=&\eta_1 \sigma -\frac{n_3-n}{24}c_1^2-
\frac{n}{8}\eta_1(\eta_1-nc_1)-\frac{1}{2}\pi^* \gamma^2
=c_2(V_{1;\gamma =0})-\frac{1}{2}\pi^* \gamma^2\nonumber\\
c_2(V_2)&=&\eta_2 \sigma -40c_1^2-45c_1t-15t^2
\eeqa

Consistent F-theory compactification on $X^4$ requires 
a number of space-time filling threebranes 
which are localized at points in the base $B$ of the elliptic four-fold. 
The number of such threebranes was determined in \cite{SVW} for the
case of a {\em smooth} Weierstrass model for the fourfold
by observing that the SUGRA 
equations have a solution only for a precise number of such threebranes, 
proportional to the Euler characteristic of the four-fold. 

In the case of an
$E_8\times E_8$ vector bundle $V$, leaving no unbroken gauge group and
corresponding to a smooth Weierstrass model for the fourfold it was 
also shown that the number of five-branes matches the number $n_3$
of three-branes on the $F$-theory side whose number is given by
\beqa
\frac{e(X^4)}{24}=n_3+\frac{1}{2}G^2
\eeqa
where $G\in H^{2,2}(X^4)$ is the four-flux \cite{DM} (cf. appendix).

For various reasons one can expect
$G$ to be associated with $\gamma$; {\em a very condensed version of this
argument can be found in the introduction to section $C$ of the appendix}.
Part of this association is the following relation
\beqa
G^2=-\pi^* \gamma^2\label{aim}
\eeqa
which in view of the assumed equality $n_5=n_3$ and
\beqa
n_5&=&c_2(Z)-c_2(V_{1; \gamma=0})+\frac{1}{2}\pi^* \gamma^2 -c_2(V_2)\\
n_3&=&\frac{e(X^4)}{24}-\frac{1}{2}G^2
\eeqa
amounts to
\beqa
e(X^4)&=&24(c_2(Z)-c_2(V_{1; \gamma=0})-c_2(V_2))\nonumber\\
  &=& 288+(1200+107n-18n^2+n^3)c_1^2+(1080-36n+3n^2)c_1t
+(360+3n)t^2\nonumber
\eeqa
{\em One would like now to see this equation directly on the $F$-theory side
thereby proving (\ref{aim})}.

This matching was extended \cite{AC} to the general 
case of heterotic string compactification on an elliptic Calabi-Yau 
threefolds together with a $SU(n_1)\times SU(n_2)$ vector bundle 
leading to an unbroken heterotic gauge group which corresponds to  
a certain locus of degenerated elliptic fibers in the fourfold. 
This concerned the case of a pure gauge group, corresponding to
having singularities of only codimension $1$ for $X^4$.
For this the Euler number of the fourfold was expressed directly in their 
Hodge numbers, which were matched via a direct spectrum comparison 
with the data of the dual heterotic model; there essential use was made 
of an index-formula, computing the number of even minus 
odd vector bundle moduli. 

Here we will adopt a different approach. 
We will express $\frac{e(X_4)}{24}$ in 
pour Calabi-Yau fourfold data without making any use of 
dual heterotic data. Now in general one will have also singularities
of codimension $2$ and even $3$. The former arise on the $F$-theory
side from intersection curves of the surface components of the discriminant
\footnote{as in the end we want to make a comparison with a dual perturbative 
(up to the five-branes) heterotic model we consider no more general 
discriminant configurations}
(the compact parts
of the seven-branes in the type IIB interpretation of $F$-theory)
: the $I_1$ surface and the $G$ surface $B_2$.
They are interpreted as matter [\ref{BIKMSV}], [\ref{KatzV}]. The idea
is that for example the collision $E_6+I_1$ leads to an $E_7$
which by the adjoint decomposition should correspond to a matter 
hypermultiplet in the ${\bf 27}$ of $E_6$.
On the heterotic side they arise from a similar condition on the
cohomology of the bundle which should lead to matter and is non-trivial
along certain curves where the spectral cover intersects the base
(for one class of matter curves). In certain cases (given in the main part 
of the paper below) of $G$ one can tune 
the class $t$ resp. the bundle so that such an intersection does not 
occur, i.e. so that one has only singularities in codimension one and
only the case of a pure gauge group. In connection with 
the appropriate conditions one is also lead
to a certain lower bound for the "instanton number" of the vector 
bundle conjectured in \cite{R}. This is described in {\em section $2$}.

In general one will have the matter curve in $B=B_2$
and for the $A$ and $D$ groups
$G$ even two of them, called $h$ and $P$
below, which intersect each other in a
codimension three locus, a number of points in $B$. 
The corresponding stratification of the discriminant will allow us
to compute the Euler number of the fourfold by adding up the 
parts with singular fibres. The corresponding computation in 6D, i.e.
for Calabi-Yau threefolds is described in {\em section 3}. Essential
is the consideration of the cusp locus $C=(f_1=0=g_1)$ 
inside the $I_1$ surface component $D_1$ (of the discriminant surface
$D=(4f^3+27g^2=0)$) which is {\em approximately} given by
$4f_1^3+27g_1^2=0$ where in $f_1, g_1$ are split off the parts of $f,g$
causing the $G$ singularity. This is exact for the $E_k$ series; for
the $D_{4+n}=I^*_n$ and $I_n$ series further $n$ powers of $z$, 
the coordinate transversal to $B_1$ in the Hirzebruch surface $B_2$, 
can be extracted 
out of $4f_1^3+27g_1^2$ and one has actually the equation of divisors
$D_1+nr=(4f_1^3+27g_1^2=0)$
where $r$ is the class of $B_1$ in $B_2$. In those cases one finds
that the naive cusp set $C_{old}=(f_1=0=g_1)$ 
(zero dimensional in the 6D case) contains actually
a number $x$ of points of the $B_1$ line (lying on one of the 
matter loci given by a divisor $h$) which are not cusp points\footnote{In
some cases ($A_3,A_4,A_5$) other singularities
arise at these points (tacnodes and even higher double points).}
and have to be taken out of the cusp set so that the true cusp set
is\footnote{note that throughout the paper cohomology classes
like $h\in H^{1,1}(B_2)$ are identified with their pullbacks under
$\pi$ so that here for example is meant $C=C_{old}-x(\pi^*h \cdot r)$}  
$C=C_{old}-xhr$. This $x$ is evaluated as the intersection 
multiplicity of $f_1$ and $g_1$ along $h$ and computed 
via their resultant.
Moreover not only elliptic fibers with 
cusp singularities $y^2=x^3$ lie in the fibers over $C$ but $C$
is also a locus of 'intrinsic' cusp singularities for the $D_1$
locus. So in 6D one has then to apply the usual Pl\"ucker formulas
to $D_1$.

Our general approach in 4D is described in {\em section 4}. Here we also
give the heterotic expectation for $24n_5$ (in the cases of $G=A_1$ and $A_2$
we give the formula for $c_2(V)$ for $V=E_7$ or $E_6$ bundle in the appendix).
In {\em section 5} we develop new "Pl\"ucker formulas" for the
now relevant case of a surface $D_1$ having a curve of cusp (or higher) 
point singularities. These formulas are {\em not just adiabatic 
extensions} of the usual formulas for singular points on a curve.
Actually the story is somewhat more complicated
as one also has to treat the case of curves of tacnode point 
singularities (where a second blow up is needed) which occur in some
cases at the collision of the $D_1$ surface 
with the $G$ surface $B_2$ along the $h$ curve. 
Then we go on to the codimension three
loci in {\em section 6}. There are two types of
codimension three loci: the ones related to enhancements of the fiber at 
the intersection of the matter (=enhancement)
curves and the intersection of the cusp curve $C$ with $B_2$, 
(because of the precise evaluation of $x$ it turns out that these
are actually proportional as cohomology classes)
and the ones related to point singularities of $D_1$. In the final 
{\em section 7} we use the techniques accumulated so far 
to actually compute 
the Euler number of $X^4$ and to show that (with suitable assumptions) 
it equals $24n_5$ from the heterotic side
where $\gamma=0$ is assumed, thereby proving \ref{aim}.

The {\em appendix} contains the explanation why one is led to expect
equ. (\ref{aim}) in a general framework and
reference material pertaining to
the relevant facts about heterotic and $F$-theory $N=1$ models.

{\bf Acknowledgements}: It is a pleasure to thank Paul Aspinwall,
Diuliu-Emanuel Diaconescu, Robert Friedman, 
William Fulton and especially Robert Hain, 
David Morrison and Jonathan Wahl for discussions. B.A. thanks the Institute 
for Advanced Study for its hospitality while doing part of the work. 

\section{The lower bound on $\eta$}

This is a necessary bound on "how much instanton number has to be turned on
to generate/fill out a certain $SU(n)$ bundle", or speaking
in terms of the unbroken gauge group $G$ (the commutator of $SU(n)$ in
$E_8$) "to have no greater unbroken gauge group than a certain $G$".
It is treated here as a warm up because it is related to the
consideration of singularities along just $B_2$ and so 
in codimension one only\footnote{up to the cusp curve in $C_1$} 
(the case that $D_1$ and $B_2$ are disjoint) 
versus singularities in even higher codimension 
(like the matter curves from the intersection of $D_1$ and $B_2$).
We will assume that $G$ is an $ADE$ group.\footnote{The $\eta$ bound
is treated in a toric framework in \cite{Peter}.}

\subsection{$F$-theory arguments}

Let us recall the situation in six dimensions. There the easiest duality
set-up is given by the duality of the heterotic string on $K3$ with
instanton numbers $(12-m,12+m)$ (and no five-branes) with $F$-theory on
the Hirzebruch surface $F_m$ \cite{MV2}. The gauge group there is 
described by the singularities of the fibration and a perturbative
heterotic gauge group corresponds to a certain degeneration over the 
zero-section $C_0$ (of self-intersection $-m$): for example to get an 
$SU(3)$ one needs a certain $A_2$ degeneration over $C_0$ available 
first for $m=3$; in general this means that the discriminant divisor 
$\Delta=12c_1(F_m)$ has a component $\delta(G)C_0$ where $\delta(G)$ is
the vanishing order of the discriminant (equivalently the Euler number of
the affine resolution tree of the singularity), giving also the relation
$m\le \frac{24}{12-\delta(G)}$ for the realization over a $F_m$ to have no
singularity worse than $G$. The last relation follows (cf. \cite{R}):
from the fact that after taking the $C_0$ component with its full 
multiplicity $\delta(G)$ out of $\Delta$ the resulting 
$\Delta^{\prime}=\Delta-\delta(G)C_0$ has transversal intersection 
with $C_0$ and so $\Delta^{\prime}\cdot C_0\ge 0$, leading with $c_1(F_m)
=2C_0+(2+m)f$ to the mentioned result.

So the instanton number $12-m$ to give a $G$ gauge group has to be
$12-m\ge 12-\frac{24}{12-\delta(G)}=(6-\frac{12}{12-\delta(G)})c_1(B_1)$ 
with $B_1$ the common $P^1$ base of the heterotic $K3$ resp. the $F_m$. 
{}From this it was conjectured in \cite{R} that a similar bound could
in four dimensions look like the generalizations of both sides of the 
six-dimensional bound, i.e. in view of the fact that the $(12-m,12+m)$
structure generalizes in four dimensions to $\eta_1=6c_1-t, \eta_2=6c_1+t$
(for this cf. the anomaly cancellation condition $c_2(V_1)+c_2(V_2)+a_fF=
c_2(Z)$ and its component $\eta_1\sigma+\eta_2\sigma=12c_1\sigma$ 
concerning the classes not pull-backed from $H^4(B)$ for the case of
an $A$ model with $W_B=0$)
\beqa
\eta_1\ge (6-\frac{12}{12-\delta(G)})c_1
\eeqa

Let us now first prove this in the $A$ model with $W_B=0$ and then 
include a non-zero $W_B$.
For this recall that the association in six dimensions of the heterotic 
$(12-m,12+m)$ with $F_m$ on the $F$-theory side generalizes \cite{FMW}
to the association of the heterotic $\eta_1=6c_1-t, \eta_2=6c_1+t$ with
the following structure of the $F$-theory base $B_3$ as a ${\bf P}^1$ 
bundle over the common (with the heterotic side) base $B=B_2$. 
Look at the ${\bf P}^1$ bundle as projectivization of a vector bundle
${\cal O}\oplus {\cal T}$ with ${\cal T}$ a line bundle over $B$ of
$c_1({\cal T})=t$ (this generalizes the twisting condition in the 
Hirzebruch surface). To make actual computations let us introduce 
homogeneous coordinates $a,b$ which are sections of ${\cal O}(1)$ and
${\cal O}(1)\otimes {\cal T}$ respectively, where ${\cal O}(1)$ is
the ${\cal O}(1)$ bundle on the ${\bf P}^1$ fibers of 
$c_1({\cal O}(1))=r$, say, and $r(r+t)=0$ as $a,b$ have no common zeroes
(the disjointness of the zero-section and the section at infinity in the
Hirzebruch surface case). Adjunction gives then 
\beqa
c_1(B_3)=c_1+2r+t
\eeqa
and the condition 
\beqa
(12c_1(B_3)-\delta(G)B_2)B_2\ge 0
\eeqa
gives with $B_2=r$ that
\beqa
12c_1+(\delta(G)-12)t\ge 0
\eeqa
resp. formulated in $\eta_1=6c_1-t$ the bound to be proved.

Now let us include the effect of a non-zero $W_B$. From six dimensions one
knows that a heterotic five-brane corresponds to a blow-up in the 
$F$-theory base. So here we have to consider the impact of the ruled 
surface $S$ (in the thereby modified $\tilde{B_3}$) over 
$W_B$ in $B=B_2$. 
Its contribution is 
\beqa
c_1(\tilde{B_3})=c_1(B_3)-S
\eeqa
leading after intersection with $B$ in the inequality above to a term
$-W_B$ on the left hand side or $+W_B$ on the right hand side. On the 
other hand one has now that $\eta_1+W_B=6c_1-t$ (we think already 
of the case $\eta_2=0$) so that the final bound is unchanged
\beqa
\eta_1+W_B\ge (6-\frac{12}{12-\delta(G)})c_1+W_B
\eeqa

\subsection{Heterotic arguments}

As the $4D$ bound was guessed from a $6D$ expression, 
let us point out that it
is also possible to see the bound from an adiabatic argument. 
Namely, assume
that $B_2$ is the adiabatic extension of a $6D$ base $B_1$, 
i.e. that $B_2$
is a Hirzebruch surface $F_n$. 
Now as remarked in \cite{R} one has from the investigation
of \cite{FMW3} that one of the inequalities $\eta\ge i\cdot c_1$ 
holds for some
$i$ in $2,\dots \, n$. Restricting to a fibre $f={\bf P}^1$ of $F_n$,
i.e. restricting to a $K3$ fibre of $Z$, one gets from $f\cdot c_1(F_n)=2$
and the $6D$ bound that the relevant $i$ is the same as occurring 
in $6D$ if one
writes there the bound as proportionality to $c_1(B_1)=2$.

Yet another argument starts from the observation that the class 
$n\sigma+\eta$ 
of 
the spectral cover should be effective, so its image $-nc_1+\eta$ 
in the base
should be effective too. Now $n$ itself is equal to
the $6-\frac{12}{12-\delta(G)}$, where $\delta(G)=11-n$, for $n=2,3$ , 
respectively
equals the next bigger integer for $n=4$ and is even greater 
(and so implying 
the lower bound from the stronger 
spectral cover effectiveness argument $\eta\ge nc_1$) for higher $n$.

\section{The three-dimensional case}

{\resetcounter}

Before coming to the actual computation of $e(X)$ let us briefly 
review the three-dimensional situation, i.e having a Calabi-Yau three-fold
$Z$ which is elliptically fibered over a two-dimensional base $B_2$. 
For that we reconsider first the case of having a smooth $Z$, 
i.e the elliptic fiber does not degenerate worse than $I_1$ (resp.
$II$ at the cusp points) over the discriminant$D$. Then 
we proceed and consider 
the case where the elliptic fiber has a $G$-singularity ($G$ will be always 
one of the $ADE$ groups and we will always be in the {\em split} case of 
[\ref{BIKMSV}]) localised over a
codimension one locus in $B_2$. In our set-up $B_2$ will be a Hirzebruch 
surface $F_n$, i.e. a $P^1$ fibration over $B_1=P^1$. Note that apart from 
$G=E_8$ where $n=12$ (so we are in the first
column of table $A.1$ of [\ref{CF}]) we restrict ourselves to $n=0,1,2$ (
the first three rows of the table mentioned). (Because of our interchange 
(compared to the usual convention) of the bundles one has strictly speaking 
to put $n=0,-1,-2$ in the formulae below.) 

The fiber enhancement is given as follows: the matter loci are specified in 
[\ref{BIKMSV}] and [\ref{KatzV}] relates matter and fiber enhancement 
(intuitively one may think for example of the $\bf 27$ matter 
locus of $E_6$ as located at
the $E_7$ fiber enhancement points given by the collision of the $E_6$ line 
$B$ and the $I_1$ curve $D_1$).

{}For background and notation of the elliptically fibered geometry 
cf. appendix.

\subsection{smooth case}
In case that $B$ is two-dimensional, $D$ is a curve in $B_2$ of class
$D= 12c_1(B_2)$. The three-dimensional Calabi-Yau $Z$ over $B_2$
is described by a smooth Weierstrass model, so one has only 
type $I_1$ (and $II$) singular fibers over $D$ which contribute to
$\chi(Z)$. The idea of an Euler number computation 
from the elliptic fibration data
is of course $e(\tilde{Z})=e({\rm sing.\, fiber})e(D)$. 
Since $D$ is a curve we
have $-D(D-c_1(B_2))=-132c_1^2$ (where $c_i=c_i(B_2)$). But $D$ 
itself will be singular at those points where the divisors 
associated to the classes $F=4c_1(B_2)$ and $G=6c_1(B_2)$ collide, i.e. at
$F\cdot G=24c_1^2$ points. At these points $D$ develops a cusp and the 
elliptic fiber will be of type II. Using the standard Pl\"ucker formula,
which takes the double points and cusps 
into account (cf. \cite{GHarris}), one gets 
$e(\tilde{D})=-132c_1^2+2(24c_1^2)$, and so we get 
\beqa 
e(Z)=e(I_1)(-84c_1^2-24c_1^2)+e(II)(24c_1^2)=-60c_1^2. 
\eeqa
\subsection{singular case}
Now assume that $Z$ has a section of $G$-singularities localised over 
the base curve in the Hirzebruch surface $F_n=B_2$. 
Consider in $F_n$ the two rational 
curves given of self-intersection $-n$ resp. $+n$ given 
by the zero section $S_0$ and the section at infinity 
$S_{\infty}=S_0+nf$ of the ${\bf P}^1$ bundle. Let us localize the $G$ 
fibers along $S_0$, where we have an eye on a 
dual perturbative heterotic\footnote{we will have $12-n$ resp. $12+n$ 
instantons on the heterotic side corresponding 
to $S_{0}$ resp. $S_{\infty}$; we put the greater number into the {\em second}
bundle where we want to span an $E_8$ bundle} description. Now, we 
can decompose the discriminant 
$D$ into two components: $D=D_1+D_2$, where $D_1$ 
denotes the component with generic $I_1$ fibers and $D_2$ has $G$ fibers. 
Each component is characterized by the order of vanishing of 
some polynomials 
as $D$ itself. Denote the class of $D_2$ by $D_2=cS_0$, resp. 
$F_2=aS_0$ and $G_2=bS_0$. With the canonical bundle of the 
Hirzebruch surface
$K_{F_n}=-2S_0-(2+n)f$ we get $D_1=(24-c)S_0+(24+12n)f$, resp. 
$F_1=(8-a)S_0+(8+4n)f$ and $G_1=(12-b)S_0+(12+6n)f$, so describing
the locus of $I_1$ singularities.

Since only singular fibers contribute to $\chi(Z)$, we get 
$e(Z)=e(D_1)e({\rm{I}}_1)+e(S_0)e(G)$.
Now $D_1$ is a curve in the base, which has cusp singularities at 
$F_1G_1=192+(6n-12)a+(4n-8)b-abn$ points, so 
applying the standard Pl\"ucker formula, we find 
$e(D_1)=-D_1(D_1+K_{F_{n}})+2F_1G_1= -1056+(46-23n)c+c^2n+2F_1G_1$. The cusps 
contribute with $e(II)F_1G_1$ to $e(Z)$; also we have to take into  
account that the $D_1$ branch will intersect the branch of 
$G$-singularities
$S_0$ in a number of points (which will modify the cusp set $F_1G_1$ for 
$G=I_n, I_n^*$). 
\beqa
e(Z)&=&\; \; \; e({\rm{I}}_1)\bigg( e(D_1)-e(D_1\cap S_0)-F_1G_1)\bigg)
\nonumber\\
& &+e({\rm{II}})F_1G_1\nonumber\\
& &+e(G)\bigg( (e(S_0)-e(D_1\cap S_0)\bigg) \nonumber\\
& &+\sum_{i\in {\cal M}}e(G^{enh}_i)e(i)\nonumber\\
&=&-480+(18n-36)a+(12n-24)b+(48-23n)c+(c^2-3ab)n
\nonumber\\
& &+\sum_{i\in {\cal M}}e(i)\bigg( e(G^{enh}_i)-e(G)-1\bigg)
\eeqa
where ${\cal M}$ is the set of components of the intersection of $D_1$ and 
$S_0$. For example, for $A_4=I_5$ one has $D_1S_0=4h_{c_1-t}+P_{8c_1-3t}$, 
so ${\cal M}$ consists of $h$ and $P$; further $e(h)=2-n, e(P)=16-3n$ and 
$e(G)=5$, $e(G_h^{enh})=6$ and $e(G_P^{enh})=7$ corresponding to the 
generic $I_5$ fibre, the $I_6$ enhancement fibre and the $D_5$ 
enhancement fiber.   

Let us now give a number of cases which illustrate the above formula. 
To do so we proceed as follows: first, we read off the necessary 
information about the base geometry from the discriminant, 
then we compute the 
the Euler characteristic of $Z$ and compare our results with the heterotic 
string side.

\subsubsection{$E_8(II^*)$ singularity}
{}From the discriminant of $Z$ 
\beqa
\Delta=z_1^{10}(g^2_{(12-n)}(z_2)+{\cal O}(z_1^2))
\eeqa
we learn that the $D_1$ locus has $(12-n)$ double points
which contribute to $e(D_1)$ and which lying on the intersection 
points of $D_1$ and $S_0$, so we have to apply the Pl\"ucker 
formula for curves which leads to a $2(12-n)$ contribution to the Euler
number of $D_1$. Further $F_1S_0=8$ and $G_1S_0=(12-n)$ and the Euler
number of $Z$ is given by
\beqa
e(Z)=-240-60n
\eeqa
Let us compare this result with the heterotic string side. There we find
$dim{\cal M}_{12+n}(E_8)+dim{\cal M}_{12-n}(SU(1))+h^{1,1}(K3)=144+29n$
which contribute to the number of hypermultiplets further we have $248$ 
vectors and $13-n=1+12-n$ tensors satisfying the gravitational anomaly 
equation $273-144-29n+248=29n_T$. This leads to the prediction for
$h^{2,1}$ resp. $h^{1,1}$ of the corresponding F-theory model
$h^{2,1}(Z)=152+28n$, $h^{1,1}(Z)=8+2+1+12-n=23-n$ giving 
$\chi(Z)=-240-60n$ which is in agreement with our computation above.

\subsubsection{$E_7^s(III^*)$ singularity}
Here the discriminant is given by 
\beqa
\Delta=z_1^9(4f^3_{8-n}(z_2)+{\cal O}(z_1))
\eeqa
telling us $D_1S_0=3(8-n)$ so that we should expect an enhancement at 
$(8-n)$ points, i.e. 
$e(D_1\cap S_0)(e(G_{enh})-e(G)-1)=(8-n)(e(II^*)-e(III^*)-1)=0$. 
But actually\footnote{as pointed out by 
Aspinwall [\ref{asp}]}
the fibre over these points is not\footnote{although a generic slice 
through the singularity might one lead 
to believe it looks like $E_8$; but the resolution of the threefold
will not give the full $E_8$ when one does the blow-up explicitly (cf. 
[\ref{Mir}])} of Kodaira type $II^*$ 
(whose affine diagram is of Euler number $10$)
but consists of a chain of $8$ $P^1$'s (which is not a Kodaira fibre)
and has Euler number
9 giving an $-1(8-n)$ contribution to the total Euler number of $Z$. 
Thus we
find (note also that $F_1S_0=(8-n)$, $G_1S_0=(12-n)$)
\beqa
e(Z)=-284-56n
\label{e6DE7}
\eeqa
{}From the heterotic side we get 
\beqa
dim{\cal M}_{12+n}(E_8)+dim{\cal M}_{12-n}(SU(2))+h^{1,1}(K3)=153+28n=n_H
\eeqa
so that $n_H=153+28n$ and $n_V=133$ satisfying the anomaly equation
$244+133=153+28n+\frac{56}{2}(8-n)$ and giving
$h^{2,1}(Z)=152+28n$ and $h^{1,1}(Z)=10$, thus equ. (\ref{e6DE7}).

\subsubsection{$E_6^s(IV^{*})$ singularity}
\beqa
\Delta=z_1^8(27q^4_{6-n}(z_2)+{\cal O}(z_1))
\eeqa
so we expect $e(D_1\cap S_0)=D_1S_0/4=(6-n)$ collisions 
between the $IV^*$ and $I_1$
fiber, further we have $F_1S_0=(8-n)$ and $G_1S_0=2(6-n)$. 
So {\em Katz/Vafa collision rules} give  
$e(D_1\cap S_0)(e(G_{enh})-e(G)-1)=(6-n)(e(III^*)-e(IV^*)-1)=0$ and we find 
\beqa
e(Z)=-300-54n
\eeqa
which can be checked on the heterotic side:
$dim{\cal M}_{12+n}(E_8)+dim{\cal M}_{12-n}(SU(3))+h^{1,1}(K3)=
160+27n=n_H$
and $n_V=78$, satisfying $244+78=160+27n+27(6-n)$
and leading to $h^{2,1}(Z)=159+27n$ and $h^{1,1}(Z)=9$.

\subsection{A subtlety concerning the cusp set}\label{xcusp}

Before we will proceed and consider some $I_n^*$ and $I_n$ examples, 
we have to make a digression concerning the cusp set in these examples. 

The reason for that is that, contrary to the case of the $E_k$ series,
now the Kodaira values $a$ and $b$ in $f=f_1+ar, \; g=g_1+br$ 
do not lead by themselves to the 
value $c$; instead they would always lead to $I_0^*$ and $I_0$. To get
actually a higher $n$ one has to tune the occurring expressions 
$f_1,\; g_1$ so that in the discriminant $n$ more powers of $z$ (the
local coordinate transversal to $B_1$) than naively 
expected (i.e. $6$ for the $D$ case and $0$ for the $A$ case) can
be extracted. 

Recall that we have ($z_1$ is the base variable of divisor $r$) 
\beqa
f(z_1,z_2)&=&\sum_{i=a}^{I}z^i_1f_{8-n(4-i)}(z_2)\nonumber\\
g(z_1,z_2)&=&\sum_{j=b}^{J}z^j_1g_{12-n(6-j)}(z_2)
\eeqa
(cf. [\ref{BIKMSV}]). 

Let us consider this in the example of $D_5$ where $(a,b,c)=(2,3,7)$ and 
\beqa
D\sim z^6(4f_1^3+27g_1^2)
\label{disc}
\eeqa
But as we have to force a $z^7$ the coefficients of $z^6$
have to cancel which leads to the conditions
$f_{4c_1-2t}\sim h^2_{2c_1-t}$ and 
$g_{6c_1-3t}\sim h^3_{2c_1-t}$; furthermore
from the split condition (to get really $E_6$ and not $F_4$)
one gets $g_{6c_1-2t}+f_{4c_1-t}h_{2c_1-t}=q_{3c_1-t}^2$.
Altogether this leads to an equation for $D$ 
\beqa
D\sim z^7[h^3_{2c_1-t}q_{3c_1-t}^2+{\cal O}(z)]
\label{discrefined}
\eeqa

Similarly for $I_5$, say, one has $(a,b,c)=(0,0,5)$ 
and again the cancellation of the leading terms
(there are now higher cancellation conditions as well) leads to
$f_{4c_1-4t}\sim h_{c_1-t}^4$ and $g_{6c_1-6t}\sim h_{c_1-t}^6$.
All the conditions including the split condition lead to a
description by four further relevant sections besides $h_{c_1-t}$,
namely $H_{2c_1-t},\, q_{3c_1-t}, \, f_{4c_1-t}, \, g_{6c_1-t}$ and
a discriminant
\beqa
D\sim z^5 h_{c_1-t}^4 P_{8c_1-3t}
\eeqa

So the fact that we have to enforce a higher power of $z$ to
be extractable leads to the cancellation conditions which come down
to $f_1r=2h_{2c_1-t},\; g_1r=3h_{2c_1-t}$ for the $I^*_n$ series
(for $n>0$) and to $f_1r=4h_{c_1-t},\; g_1r=6h_{c_1-t}$ for the 
$I_n$ series (for $n>0$). 
This fact then, that $f_1$ and $g_1$ have a component $h$ in common,
changes the actual cusp set from $f_1g_1$ to 
$f_1g_1-x\cdot h$ where $x$ is the intersection multiplicity of
$f_1$ and $g_1$ at $h$ (as computed by the vanishing order of the
resultant) which counts the number of times $h$ 
lies in the intersection product. So to compute $x$ we have
to determine the $f_1$ and $g_1$ polynomials, i.e. express them in local
data near the collision point $D_1r$. This can be done by using the 
more general Weierstrass equation 
\beqa
y^2+a_1xy+a_3y^3=x^3+a_2x^2+a_4x+a_6
\eeqa
where the $a_i$'s are locally defined polynomial functions on the base 
as $f$ and $g$ (for details see [\ref{BIKMSV}]). 
Further one can express the $f$ and $g$ 
polynomial in terms of the $a_i$'s 
\beqa
f&=&-\frac{1}{48}((a_1^2+4a_2)^2-24(a_1a_3+2a_4))\nonumber\\
g&=&-\frac{1}{864}(-(a_1^2+4a_2)^3+36(a_1^2+4a_2)(a_1a_3+2a_4)
-216(a_3^2+4a_6))
\eeqa       
The local structure (orders in $z$) 
of the $a_i$'s is given by [\ref{BIKMSV}] 
(we are always in the split case)
\begin{center}
\begin{tabular}{c||c|c|c|c|c}
$G$ & $a_1$ & $a_2$ &$a_3$ &$a_4$ &$a_6$\\ \hline
\hline
$I_{2}$ & $0$ & $0$ & $1$ & $1$ & $2$  \\ 
\hline
$I_{2k+1}$ & $0$ & $1$ & $k$ & $k+1$ & $2k+1$  \\ 
\hline
$I_{2k}$ & $0$ & $1$ & $k$ & $k$ & $2k$  \\ 
\hline
$I^*_0$ & $1$ & $1$ & $2$ & $2$ & $4$ \\
\hline
$I^*_1$ & $1$ & $1$ & $2$ & $3$ & $5$ \\
\hline
$I^*_2$ & $1$ & $1$ & $3$ & $3$ & $5$ \\
\end{tabular}
\end{center}
The $h$-locus, related to the relevant enhancement, is
given for the $I$ series by $a_1$ (with the exception of $I_2$ where
not $h_{c_1-t}$ but $H_{2c_1-2t}$ is relevant and where the
corresponding enhancement locus is given by $b_2=a_1^2+4a_2$)
and for the $I^*$ series
by $a_2$ for $n>0$ and by $a_4$ for $n=0$. This is actually 
refined with a corresponding $z$ power according to $t=a_{2,1}=a_2/z$ for 
example for $D_5$ (cf. [\ref{BIKMSV}]).

Concerning the value of $x$ one finds that for the $I_n$ series
(with two exception: $x=3$ for $n=2$, but this would give $x=6$ if $H$ were
$h^2$ thus fitting the pattern of the $I$ series, and $x=8$ for $n=3$) 
with $n=4,5,6$ it is given by $x=3n$ 
and for $I^*_0,\; I^*_1, \; I^*_2$ it is given by $x=0,\; 2,\; 3$.

{}Furthermore we have to note that there is another twist in the story
which comes from the fact that $D_1$ will have a tacnode or an even higher
double point
when colliding with $r=S_0$ for the cases $I_4, \; I_5, \; I_6$ at the points 
on $D_1=r=S_0$ of $h$, 
that is a singular point of the form $t^m+z^2=0$ with $m=4$ 
(tacnode, for $I_4$) resp. $m=6$ for $I_5, I_6$
(after suitable coordinate change).

\begin{center}
\begin{tabular}{c|c|c|c}
\hline
$\rm{Group}$ & $\rm{f_1,g_1}$ & $\rm{sing.\ \ at}\ \ t$ &
 $x$\\ \hline
\hline
$I_2$ & $-\frac{1}{48}(t^2-72z)$ & $\rm{-}$ & $3$ \\ 
$$ & $-\frac{1}{864}(-t^3+108tz-1080z^2)$ & $$ & 
$$ \\ 
\hline
$I_3$ & $-\frac{1}{48}((t^2+4z)^2-24(tz+2z^2))$ & $\rm{-}$ & $8$ \\ 
$$ & $-\frac{1}{864}(-(t^2+4z)^3+36(t^2+4z)(tz+2z^2)-216(z^2+4z^3))$ &
 $$ & 
$$ \\ 
\hline
$I_4$ & $-\frac{1}{48}((t^2+4z)^2-24(tz^2+2z^2))$ & $t^4+v^2$ & 
$12$ \\ 
$$ & $-\frac{1}{864}(-(t^2+4z)^3+36(t^2+4z)(tz^2+2z^2)-216(5z^4))$ & $$ & 
$$ \\ 
\hline
$I_5$ & $-\frac{1}{48}((t^2+4z)^2-24(tz^2+2z^3))$ & $t^6+v^2$ &
 $15$ \\ 
$$ & $-\frac{1}{864}(-(t^2+4z)^3+36(t^2+4z)(tz^2+2z^3)-216(z^4+4z^5))$ &
 $$ & 
$$ \\ 
\hline
$I_6$ & $-\frac{1}{48}((t^2+4z)^2-24(tz^3+2z^3))$ & $t^6+v^2$ &
 $18$ \\ 
$$ & $-\frac{1}{864}(-(t^2+4z)^3+36(t^2+4z)(tz^3+2z^3)-216(5z^6))$ & $$ & 
$$ \\ 
\hline
\hline
$I_0^*$ & $-\frac{1}{48}((z^2+4z)^2-24(z^3+2tz^2))/z^2$ & $\rm{-}$ &
 $0$ \\ 
$$ & $-\frac{1}{864}(-(z^2+4z)^3+36(z^2+4z)(z^3+2tz^2)-1080z^4)/z^3$ &
 $$ & 
$$ \\
\hline
$I_1^*$ & $-\frac{1}{48}((z^2+4tz)^2-72z^3)/z^2$ & $\rm{-}$ &
 $2$ \\ 
$$ & $-\frac{1}{864}(-(z^2+4tz)^3+108(z^2+4tz)z^3-216(z^4+4z^5))/z^3$ &
 $$ & 
$$ \\
\hline
$I_2^*$ & $-\frac{1}{48}((z^2+4tz)^2-24(z^4+2z^3))/z^2$ &
 $-$ & $3$ \\ 
$$ & $-\frac{1}{864}(-(z^2+4tz)^3+36(z^2+4tz)(z^4+2z^3)
-216(z^6+4z^5))/z^3$ &
 $$ & 
$$ \\
\hline
\end{tabular}
\end{center}

\subsubsection{$D_4^s(I^*_0)$ singularity}
\beqa
\Delta=z_1^6((h^2_{4-n}+q^2_{4-n})(h^2_{4-n}+\omega q^2_{4-n})
(h^2_{4-n}+\omega^2q^2_{4-n})+{\cal O}(z))
\eeqa
Here we find $D_1S_0/6=(4-n)$ intersection points between the $I_0^*$
and the $I_1$ locus. Further we have $F_1S_0=2(4-n)$ and $G_1S_0=3(4-n)$.
Since we have no additional corrections from the cusps ($x=0$) 
and assuming the {\em Katz/Vafa collision rules} applies in the form
$e(G^{enh}_i)=e(G)+1$ we find
\beqa
e(Z)=-336-48n
\eeqa
Now let us see which prediction comes from the heterotic side. There we
find for the number of hyper-multiplets $3\cdot 8(4+n)-28=68+24n$ and 
additional ones coming from 
$dim_Q({\cal M}_{inst}^{(n_1+n_2)})+h^{1,1}(K3)=88+20$ giving a total 
$n_H=176+24n$ and with $n_V=28$ vectors we find that
the anomaly equation $244+28=176+24n+3\cdot 8(4-n)$ is satisfied. 
Thus we find
$h^{2,1}(Z)=175+24n$ and $h^{1,1}(Z)=7$ and so $e(Z)=-336-48n$ 
in agreement with our F-theory computation.

\subsubsection{$D_5^s(I^*_1)$ singularity}
\beqa
\Delta=z_1^7(h^3_{4-n}q^2_{6-n}+{\cal O}(z_1))
\eeqa
We have $D_1S_0=3(4-n)+2(6-n)$ and $F_1S_0=2(4-n)$, $G_1S_0=3(4-n)$ and
further we have to take into account the change of the cusp set 
since $x=2$, i.e. we find $C=F_1G_1-2(4-n)$. Assuming 
the {\em Katz/Vafa collision rules}
we get 
\beqa 
e(Z)=-312-52n
\eeqa
Now the heterotic side gives 
$dim_Q({\cal M}_{inst}^{(n_1+n_2)})+h^{1,1}(K3)=66
+20$ hypers resp. $(4+n)16+(6+n)10-45=79+26n$ hypers so $n_H=165+26n$ and 
$n_V=45$ satisfying the anomaly cancellation
$244+45=165+26n+16(4-n)+10(6-n)$ and leading to $h^{2,1}(Z)=164+26n$ 
resp. $h^{1,1}(Z)=8$ thus $e(Z)=-312-52n$.

Let us give for later reference the equation for the $D_1$ part of the
discriminant; here we will already use the 4D notation so that for example
the degree $4-n$ becomes the class $2c_1-t$.

Now $D_1$ is given by
$h_{2c_1-t}^3q_{3c_1-t}^2+{\cal O}(z)=0$. Let us make the 
accompanying\footnote{one sees
from $r^2=-rt$ that for ascending powers $i$ of $z$ also 
the $t$ coefficient rises and so the degree of the accompanying term
changes as $12c_1-(12-i)t$, keeping always (including the overall
$z^7$) $12c_1-(12-i)t-it=12c_1-12t=Dr$} term of the power $z$ explicit in 
$h_{2c_1-t}^3q_{3c_1-t}^2+S_{12c_1-4t}z+{\cal O}(z^2)$.
Explicitely one finds with (using the notation $f_i:=f_{4c_1-it}, \;
g_i:= g_{6c_1-it}$ and denote our former $f_1, g_1$ by $F_1, G_1$ if
there is change of confusion) 
\beqa
F_1&=&\frac{1}{48}(- h^2+zf_1+z^2f_0)\nonumber\\
G_1&=&\frac{1}{864}(h^3+z(q^2-\frac{3}{2}f_1h)+z^2g_1+z^3g_0)
\label{D5f1g1}
\eeqa
that
\beqa
32\cdot 864\; \Big( 4F_1^3+27G _1^2\Big) &=&z\Bigg( 2h^3q^2\nonumber\\
& &\;\;\;\;  +(-\frac{3}{4}f_1^2h^2+2g_1h^3+3f_0h^4-3f_1hq^2
+q^4)z\nonumber\\
& &\;\;\;\;  +(f_1^3-3f_1g_1h-6f_1f_0h^2+2g_0h^3+2g_1q^2)z^2\nonumber\\
& &\;\;\;\;  +(3f_1^2f_0+g_1^2 -3f_1g_0h-3f_0^2h^2+2g_0q^2)z^3 \nonumber\\
& &\;\;\;\;  +(3f_1f_0^2+2g_1g_0)z^4\nonumber\\
& &\;\;\;\;  +(f_0^3+g_0^2)z^5\Bigg)
\eeqa
so that 
\beqa
S_{12c_1-4t}=h^2(-\frac{3}{4}f_1^2+2g_1h+3f_0h^2)-3f_1hq^2+q^4
\eeqa

\subsubsection{$D_6^s(I^*_2)$ singularity}

As explained below one finds the following discriminant structure
(with $\Delta=32\cdot 864 (4f_1^3+27g_1^2)$)
\beqa
\Delta=z_1^8(h^2_{4-n}P^2_{8-n}+{\cal O}(z_1))
\eeqa
Note that in this case we take
for $V_1$ a product bundle $SU(2^{(1)})\times SU(2^{(2)})$
with $\eta_1^{(1)}=2c_1$ and $\eta_1^{(2)}=4c_1-t$ (this is the case $r=0$ 
of [\ref{BIKMSV}]).

We have $D_1S_0=2(4-n)+2(8-n)$ and $F_1S_0=2(4-n)$, $G_1S_0=3(4-n)$ and
further we have to take into account the change of the cusp set 
since $x=3$, i.e. we find $C=F_1G_1-3(4-n)$. Assuming 
the {\em Katz/Vafa collision rules} we get 
\beqa 
e(Z)=-276-57n
\eeqa
Now the heterotic side gives 
$dim_Q({\cal M}_{inst}^{(n_1+n_2)})+h^{1,1}(K3)=36
+20$ hypers resp. $(4+n)16+(8+n)12-66=94+28n$ hypers so $n_H=150+28n$ and 
$n_V=45$ satisfying the anomaly cancellation
$244+28=150+28n+32\frac{1}{2}(4-n)+12(8-n)$ and 
leading to $h^{2,1}(Z)=149+28n$ 
resp. $h^{1,1}(Z)=9$ thus predicting $e(Z)=-280-56n$.

The further necessary contribution $-1(4-n)$ is explainable as follows:
the enhanced fibre over these points is not (although a generic slice 
through the singularity might one lead 
to believe it looks like $E_7$; but the resolution of the threefold
will not give the full $E_7$ when one does the blow-up explicitly a la
Miranda) of Kodaira type $III^*$ 
(whose affine diagram is of Euler number $9$)
but consists of a chain of $7$ $P^1$'s (which is not a Kodaira fibre)
and has Euler number 8 giving an $-1(4-n)$.

One finds with
\beqa
F_1&=&\frac{1}{48}(- h^2+zf_1+z^2f_0)\nonumber\\
G_1&=&\frac{1}{864}(h^3+zg_2+z^2g_1+z^3g_0)
\eeqa
that the condition to have $c=8$ leads to 
\beqa
g_2=-\frac{3}{2}hf_1
\eeqa
and to get the 'split' $SO(12)$ situation (with parameter $r=0$) 
one has [\ref{BIKMSV}] to introduce $q=q_{4c_1-t}$ and
$u=u_{2c_1}$ and to impose the conditions
\beqa
f_1&=&q+hu\nonumber\\
g_1&=&\frac{3}{4}qu
\eeqa
(note that [\ref{BIKMSV}] have $a_{4,3}\sim F_1,\; a_{6,5}\sim G_1$). 
This gives
\beqa 
\Delta \sim z^8(-\frac{3}{4}h^2(f_1^2-\frac{8}{3}g_1h-4f_0h^2)+{\cal O}(z))
\eeqa
The identification $a_{4,3}^2-4a_{2,1}a_{6,5}$ 
of the second enhancement locus given in [\ref{BIKMSV}]
shows that for $f_0=\frac{1}{4}u^2$ one has
$f_1^2-\frac{8}{3}hg_1-4f_0h^2=(q+hu)^2-2hqu-u^2h^2=q^2=:P^2$.

\subsubsection{$A_{k-1}^s(I_k)$ singularity}

We consider now the $I_k$ series for $k=2,3,4,5,6$ where the leading
term of the discriminant
is given by $z^2H^2_{4-2n}P_{16-6n}$ for $I_2$, by $z^3h^3_{2-n}P_{18-6n}$
for $I_3$ and by
$z^kh^4_{2-n}P_{16-(8-k)n}$ for $I_k$ with $k=4,5,6$. One has always
$FS_0=4(2-n),\; GS_0=6(2-n)$ and $D_1S_0=4(2-n)+(16-(8-k)n)$ and the
cusp set is given by $C= F_1G_1-x(2-n)$ with $x$ the intersection 
multiplicity of $f_1$ and $g_1$ at $h=0$. Including the additional 
singularity contributions (tacnode, etc.) 
at the $h$-points of $D_1\cap r$ ($r=B_1=S_0$)
for $I_4,I_5,I_6$ considered below one gets 
\begin{center}
\begin{tabular}{c|c}
\hline
$G$ & $\chi(Z)$ \\
\hline
\hline
$I_2$ & $-420-24n$ \\
\hline
$I_3$ & $-384-36n$ \\
\hline
$I_4$ & $-352-44n$ \\
\hline
$I_5$ & $-320-50n$ \\
\hline
$I_6$ & $-288-54n$ \\
\hline
\end{tabular}
\end{center}
Note that the Euler numbers (cf. [\ref{CF}]) 
match with the heterotic expectations
for the spectrum. 

Now let us look at the tacnode (and higher double point) singularities
of $D_1$ at $h$ mentioned above (cf. in the following the explicit 
discriminant forms given in the appendix).

{\bf $I_4$}

Note that on $h$ the equation of $D_1$ is given by ($e:=f_2+H^2$)
\beqa              
-\frac{3}{4}e^2(h^4-2h^2Hz+(H^2-\frac{4}{3}e)z^2)+{\cal O}(z^3)=0
\eeqa
So the fact that we do not have a complete square structure
in the leading terms (because the expression $\frac{4}{3}e$) shows that we
have a generic tacnode structure at $h$, in contrast 
to the cases $I_5$ and $I_6$. So the $6D$ Euler number contribution
will be $+4-1=3$ (taking into account that one has to go back to the 
singular model) at each of the $(2-n)$ intersection points of the $h$
component of the intersection of $D_1$ with the $B_1$ line.

{\bf $I_5$}

Note that near $h$ the equation of $D_1$ is given by (with $e:=f_1H+q^2$)
\beqa
-3Hq^2(h^4-2h^2Hz+H^2z^2)+(-H^2(2g_1H+3f_1^2)+\frac{9}{4}e^2)z^3
+(f_1^3+3g_1e)z^4+g_1^2 z^5=0\nonumber
\eeqa
So the complete square structure of the leading terms shows that we do not
have a generic tacnode structure, just as for $I_6$ but in contrast 
to the case $I_4$.
If we replace the variable $z$ by $w:=Hz-h^2$
the terms up to third order become (everything up to coefficients)
$w^2+z^3\rightarrow h^6+3h^4w+w^2+w^3\sim h^6+3h^4w+w^2$ 
near $(h,w)=(0,0)$ which
goes with $w:=v-\frac{3}{2}h^4$ to the normal form $h^6+v^2$.
So the $6D$ Euler number contribution
will be
$+6-1=5$ (taking into account that one has to go back to the 
singular model) at each of the $(2-n)$ intersection points of the $h$
component of $D_1\cap r$. One finds that one has to adopt a refined 
analysis to get a missing contribution $+1(2-n)$.

{\bf $I_6$}

Note that on $h$ the $D_1$ equation is given by 
($e:=\frac{1}{2}f_1+{\cal F}H\; e':=f_1+{\cal F}H$)
\beqa
-3e^2(h^4-2h^2Hz+H^2z^2)
+f_1(\frac{9}{2}{\cal F}He'+f_1^2)z^3+\frac{9}{2}{\cal F}^2e'^2z^4
\eeqa
So the complete square structure of the leading terms shows that we 
do not have a generic tacnode structure, 
just as for $I_5$ but in contrast to the case $I_4$. With $w:=Hz-h^2$
the terms up to third order become (everything up to coefficients)
$w^2+z^3\rightarrow h^6+3h^4w+w^2+w^3\sim h^6+3h^4w+w^2$ 
near $(h,w)=(0,0)$ which
goes with $w:=v-\frac{3}{2}h^4$ to the normal form $h^6+v^2$.
So the $6D$ Euler number contribution
will be $+6-1=5$ (taking into account that one has to go back to the 
singular model) at each of the $(2-n)$ intersection points of the $h$
component of the intersection of $D_1$ with the $B_1$ line.

\section{The four-dimensional case}

{\resetcounter}

In this section we start after the foregoing introductory sections 
with the Euler number computation in the
four-dimensional case. Here we give the smooth case and in the general
case the relation to the heterotic situation. In general we will 
have to consider two types of contributions: singular fibers
(corresponding in codimension one to $G$ over $B_2$ and $I_1$ over $D_1$,
this is the generic situation in the discriminant surface inside $B_3$; this
is enhanced at the matter curves in $B_2$ and at the cusp curve $C$ of
$D_1$ in codimension two, and finally further enhanced at the
intersection of the matter curves and the intersection of the cusp
curve (i.e. the curve of {\em cuspidal type $II$ fibers above}) 
with $B_2$ in codimension three) 
on the one hand and 'intrinsic' contributions to $e(D_1)$ from its 
various singularities (the curve of intrinsic cusp 
{\em singularities of $D_1$}, which will always be present, and a
curve (actually one of the matter curves, the $h$ curve) of tacnode
resp. higher double point singularities in codimension one inside
$D_1$ and various complicated point singularities at the points
mentioned above as well as at further points detected by an analysis
of the discriminant equation). 

In {\em section 5} we will derive the contributions of the intrinsic 
singularity curves. In {\em section 6} we begin the discussion of the
codimension loci. In {\em section 7} we present in a case by case discussion
the various examples and further refine investigation of the
singularity contributions.

\subsection{smooth case}

Now if $B$ is three-dimensional, the discriminant $D$ is a surface in 
$B_3$ whose
class is given by $D=12c_1(B_3)$ resp. $G=4c_1(B_3)$ and $F=6c_1(B_3)$.
In analogy to the Calabi-Yau threefold case we will compute $e(X)$ 
from 
$e(sing.fiber)e(D)$. For a smooth $D$ we can obtain from the exact 
sequence 
$0\rightarrow T_D\rightarrow T_{B_3|_D}\rightarrow N_{D|_{B_3}}
\rightarrow 0$
the adjunction formulas (note that $N_{D|_{B_3}}={\cal O}(D)|_{D}$) 
\beqa
c_1(B_3)|_D&=&c_1(D)+D|_D \nonumber\\
c_2(B_3)|_D&=&c_2(D)+c_1(D)D|_D
\eeqa
which leads to the Euler characteristic of a non-singular $D$ 
\beqa
e(D)=c_2(B_3)D-c_1(B_3)D^2+D^3
\eeqa
But $D$ will be singular along a curve $C=FG$ and we expect a 
Pl\"ucker correction to $e(D)$. 
For $C=FG$ we can derive the Euler characteristic of $C$ from the 
above exact sequence by restricting to $C$ and, with the 
normal bundle of $C$ in $B_3$ given by $N_C|_{B_3}=({\cal O}(F)
\oplus{\cal O}(G))|_C$, we get
\beqa
e(C)=c_1(B_3)FG-(F+G)FG= -216c_1^3(B_3)= -1296c_1^2-432t^2
\eeqa
Using the {\em generalised Pl\"ucker formulas} derived in the next section
we finally get the corrected Euler characteristic of $D$ 
\beqa
e(D)&=&c_2(B_3)D-c_1(B_3)D^2+D^3+2(e(C)-DC)\nonumber\\
&=&288+576c_1^3(B_3)=288+3456c_1^2+1152t^2 
\eeqa
and so
\beqa
e(X)&=&e(I_1)(e(D)-e(C))\nonumber\\
    & &+e(II)e(C)\nonumber\\
    &=&288+360c_1(B_3)^3=288+2160c_1^2+720t^2
\eeqa
\subsection{singular case}
After reproducing the Euler characteristic for smooth $X$, 
we will consider
the case of having a section of $G$-singularities located over a surface
$D_2$ in $B_3$. Let us localize the $G$ fibers along the zero section of 
the ${\bf P}^1$ bundle $B_3$ over $B_2$, 
whose class we denoted above by $r$.
Following the procedure from above, we decompose the discriminant $D$ into
$D=D_1+D_2$ where again $D$ denotes the component with 
$I_1$ fibers and $D_2$ 
carrys $G$ fibers. With $D_2=cr$, $F_2=ar$ resp. $G_2=br$ and
the canonical bundle of the base $K_{B_3}=-c_1-2r-t$, we get the classes
$D_1=12c_1+(24-c)r+12t$, $F_1=4c_1+(8-a)r+4t$ and $G_1=6c_1+(12-b)r+6t$ 
which
describe our $I_1$ locus.

As we want to check our results on the Euler number of the $F$-theory
four-fold via $n_3=n_5$ with a corresponding heterotic computation
let us now assume, as we want to use the computations of $n_5$
from the the spectral cover method for $SU(n)$ bundles \cite{FMW},
that heterotically an $SU(n)\x E_8$ bundle $(V_1,V_2)$ 
is given and let us look for the first few non-trivial 
(the case $"n=1"$ of $G=E_8$ is treated below also; furthermore some
other cases of $G$, mainly in the $I_n$ series, will be discussed; this
requires in the case of $I_2, I_3$ the use of $E_7, E_6$ bundle $V$, 
whose second Chern class is computed in appendix (B) from parabolic methods)
cases where the gauge group $G$ is simple (let now $V:=V_1$). 
\begin{center}
\begin{tabular}{|cc|ccc|c|cc|cc|}
\hline
$V$&$G$    &$a$&$b$& $c$ & matter curve(s) & $fib_{enh}$&
 matter&het&{\rm het. loc.}\\
\hline
$SU(2)$&$E_7$  &$3$&$5$& $9$ &$f_{1;4c_1-t}$ &"$E_8$" &
$(\frac{1}{2}){\bf 56}$&$H^1(Z,V)$&$a_2$\\ \hline
$SU(3)$&$E_6$  &$3$&$4$& $8$ &$q_{3c_1-t}$&$E_7$&${\bf 
27}$&$H^1(Z,V)$&$a_3$\\ \hline
$SU(4)$&$D_{5}$&$2$&$3$& $7$ &$h_{2c_1-t}$&$E_6$&${\bf 
16}$&$H^1(Z,V)$&$a_4$\\
$$     &$$     &$$ &$$ & $$&$q_{3c_1-t}$&$D_6$&${\bf 
10}$&$H^1(Z,\Lambda^2V)$&$a_3$\\ \hline
$SU(5)$&$I_5$  &$0$&$0$& $5$ &$h_{c_1-t}$ &$D_5$&${\bf 
10}$&$H^1(Z,V)$&$a_5$\\
$$&$$&$$&$$&$$&$P_{8c_1-3t}$&$I_6$&${\bf 5}$ 
&$H^1(Z,\Lambda^2V)$&$R(a_i)$\\
\hline
\end{tabular}
\end{center}
with $R(a_i):=a_0a_5^2-a_2a_3a_5+a_3^2a_4$.
Here the matter was read off from the Tate formalism [\ref{BIKMSV}], 
and then the enhancement pattern from [\ref{KatzV}]. Note that,
as remarked already in \cite{FMW}, this matches precisely with the
heterotic expectations. 
\beqa
{\bf 248}&=&({\bf 2}, {\bf 56})\o+
            ({\bf 1},{\bf 133})\o+ ({\bf 3},{\bf 1})\nonumber\\
         &=&({\bf 3}, {\bf 27})\o+ ({\bf \bar{3}}, {\bf \overline{27}})\o+
            ({\bf 1},{\bf 78})\o+ ({\bf 8},{\bf 1})\nonumber\\
         &=&({\bf 4}, {\bf 16})\o+ ({\bf \bar{4}}, {\bf \overline{16}})\o+
            ({\bf 6},{\bf 10})\o+({\bf 1},{\bf 45})\o+ 
            ({\bf 15},{\bf 1})\nonumber\\
         &=&({\bf 5}, {\bf 10})\o+ ({\bf \bar{5}}, {\bf \overline{10}})\o+
            ({\bf 10}, {\bf \bar{5}})\o+({\bf \overline{10}}, {\bf 5})\o+
            ({\bf 1},{\bf 24})\o+ ({\bf 24},{\bf 1})\nonumber
\eeqa
There $H^1(Z,V)$ was localized on the curve $a_n=0$ 
(meaning $x=\infty$, the zero point in the group law; 
cf. sect. (\ref{spectralcovermethod}),
note that $a_i$ is of class $\eta-ic_1$) and $H^1(Z,\Lambda^2V)$ 
on the common zeroes of $P$ and $Q$ in the representation 
$w=P(x)+yQ(x)=0$ of the spectral equation (meaning that $y$ and $-y$ (the
inverse bundle) are spectral points in $E_b$). For example for
$G=A_4=I_5$ one has $P=a_0+a_2x+a_4x^2, \, Q=a_3+a_5x$.

So, after we will have computed $e(X^4)$ in the following sections, we
will compare with the heterotic expectation derived in the introduction.
\beqa
24n_5=288+(1200+107n-18n^2+n^3)c_1^2+(1080-36n+3n^2)c_1t
+(360+3n)t^2
\nonumber
\eeqa
with $n=0,2,3,4,5$ for $G=E_8, E_7, E_6, D_5, I_5$.

\section{"Pl\"ucker formulas" for surfaces with curves of singularities}

{\resetcounter}

As we will have to compute a number of times the Euler number of 
a surface component of the discriminant surface in $B_3$ we give
here the general computation. So let $D$ be a surface in $B_3$
with a curve of singularities along the curve $C$. Our applications 
below will include a curve of cusps resp. tacnodes and higher double
points. The cusp curve will always be present, while the tacnodes and 
higher double points occur for $G=A_3=I_4$ resp. $G=I_5, I_6$.
In the actual applications the singularity
will be worse at special points on the curve, a possibility which we
exclude here.

In {\em subsection one}
we derive the contribution to the Euler number 
in the general case of a curve of singularities 
of multiplicity $k$, resolved by one blow-up. In {\em subsection two}
we prove some main formulae used in this derivation and give an
outlook on a closely related application of this technical set-up.
In {\em subsection three} we specialise to the case of the {\em cusp curve};
here we find the contribution for the smooth case already used in
section 4. In {\em subsection four} we proceed to the case of a {\em tacnode
curve} which makes two successive blow-up's necessary. {\em Subsection five}
treats the case of an even higher double point with three blow-up's. 

\subsection{The general case of a curve of singularities 
of multiplicity $k$, resolved by one blow-up}

Let $k$ denote the multiplicity of a surface
$D$ (which will be our surface $D_1$ in the applications) 
along $C$ (being 2 in the two mentioned examples). 
We assume at first that the singularity is resolved after one blow-up.
So one first blows up $C$ in $B_3$, producing
a three-fold $\pi:\tilde{B}_3\ra B_3$. Then one
has for the total transform $\tilde{D}$
\beqa
\tilde{D}=\bar{D}+kE
\eeqa
with the proper transform $\bar{D}$ and the exceptional divisor $E$,
a ruled surface over $C$.
One has the relations (which are proved below)
\beqa
c_1(\tilde{B_3})&=&\pi^* c_1(B_3)-E\nonumber\\
c_2(\tilde{B_3})&=&\pi^* (c_2(B_3)+C)-\pi^* c_1(B_3)\cdot E
\eeqa
Furthermore the fact that, after blowing up a (itself nonsingular) 
point on a surface, the self-intersection of the exceptional $P^1$
is $-1$ generalizes essentially to a relation $E^2=-\pi^* C$ up to
a correction term $aF$ where $F$ is the fibre of the 
ruled surface $E$ over $C$ (so $E\cdot F=-1$)
and $a$ a number determined by the 
exterior geometry of $C$ in $B_3$ (again these relations are proved below)
\beqa
E^2&=&-\pi^* C-E^3\; F\nonumber\\
E^3&=&-\int_C c_1(N_{B_3}C)=-(c_1(B_3)C-e(C))
\eeqa
With these formulas one gets from the usual formula for a smooth surface
$c_2(\bar{D})=
c_2(\tilde{B}_3)\bar{D}-c_1(\tilde{B}_3)\bar{D}^2+\bar{D}^3$ 
that 
\beqa
c_2(\bar{D})=c_2(B_3)D-c_1(B_3)D^2+D^3+\Delta_k
\eeqa
with the correction term (of course $\Delta_1=0$ and one has a factor $(k-1)$)
\beqa
\Delta_k&=& (k-1)[-(3k+1)CD+kc_1(B_3)C-k^2E^3]\nonumber\\
              &=& (k-1)[-(3k+1)CD+k(k+1)c_1(B_3)C-k^2e(C)]
\eeqa
This is seen as follows
\beqa
c_2\bar{D}&=&c_2(\tilde{B}_3)\bar{D}-c_1(\tilde{B}_3)\bar{D}^2
+\bar{D}^3\nonumber\\
&=&c_2(B_3)D+CD-kc_1(B_3)C\nonumber\\
& &-c_1(B_3)D^2+k^2c_1(B_3)C+2kCD+k^2E^3\nonumber\\
& &+D^3-3k^2DC-k^3E^3
\eeqa
In particular for our case of interest $k=2$ one has
\beqa
\Delta_2=-7DC+2c_1(B_3)C-4E^3
\eeqa

\subsection{The Chern classes of Blow-ups}

Let us now prove the relations used above
\beqa
(*)\ \ \  c_2(\ti B_3)&=&f^*(c_2(B_3)+C)-f^*c_1(B_3)E \nonumber\\
(**)\ \ \ E^2&=&-\pi^*C-E^3  F \nonumber
\eeqa
Consider the following blow-up diagram for $X$ 
a non-singular
variety in $Y$
\begin{center}
\begin{tabular}{ccc}
$\;\; {\ti X}$&$\buildrel j \over \longrightarrow $&${\ti Y}$ \\
{\tiny $g$}$ \biggl \downarrow$ & &$\;\; \biggr \downarrow $ 
{\tiny $f$} \\
$\; X$&$\buildrel i \over \longrightarrow$&$Y$\\
\end{tabular}
\end{center}
Further let $N$ be the normal bundle to $X$ in $Y$ with rank$N=d$, the
codimension of $X$ in $Y$, and identify 
$\ti X$ with $P(N)$, so $N_{\ti X}{\ti Y}= {\cal O}(-1)$. Then from the 
above diagram one derives \cite{ful} the relation
\beqa
c_2(\ti Y)-f^* c_2(Y)=-j_*((d-1)g^*c_1(X)+\frac{d(d-3)}{2}{\cal O}(1)
+(d-2)g^*c_1(N))
\eeqa
\subsubsection{The case $d=2$}

For $d=2$ one has
\beqa
c_2(\ti Y)-f^* c_2(Y)&=&-j_*g^*c_1(X)-[\ti X][\ti X]\nonumber\\
                     &=&-j_*g^*c_1(X)+f^*i_*[X]-j_*g^*c_1(N_X Y)
\nonumber\\
                     &=&f^*i_*[X]-j_*g^*(c_1(X)+c_1(N_X Y))\nonumber\\
                     &=&f^*i_*[X]-j_*g^*(c_1(Y)|_X)\nonumber\\
                     &=&f^*i_*[X]-f^*c_1(Y)[\ti X]
\eeqa
note in the second line we made use of 
\beqa
f^*i_*[X]&=&j_*c_1(F)\nonumber\\
         &=&j_*g^*c_1(N_X Y)-[\ti X][\ti X]
\eeqa
with $F=g^*N/N_{\ti X}{\ti Y}$ and in for the last line 
one has $j_*g^*i^* u=f^* u [\ti X]$.
So identifying ${\ti Y}={\ti B_3}$ and $[\ti X]=E$ resp. 
$i_*[X]=C$ we arrive
at our expression $(*)$.

For the second relation let us consider again
\beqa
f^*i_*[X]&=&j_*c_1(F)\nonumber\\
         &=&j_*g^*c_1(N_X Y)-[\ti X][\ti X]\nonumber\\
         &=&{\rm deg}N j_*g^*[pt]-[\ti X][\ti X]\nonumber\\
         &=&{\rm deg}N\; l-[\ti X][\ti X]
\eeqa
where $l=g^*[pt]$ denotes the class of a fiber in $g:P(N)\rightarrow X$. 
Then from $ [\ti X]^2-[\ti X]g^*c_1(N))|_{[\ti X]}=0$ we get 
\beqa
[\ti X]^3&=&[\ti X]|_{[\ti X]}g^*c_1(N)\nonumber\\
         &=&[\ti X]|_{[\ti X]}{\rm deg}N g^*[pt]\nonumber\\
         &=&{\rm deg}N\; l\cdot [\ti X]|_{[\ti X]}\nonumber\\
         &=&-{\rm deg}N
\eeqa
so that $f^*i_*[X]=-[\ti X]^3 \; l-[\ti X]^2$ which is what we were 
looking for in $(**)$.

\subsubsection{The case $d=3$}

Let us give an outlook on a further application of this technique.
In connection with our main theme it is also of interest to compute
$c_2(X^4)$. First this gives in principle an alternative way to
compute the Euler number of $X$ by making use of the relation 
$c_2^2(X^4)=480+e(X^4)/3$ (cf. \cite{SVW}). Secondly $c_2(X)$ is of
interest because of the congruence relation between the four-flux and
$c_2(X)/2$ (cf. \cite{W4flux} and appendix (C)).

Consider now a Calabi-Yau 4fold $Z$ embedded (via its Weierstrass
representation) into an 5 dimensional 
ambient space $Y$, then it follows from adjunction 
(since $Z$ is a smooth 
divisor in $Y$) that
\beqa
c_1(Y)|_Z&=&c_1(Z)+Z|_Z \nonumber\\
c_2(Y)|_Z&=&c_2(Z)+c_1(Z)Z|_Z
\eeqa
and thus
\beqa
c_2(Z)=c_2(Y)|_Z
\eeqa
further recall that $c(Y)=c(B)(1+r)(1+r+2c_1)(1+r+3c_1)$ from which we 
get
\beqa
c_1(Y)&=&6c_1+3r\nonumber\\
c_2(Y)&=&11c_1^2+c_2+13rc_1+3r^2
\eeqa
and setting $r^2=-3rc_1$ (i.e. restricting to $Z$) then leads to the 
expression in the smooth case
\beqa
c_2(Z)=c_2(Y)Z=11c_1^2+c_2+4rc_1
\eeqa 

Now let us consider the simplest more complicated case, that of an
singularity of codimension one which is $A_1$. In order to do so let
us first analyse the change of $c_2$ of the ambient space. This is
computed as follows
\beqa
c_2(\ti Y)-f^* c_2(Y)&=&-j_*g^*(2c_1(X)+c_1(N_X Y))\nonumber\\
                     &=&-j_*g^*(c_1(X)+c_1(Y)|_X)\nonumber\\
                     &=&-j_*g^*c_1(X)-j_*g^*i^*c_1(Y)\nonumber\\
                     &=&-j_*g^*c_1(X)-f^*c_1(Y)[\ti X]
\eeqa
Now we have to compute using $\bar{Z}=\ti{Z}-2[\ti X]$ 
\beqa
c_2(\bar{Z})=
c_2(\ti{Y})|_{\bar{Z}}&=&f^*c_2(Y)\bar{Z}-f^*c_1(Y)[\ti X]\bar{Z}
-j_*g^*c_1(X)\bar{Z}\nonumber\\
&=&f^*c_2(Y)\ti{Z}-2f^*c_2(Y)[\ti X]-f^*c_1(Y)[\ti X]\ti{Z}
   +2f^*c_1(Y)[\ti X]\nonumber\\
& &-j_*g^*c_1(X)\bar{Z}\nonumber\\
&=&c_2(Y)Z-2f^*c_1(Y)[\ti X]^2-j_*g^*c_1(X)\bar{Z}\nonumber\\
&=&c_2(Y)Z-2c_1(Y)r-j_*g^*c_1(X)\bar{Z}\nonumber\\
&=&11c_1^2+c_2+4rc_1+2(6rc_1-9rc_1)-j_*g^*c_1(X)\bar{Z}\nonumber\\
&=&11c_1^2+c_2-2rc_1-j_*g^*c_1(X)\bar{Z}
\eeqa
showing the crucial deviation term $-j_*g^*c_1(X)\bar{Z}$ relative to
the smooth case.

\subsection{Cusp curve}

So for example for the cusp curve case (where
also $c_2(\bar{D})=c_2(D)$ as, in contrast to the double 
point case, no points are identified in blowing down $\bar{D}$ 
back to its singular version $D$) one gets that
\beqa
\Delta_{cusp}= -7CD+6c_1(B_3)C-4c_1(C)
\eeqa

Note also that for the cases where the cusp curve is given by the
uncorrected $F_1G_1$ (so this includes the smooth case, the 
pure gauge group case of singularities only in codimension 1, 
where still $D_1$ and therefore $C$ is separated from $B_2$, 
and furthermore the $E$ series in general)
\beqa
-7CD+6c_1(B_3)C-4c_1(C)&=&-2CD+2c_1(C)+6c_1(B_3)C-5CD-6c_1(C)\nonumber\\
  &=& -2CD+2c_1(C)+6c_1(N\, C)-5CD\nonumber\\
  &=& -2CD+2c_1(C)+(6(F_1+G_1)-5D_1)C\nonumber\\
  &=& -2CD+2c_1(C)+((6(F+G)-5D)-(6(a+b)-5c)r)C\nonumber\\
  &=& -2CD+2c_1(C)+dCr
\eeqa
(note that this $D$ is $D_1$ in our application)
where the term $d:=5c-6(a+b)$ equals $-4,-3,-2$ for 
$E_8,E_7,E_6$ and of course zero for the smooth case. 
This shows explicitely the deviation 
$-2CD+dCr$ used above in the smooth case to the 
naive adiabatic extension $2c_1(C)$ of the one-dimensional 
Pl\"ucker formula.

\subsection{Tacnode curve}

Now we come to the more complicated case of the tacnode, where
we need a second blow-up, as the first blow-up just 
brings one 
to the case of an ordinary double point (having distinct tangents 
as opposed to the tacnode). This second blow-up is along the 
well-defined (as the two tangent directions of the tacnode points 
of $D$ along $C$ coincide) proper transform $\bar{C}$ of $C$ 
under the first blow-up. Note that 
$\bar{C}=E_{(1)}\bar{D}=E_{(1)}(\tilde{D}-2E_{(1)})$.
Note also that at the end of the procedure we have to go back
to the singular model $D$ and to get its Euler number we still have 
to subtract $e(C)$ as in the second resolution step the double points
became separated, i.e. (with $c_2(D)^{ord}=c_2(B_3)D-c_1(B_3)D^2+D^3$)
\beqa
c_2(\bar{\bar{D}})&=&c_2(D)^{ord}+\Delta_{tacn}\nonumber\\
c_2(D)&=&c_2(D)^{ord}+\Delta_{tacn}-e(C)
\eeqa
In other words the corrections $\Delta_{cusp}, \Delta_{tacn}$ refer
in our conventions to the desingularized model (just as in the ordinary
Pl\"ucker formulas).

Here one gets (up to codimension 3 contributions)
\beqa
\Delta_{tacn}= -21CD+26c_1(B_3)C-20e(C)
\eeqa

To prove this let us follow the two steps of the resolution.
Clearly in the second resolution step we are again back in the case
of a curve of ordinary double points.
\beqa
c_2(\bar{\bar{D}})&=&c_2(\tilde{\tilde{B}}_3)\bar{\bar{D}}-
c_1(\tilde{\tilde{B}}_3)\bar{\bar{D}}^2+\bar{\bar{D}}^3\nonumber\\
&=&c_2(\tilde{B}_3)\bar{D}-c_1(\tilde{B}_3)\bar{D}^2+\bar{D}^3\nonumber\\
& &-7\bar{D}\bar{C}+2c_1(\tilde{B}_3)\bar{C}-4E_{(2)}^3\nonumber\\
&=&c_2(B_3)D-c_1(B_3)D^2+D^3-7DC+2c_1(B_3)C-4E_{(1)}^3\nonumber\\
& &-7(\tilde{D}-2E_{(1)})\bar{C}+2(\pi^*c_1(B_3)-E_{(1)})\bar{C}-4E_{(2)}^3
\eeqa
giving this time
\beqa
\Delta_{tacn}&=&-7DC+2c_1(B_3)C-4E_{(1)}^3\nonumber\\
& &-7\tilde{D}\bar{C}+2\pi^*c_1(B_3)\bar{C}+12E_{(1)}\bar{C}-4E_{(2)}^3
\eeqa
Now, concerning the four new terms in the second line, 
note that one has, concerning the first three of them, that 
\beqa
\tilde{D}\bar{C}&=&\tilde{D}E_{(1)}(\tilde{D}-2E_{(1)})=2DC\nonumber\\
\pi^*c_1B_3\bar{C}&=&\pi^*c_1B_3E(\tilde{D}-2E_{(1)})
=2c_1(B_3)C\nonumber\\
E_{(1)}\bar{C}&=&E_{(1)}^2(\tilde{D}-2E_{(1)})=-CD-2E_{(1)}^3
\eeqa
On the other hand concerning the last new term $E_{(2)}^3$ 
one has again that
$E_{(2)}^3=-c_1(N_{\bar{C}|\tilde{B}_3})$, whereas 
$E_{(1)}^3=-c_1(N_{C|B_3})$. Now, to express the former in terms 
of the latter, note that the short exact sequence
\beqa
0\rightarrow N_{\bar{C}|E_{(1)}}\rightarrow N_{\bar{C}|\tilde{B_3}}
\rightarrow N_{E_{(1)}|\tilde{B_3}}\rightarrow 0
\eeqa
gives 
\beqa
c_1(N_{\bar{C}|\tilde{B}_3})=
c_1(N_{\bar{C}|E_{(1)}})+c_1(N_{E_{(1)}|\tilde{B_3}})
\eeqa
where the first term on the right hand side
is evaluated as $\bar{C}^2$ in $E_{(1)}$, i.e. as
\beqa
c_1(N_{\bar{C}|E_{(1)}})=\bar{D}^2E_{(1)}=
(\tilde{D}-2E_{(1)})^2 E_{(1)}=4DC+4E_{(1)}^3
\eeqa
Similarly
the second term is $c_1(T)=E_{(1)}|_{E_{(1)}}$ 
of the tautological bundle $T$ over $E_{(1)}$, restricted to 
$\bar{C}=\bar{D}|_{E_{(1)}}$,
i.e. 
\beqa
c_1(N_{E_{(1)}|B_3})=\bar{D}E_{(1)}^2=(\tilde{D}-2E_{(1)})E_{(1)}^2=
-DC-2E_{(1)}^3
\eeqa
So that one gets
\beqa
E_{(2)}^3&=&-c_1(N_{\bar{C}|\tilde{B}_3})\nonumber\\
    &=&-(c_1(N_{\bar{C}|E_{(1)}})+c_1(N_{E_{(1)}|\tilde{B_3}}))\nonumber\\
         &=&-(4DC+4E_{(1)}^3-DC-2E_{(1)}^3)\nonumber\\
         &=&-3DC-2E_{(1)}^3
\eeqa

So finally
\beqa
\Delta_{tacn}&=&-7DC+2c_1(B_3)C-4E_{(1)}^3\nonumber\\
             & &-14DC+4c_1(B_3)C-12CD-24E^3-4E_{(2)}^3\nonumber\\
             &=&-33CD+6c_1(B_3)C-28E_{(1)}^3-4(-3DC-2E_{(1)}^3)\nonumber\\
             &=&-21CD+6c_1(B_3)C-20E_{(1)}^3\nonumber\\
             &=&-21CD+26c_1(B_3)C-20e(C)
\eeqa

\subsection{curve of higher double points}

If a third blow-up is necessary like for the case of a curve of
singularities of type $t^6+v^2$ one gets
\beqa
c_2(\bar{\bar{\bar{D}}})&=&c_2(\tilde{B_3})\bar{D}-c_1(\tilde{B_3})\bar{D}^2
+\bar{D}^3\nonumber\\
 & &-21\bar{D}\bar{C}+6c_1(\tilde{B_3})\bar{C}-20E_1^3\nonumber\\
 &=&c_2(B_3)D-c_1(B_3)D^2+D^3-7DC+2c_1(B_3)C-4E_1^3\nonumber\\
 & &-21(\tilde{D}-2E_1)+6\pi^*c_1(B_3)\bar{C}-6E_1\bar{C}-20E_1^3\nonumber\\
 &=&c_2(B_3)D-c_1(B_3)D^2+D^3\nonumber\\
 & &-25DC+12c_1(B_3)C-32E_1^3
\eeqa

In the concrete application in the $I$ series discriminant one has
along the $h$ curve an equation for $D_1$ of the form
$(x^2-z)^2+z^3$ where we have written $x$  for $h$. One finds
in the explicit resolution process further contributions at
codimension three loci inside the $h$ curve
which we will not need to write down.

\section{On the codimension 3 loci}

{\resetcounter}

Consider the cases $D_5=I_1^*$ and $I_5$.
There are two new features compared to the $E_k$ series: 
first that without further tuning the
$I^*_n$ and the $I_n$ series would remain at $n=0$ 
(cf. sect. (\ref{xcusp})), and secondly
the existence of {\em two} matter curves.

What we want to see in the following is that actually the
cohomology classes of the two codimension 3 loci, i.e.
of $Cr$ and the intersection of the matter curves $h$ and $P$, are 
proportional; more precisely that $Cr$ is a multiple of $hP$.

Now one has
\beqa
(4f_1^3+27g_1^2=0)=D_{old}=D_1+nr
\eeqa
where $n$ is the subscript in the $I^*_n$ and the $I_n$ series, i.e.
the number of powers of $z$ one can extract from the left hand side.

Furthermore one has the decomposition of $D_1r$ into the 
matter (=enhancement) curves ($P$ means here our $q$ in the $I^*_1$ case)
\beqa
D_1 r=\pi h +\rho P
\eeqa
where $D_1r$ is also given by
\beqa
D_{old}r = D_1r-nt
\eeqa
One has a corresponding decomposition 
\beqa
(f_1=0=g_1)=C_{old}=C+xhr
\eeqa
where $x=ord_h res(f_1,g_1)$, so that one also gets (with
$\alpha =ord_h f_1, \beta =ord_h g_1$)
\beqa
Cr=\alpha h \cdot \beta h+xht
\eeqa

\subsection{The $I_n$ series}

For the cases $n=4,5,6$ which show the general $I$ series pattern
one finds the following.
There is $f_1r=4h, g_1r=6h$ and one finds
\beqa
x&=&3n\\
Cr&=&3h(8h + nt)=3hP_{8c_1-(8-n)t}
\eeqa
Let us now understand why $Cr$ is indeed a multiple of $hP$, considered
as cohomology classes, i.e. why the used cohomological 
relation $8h + nt=P$ is
not accidental. For $I_0$ one has that 
$D_{old}r=12c_1-12t=12h_{c_1-t}$. For $I_5$ one has 
$D_{old}r=D_1r-5t=4h_{c_1-t}+P_{8c_1-3t}-5t$ 
and therefore $12h=4h+P-5t$ or $8h+5t=P$. Similarly for $I_n$.

For the case $n=2$ one has $f_1r=2H, g_1r=3H$ and with $x=3$ one finds
$Cr=2H\cdot 3H +3Ht=3H(2H+t)=3H(4c_1-3t)=\frac{3}{2}HP$. For $n=3$
with $f_1r=4h, g_1r=6h$ and $x=8$ one finds 
$Cr=8h(3h + t)=8h(3c_1-2t)=\frac{8}{3}h(9c_1-6t)$.

\subsection{The $I^*_n$ series}

The case $D_4=I^*_0$ is somewhat exceptional as here one has
$f_1r=2h_{2c_1-t}, g_1r=3P_{2c_1-t}$ with different polynomials
of the same degree (whereas for $n>0$ one has $g_1r=3h$ and $P$
will represent a different cohomology class giving the other matter
curve) and a ${\bf Z_3}$ symmetry related to $D_4$ triality.
This manifests itself in the discriminant as follows
\beqa
D_1r=12c_1-6t=A^{(0)}_{4c_1-2t}+A^{(1)}_{4c_1-2t}+A^{(2)}_{4c_1-2t}
\eeqa
where $A^{(i)}_{4c_1-2t}=(h_{2c_1-t}^2+\omega ^iP_{2c_1-t}^2=0)$ with 
$i=0,1,2$ and $\omega ^3=1$.
Then one has with $x=0$
\beqa
Cr=2h\cdot 3P
\eeqa
Now note that the locus of simultaneous vanishing of $h$ and $P$
is also the locus of intersection of the $A^{(i)}$.

For $I^*_n$ with $n>0$ is $f_1r=2h_{2c_1-t}, g_1r=3h_{2c_1-t}$.
For $D_5=I^*_1$ one has 
\beqa
D_1r=12c_1-5t=3h_{2c_1-t}+2q_{3c_1-t}
\eeqa
and with $x=2$ one finds
\beqa
Cr=2h\cdot 3h+2ht=2h(3h+t)=2h\cdot 2q
\eeqa
where $2h$ occurs in $Cr$ actually on the level of divisors. 

For $D_6=II_2^*$ one has (for the parameter $r=0,1,2,3,4$ being 0
(cf. the discussion of $D_6$ in the six-dimensional case and
[\ref{BIKMSV}])) that
\beqa
D_1r=(12c_1+16r+12t)r=12c_1-4t=2h_{2c_1-t}+2P_{4c_1-t}
\eeqa
and with $x=3$ one has
\beqa
Cr=2h\cdot 3h+3ht=3h(2h+t)=3hP_{4c_1-t}
\eeqa
So one has that for the $I_n^*$ cases with $n=0,1,2$
\beqa
x&=&6\frac{n}{2+n}\\
Cr&=&h(6h+xt)=\frac{12}{2+n}h\cdot ((2+n)c_1-t)
=\frac{12}{2+n}hP_{(2+n)c_1-t}
\eeqa
Let us again see why $Cr$ arises as a multiple of $hP$.
For $I_0^*$ one has $D_{old}r=12c_1-6t=6(2c_1-t)$. For $I_1^*$
one has $D_{old}r=D_1r-t=3h+2q-t$ and therefore 
$6h=3h+2q-t$ or $3h+t=2q$ as we wanted to prove. 
Similarly for $I_2^*$ one has 
$D_{old}r=D_1r-2t=2h_{2c_1-t}+2P_{4c_1-t}-2t$ and so
$6h=2h+2P-2t$ or $4h+2t=2P$ resp. $2h+t=P$.

Let us finally, for example in the case of $D_5$, come to the question
whether actually the two {\em sets} $C\cap r$ and $h\cap q$ coincide. 
From equ. (\ref{D5f1g1}) it follows that 
$48F_1h+864G_1=z(q^2-\frac{1}{2}f_1h)+{\cal O}(z^2)$ so if we approach
$C\cap r$ coming {\em from the outside} of $B_2=r=(z=0)$ we find 
$0=q^2-\frac{1}{2}f_1h +{\cal O}(z)$ which goes in the limit
$z\rightarrow 0$ to the condition $0=q^2-\frac{1}{2}f_1h$
resp. $0=q^2$ for $C\cap r$ as it will lie in the divisor $h$ anyway.

\section{The explicit computation of $e(X^4)$}

{\resetcounter}

Now, finally, we come to our main computation announced in the 
introduction. This Euler number computation  for the various
cases is treated in {\em subsection (7.2)}. In {\em subsection (7.1)}
we make contact with a formula given in \cite{KLRY}
(for the case of singularities in codimension one only) which was guessed
there from a list of values based on a computer analysis in a toric framework.
The case of pure codimension one is also of interest because in this
case the expression $\pi_* (\gamma^2)=-\lambda^2 N \eta (\eta -N c_1)$
for an $SU(N)$ bundle 
will vanish as $\eta -Nc_1=(6-N)c_1-t=0$ for\footnote{The case
$G=A_4$ is only a 'pseudo-separation case' between $B_2$ and $D_1$ as
only one ($h$) of the two matter curves is turned off cohomologically,
but note that over the other matter curve $P$ the enhancement is
additive, leading from $I_5$ to $I_6$, so that the Euler number
computation is not effectively disturbed, cf. section (7.1); note that
by contrast in the case $G=D_5$ the choice $N=4$ and so $t=2c_1$ turns
off again the $h_{2c_1-t}$ matter curve and again one has the additivity of
the enhancement over the other matter curve $q$ but this time there is
an intrinsic codimension three locus left over (see below).}
$G=E_8, E_7, E_6, A_4$ 
and $t=(6-N)c_1$ with $N=0,2,3,5$ . 
Finally in {\em subsection (7.3)} we note an observation relating
Euler number values in neighbouring cases of certain Higgs chains.

\subsection{Euler number formula for codimension one}

A byproduct of our analysis is the
proof of an Euler number formula for elliptic Calabi-Yau fourfolds  
for which the elliptic fiber degenerates over the generic codimension one 
locus $B_2$ in the Calabi-Yau base $B_3$. This formula was first written 
down in \cite{KLRY} based on a toric computer analysis. The formula
suitably rewritten 
reads
\beqa
e(X_4)=288+360\int_{B_3}c_1^3(B_3)-r(G)c(G)(c(G)+1)\int_{B_2}
c_1^2(B_2) 
\eeqa
where $r(G)$ and $c(G)$ are the rank resp. Coxeter number of the gauge group
$G$. Now (with $B_3=F_{k,m,n}$ the generalized Hirzebruch
surface of base $B_2=F_k$; below we consider yet another example) 
using the fact that
$c_1^3(B_3)=6c_1^2+2t^2$ 
and that we can express
$t$ in data of $F_{k,m,n}$ so that $t=m[b]+n[f]$ with $t^2=2mn-m^2k$ where 
$[b]^2=-k$ and also noting that $t^2=2n^2=\frac{n^2}{4}c_1^2$ from
implementing the codimension one condition $m=n$ and $k=0$, 
we can rewrite the above formula as
\beqa
e(X_4)=288+(180(12+n^2)-r(G)c(G)(c(G)+1))c_1^2(B^{\prime})    
\eeqa
The case of purely codimension one (fiber) singularity
(i.e. especially without matter curves; so this is a 'separation case'
what concerns the relative position of the two discriminant components
$B_2$ and $D_1$) is realizable for $G=E_8, E_7, E_6, D_4, A_2$ over $B_2=F_0$
with $n=m=12,8,6,4,3$.
In \cite{KLRY} the authors were restricted to reflexive polyhedra and
thus excluding the $E_7$ case. However, using 
naively the above formula leads to a prediction for $E_7$ which will be 
vindicated by our computation which therefore gives an independent check 
of this formula. Moreover, we also find agreement for the cases
indicated in the following table (always assuming the
(pseudo-)separation case)
\begin{center}
\begin{tabular}{|c|c||c|c|}
\hline
$G$   & $e(X_4)$ & $G$ & $e(X_4)$  \\
\hline
\hline
$A_1$ & $288+2874c_1^2$ & & \\
\hline
$A_2$ & $288+2856c_1^2$ & $D_4$ & $288+4872c_1^2$ \\
\hline
$A_3$ & $288+2820c_1^2$ & $E_6$ & $288+7704c_1^2$ \\
\hline
$A_4$ & $288+2760c_1^2$ &  $E_7$ & $288+11286c_1^2$\\
\hline
$A_5$ & $288+2670c_1^2$  & $E_8$ & $288+20640c_1^2$\\
\hline
\end{tabular}
\end{center}
Let us make some remarks:\\
{\bf A series}\\
The codimension one condition is established by setting $t=c_1$ or equivalently
$n=2$ ({\em pseudo}-separation, cf. the last footnote). 
Note that in \cite{KLRY} the $A_2$ singularity was 
specified by $n=3$, i.e. $t=3/2c_1$ and therefore one 
has $e(X_4)=288+3756c_1^2$ which matches our computation too. \\ 
{\bf D series}\\
In the $D$ series we find only for $G=D_4$ a codimension one condition which 
is $t=2c_1$ resp. $n=4$.\\ 
{\bf E series}\\
The codimension one condition is here established by setting 
$t=3c_1, 4c_1, 6c_1$ resp. $n=6,8,12$ for $E_6, E_7, E_8$.

As a last point we remark that for the choice of $B_2=P^2$ of table
(6.3) of \cite{KLRY} we find also agreement with our formulae given below.

\subsection{The cases}

\subsubsection{$E_8(II^*)$ singularity}

Now, from our above analysis we see that 
$e(D_1)^{ord}=c_2(B_3)D_1-c_1(B_3)D_1^2+D_1^3$ 
receives two corrections (in later cases more from codimension three 
contributions). The first one is coming from the fact that $D$ has a 
cusp curve $C=F_1G_1$ but we 
have also to take into account that $D$ is "double" along $T=D_1\cap r$ 
(of class $6c_1-t$ related to $g_1(z=0)$) and 
has to be resolved (note that this is a 'real' resolution in contrast
to the case of the cusp curve where the resolution is only an
intermediate computational step to get the contribution of the
singular geometry).

Let us first compute the contributions from {\em fiber} singularities
(for the general set-up of the computation involving fiber and
intrinsic singularities cf. the introduction to section 4)
\beqa
e(X)&=&1(e(D_1)-e(C))\nonumber\\
 & &+2e(C)\nonumber\\
 & &+10e(B_2)
\eeqa
Then, the intrinsic singularities of $D_1$ are computed as (where
$C\cap r$ is a special codimension three locus)
\beqa
e(D_1)=c_2(B_3)D_1-c_1(B_3)D_1^2+D_1^3+\Delta_{\rm double}
+\Delta_{\rm cusp}+\Delta_{C\cap r}
\eeqa
which can be derived as follows. One has
\beqa
c_2(\bar{\bar{D_1}})&=&c_2(\tilde{\tilde{B}}_3)\bar{\bar{D_1}}-
c_1(\tilde{\tilde{B}}_3)\bar{\bar{D_1}}^2+\bar{\bar{D_1}}^3\nonumber\\
&=&c_2(\tilde{B}_3)\bar{D_1}-c_1(\tilde{B}_3)\bar{D_1}^2+\bar{D_1}^3
-7\bar{D_1}\bar{C}+2c_1\tilde{B}_3\bar{C}-4E_{(2)}^3\nonumber\\
&=&(\pi^*(c_2(B_3)+T)-\pi^*c_1(B_3)E_{(1)})(\ti{D_1}-2E_{(1)})\nonumber\\
& &-(\pi^*c_1(B_3)-E_{(1)})(\ti{D_1}-2E_{(1)})^2\nonumber\\
& &+(\ti{D_1}-2E_{(1)})^3\nonumber\\
& &-7(\ti{D_1}-2E_{(1)})\bar{C}+2(\pi^*c_1(B_3)-E_{(1)})\bar{C}-4E_{(2)}^3
\nonumber\\
&=&c_2(B_3)D_1-c_1(B_3)D_1^2+D_1^3-7D_1T+2c_1(B_3)T-4E_{(1)}^3\nonumber\\
& 
&-7\ti{D_1}\bar{C}+14E_{(1)}\bar{C}+2\pi^*c_1(B_3)\bar{C}-2E_{(1)}\bar{C}-
4E_{(2)}^3
\eeqa
using $\bar{C}=\ti{C}-\#(C\cap r)\; F$ with $F$ the fiber
of $E_{(1)}$ we find
\beqa
\ti{D_1}\bar{C}&=&\ti{D_1}(\ti{C}-\#(C\cap r)\; F)=\ti{D_1}\ti{C}
=D_1C \nonumber\\
E_{(1)}\bar{C}&=&E_{(1)}(\ti{C}-\#(C\cap r)\; F)= \#(C\cap r) \nonumber\\
\pi^*c_1(B_3)\bar{C}&=&\pi^*c_1(B_3)(\ti{C}-\#(C\cap r)\; F)
=c_1(B_3)C \nonumber\\
E_{(2)}^3&=& -(c_1(\tilde{B_3})C-e(C))=-(c_1(B_3)C-e(C))+\#(C\cap r)
\eeqa
which then leads to 
\beqa
c_2(\bar{\bar{D_1}})=c_2(B_3)D_1-c_1(B_3)D_1^2+D_1^3+
\Delta_{\rm double}+\Delta_{cusp}+\Delta_{C\cap r}
\eeqa
So altogether
\beqa
c_2(\ti{\ti{D_1}})=168+1702c_1^2+1760c_1t+576t^2+(12-4)\#(C\cap r)
\eeqa
and using
\beqa
c_2(B_3)D_1-c_1(B_3)D_1^2+D_1^3&=&168 + 5434c_1^2 + 4670 c_1t + 1588t^2
\nonumber\\
e(C)&=&-684c_1^2-648c_1t-216t^2     \nonumber\\
10e(B_2)&=&120-10c_1^2                \nonumber\\
\Delta_{\rm double}&=&-2(29c_1-2t)(6c_1-t) \nonumber\\
\Delta_{\rm cusp}&=&-3384c_1^2-2992c_1t-1008t^2 \nonumber\\
\Delta_{C\cap r} &=& (12-4)\#(C\cap r)\nonumber\\
\#(C\cap r) &=& Cr=4c_1(6c_1-t)
\eeqa
we find 
\beqa
e(X)=288+1008c_1^2+1112c_1t+360t^2+(12-4)Cr
\eeqa
which leads to agreement when comparing with the heterotic side where one has
\beqa
24n_5=288+1200c_1^2+1080c_1t+360t^2
\eeqa

Finally we remark that this formula also reproduces the computation
for the case $B_2=F_0$ and $t=0[b]+12[f]$ in table (6.4) of \cite{KLRY}.

\subsubsection{$E_7(III^*)$ singularity}

Here we have again to take into account the subtlety concerning the
fibre enhancement along the matter (=enhancement) curve $T$ of class
$4c_1-t$ related to $f_1(z=0)$
mentioned in the six-dimensional analysis.
\beqa
e(X)&=&1((e(D_1)-e(C)-e(T)+\#(C\cap r))\nonumber\\
       & &+2(e(C)-Cr)\nonumber\\
       & &+9(e(B_2)-e(T))\nonumber\\
       & &+9(e(T)-\#(C\cap r))\nonumber\\
       & &+A\#(C\cap r)\nonumber\\
       &=&e(D_1)+e(C)+9e(B_2)-e(T)+(A-10)\#(C\cap r)
\eeqa
with
\beqa
e(D_1)&=&c_1(B_3)D_1-c_1(B_3)D_1^2+D_1^3+\Delta_{\rm cusp}\nonumber\\
&=&180+2133c_1^2+1646c_1t+583t^2\nonumber\\
e(C)&=&-762c_1^2-629c_1t-217t^2     \nonumber\\
9e(B_2)&=&+108-9c_1^2\nonumber\\
e(T)&=&-(3c_1-t)(4c_1-t)\nonumber\\
C\cap r  &=& (6c_1-t)(4c_1-t) \ \ \;  Cr=\#(C\cap r)
\eeqa
so we find 
\beqa
e(X)=288+1134c_1^2+1110c_1t+357t^2+A\#(C\cap r)
\eeqa 
Now comparing with the heterotic side where one has 
\beqa
24n_5=288+1350c_1^2+1020c_1t+366t^2
\eeqa
one is lead from $\Delta_e=24n_5-e(X)$ where
\beqa
\Delta_e=(9-A)(6c_1-t)(4c_1-t)=(9-A)\#(C\cap r)
\eeqa
to a prediction $A=9$ for the Euler number $A$ of the fiber
configuration over the codimension three locus $C\cap r$.

\subsubsection{$E_6(IV^*)$ singularity}
Now one has with the matter (=enhancement curve) $T$ of class $q_{3c_1-t}$
related to $g_1(z=0)=q^2$
\beqa
e(X)&=&1((e(D_1)-e(C)-e(T)+\#(C\cap r))\nonumber\\
       & &+2(e(C)-\#(C\cap r))\nonumber\\
       & &+8(e(B_2)-e(T))\nonumber\\
       & &+9(e(T)-\#(C\cap r))\nonumber\\
       & &+A\#(C\cap r)\nonumber\\
       &=&e(D_1)+e(C)+8e(B_2)+(A-10)\#(C\cap r)
\eeqa
with
\beqa
e(D_1)&=&c_1(B_3)D_1-c_1(B_3)D_1^2+D_1^3+\Delta_{\rm cusp}\nonumber\\
&=&192+2300c_1^2+1552c_1t+596t^2\nonumber\\
e(C)&=&-822c_1^2-602c_1t-220t^2     \nonumber\\
8e(B_2)&=&+96-8c_1^2\nonumber\\
e(T)&=&-(2c_1-t)(3c_1-t)\nonumber\\
\#(C\cap r)&=&(3c_1-t)(4c_1-t), \ \ \; 
Cr=2\#(C\cap r)
\eeqa
(where in the last line one has to take into account that $G_1r=2q$
from the split condition $g_1(z=0)=q^2$)
so we find 
\beqa
e(X)=288 +1470c_1^2+950c_1t+376t^2+A\#(C\cap r)
\eeqa
if we compare this with
\beqa
24n_5=288+1386c_1^2+999c_1t+369t^2
\eeqa
we find a prediction for $A$ from the vanishing of
\beqa
\Delta_e=(7-A)(3c_1-t)(4c_1-t)=(7-A)\#(C\cap r)
\eeqa

\subsubsection{$D_4(I_0^*)$ singularity}

Let us start with the relevant cohomology classes resp. divisors
\beqa
F_1&=& 4c_1+ 6r+ 4t\Rightarrow F_1r=2h\nonumber\\ 
G_1&=& 6c_1+ 9r+ 6t\Rightarrow G_1r=3P\nonumber\\ 
D_1&=&12c_1+18r+12t\Rightarrow D_1r=A^{(0)}+A^{(1)}+A^{(2)}
\eeqa
where $h=2c_1-t$ and $P=2c_1-t$ and 
$A^{(i)}_{4c_1-2t}=(h_{2c_1-t}^2+\omega ^iP_{2c_1-t}^2=0)$ with 
$i=0,1,2$ and $\omega ^3=1$ and the last decomposition holds
not only on the level of cohomology classes but {\em actually 
on the level of divisors}
as seen from the equation $D_1=(A^{(0)}A^{(1)}A^{(2)}+{\cal O}(z)=0)$.

Now we have
\beqa
F_1G_1=C_{old}=C_{new}
\eeqa
So
\beqa
C&=&24(c_1+t)(c_1+2t)+6t(3r+4t)\nonumber\\
Cr&=&6hP=6(2c_1-t)^2
\eeqa
further we have
\beqa
e(D_1)&=&c_2(B_3)D_1-c_1(B_3)D_1^2+D_1^3+\Delta_{cusp}\nonumber\\
&=&216+7062c_1^2+3642c_1t+1764t^2+\Delta_{cusp}
\eeqa
One has
$\Delta_{cusp}=-4512c_1^2-2292c_1t-1128t^2$
where we used
$e(C)=(c_1-F_1-G_1)F_1G_1=-960c_1^2-498c_1t-240t^2$ 
and altogether 
$e(D_1)=216+1584c_1^2+852c_1t+396t^2$
and so
\beqa
e(X_4)&=&e(D_1)+e(C)+6(12-c_1^2)\nonumber\\
&=&288+1584c_1^2+852c_1t+396t^2
\eeqa
This is in agreement with the computation $e(X_4)=39264$ of \cite{KLRY}
for $B_3=F_{0,4,4}$ which means $B_2=F_0$ and so $c_1=(2,2)$ and
$t=(4,4)$.
The other choice $t=(0,4)$ and $c_1=(2,2)$ with $e(X_4)=19680$ given
there leads if one includes in the above formula an fiber enhancement
$k$ over the matter curves $e(X_4)=19776+3\cdot 16(k-6)$
to a prediction $k=4$, i.e. an effect similar to the cases of $E_7$
and $D_6$ where one did not have the naive additivity of the collision rules.

\subsubsection{$D_5(I_1^*)$ singularity}

Now one has
\beqa
F_1&=& 4c_1+ 6r+ 4t\Rightarrow F_1r=2h\nonumber\\ 
G_1&=& 6c_1+ 9r+ 6t\Rightarrow G_1r=3h\nonumber\\ 
D_1&=&12c_1+17r+12t\Rightarrow D_1r=3h+2q
\eeqa
where $h=2c_1-t$ and $q=3c_1-t$ and the last decomposition holds
not only on the level of cohomology classes but {\em actually 
on the level of divisors}
as seen from the equation $D_1=(h^3q^2+{\cal O}(z)=0)$.

Now just as
\beqa
(4f_1^3+27g_1^2=0)=D_1+r
\eeqa
we will have a decomposition, again {\em actually 
on the level of divisors},
\beqa
F_1G_1=C_{old}=C_{new}+2hr
\eeqa
where $C_{new}$ (which we denote simply by $C$ in the following)
is the true cusp curve of $D_1$ we are interested in.

So
\beqa
C_{old}&=&6(4(c_1+t)^2+3r(4c_1+t))=24(c_1+t)^2+18r(4c_1+t)\nonumber\\
C&=&f_1g_1-2(2c_1-t)r=24(c_1+t)^2+(68c_1+20t)r\nonumber\\
Cr&=&2h\cdot 3h+2ht=2h(3h+t)=2h\cdot 2q=4(2c_1-t)(3c_1-t)\nonumber\\
\#(C\cap r)&=&\#(h \cap q)=hq
\eeqa
where in the third line both factors occur not only as cohomology classes
but even as divisors: for the part $2h$ this follows from the construction
of $C$ and for the part $2q$ we saw this in section 6.
Note also that not only $Ch\subset Cr$ but that they are actually 
equal as sets. The precise multiplicity is given by $C2h=Cr$ as
the part $2h$ occurs in the $Cr$ not only as cohomology class
but even as divisor. So $Ch=2pq$.

We proceed now in two steps: first we compute the intrinsic
singularity corrections in $e(D_1)$, then we collect the fiber enhancements.

Now $e(D_1)$ is given by 
(up to contributions from point singularities of $D_1$ considered below)
\beqa
e(D_1)^{ord}+\Delta_{cusp}&=&c_2(B_3)D_1-c_1(B_3)D_1^2+D_1^3
+\Delta_{cusp}\nonumber\\
&=&204+6655c_1^2+4004c_1t+1684t^2+\Delta_{cusp}
\eeqa
One has $\Delta_{cusp}=-7CD_1+6c_1(B_3)C-4e(C)$
where the first two terms are
$-7CD_1+6c_1(B_3)C=-7872c_1^2-4724c_1t-1988t^2$.
Now $F_1G_1=C_{old}=2hr+C$ so 
\beqa
2\chi_{ar}(C_{old})=2\chi_{ar}(2h)+e(C)-2\chi_{ar}(2hC)
\eeqa
Now 
$2\chi_{ar}(C_{old})=(c_1+2r+t)C_{old}-(F_1+G_1)C_{old}
=-(960c_1^2+498c_1t+240t^2)$
giving with $2\chi_{ar}(2h)=-2h(2h-c_1)=-(12c_1^2-14c_1t+4t^2)$ 
that
\beqa
e(C)&=&2\chi_{ar}(C_{old})-2\chi_{ar}(2h)+2(C2h)\nonumber\\
    &=&2\chi_{ar}(C_{old})-2\chi_{ar}(2h)+8hq\nonumber\\
    &=&-(900c_1^2+552c_1t+228t^2)
\eeqa
and so
\beqa
\Delta_{cusp}=-4272c_1^2-2516c_1t-1076t^2
\eeqa
and altogether
\beqa
e(D_1)^{ord}+\Delta_{cusp}=204+2383c_1^2+1488c_1t+608t^2
\eeqa
Let us now come to the discussion of the codim 3 contributions.
$D_1$ is given by
$h_{2c_1-t}^3q_{3c_1-t}^2+S_{12c_1-4t}z+{\cal O}(z^2)$ where
\beqa
S_{12c_1-4t}=h^2(-\frac{3}{4}f_1^2+2g_1h+3f_0h^2)-3f_1hq^2+q^4
\eeqa
Let us now investigate what happens if, when we are lying on one
of the two matter curves $h$ or $q$, in addition $S$ vanishes,
so that the leading terms for the 
equation of $D_1$ becomes there $h^3+Sz+z^2$ resp.
$q^2+Sz+z^2$ and we can expect a singularity at $h\cap S$ resp. $q\cap S$.

Now, on closer inspection, one notices that $h\cap q\subset S$  
and $h\cap S\subset q$; 
therefore $h\cap S=h\cap q$ and $h\cap S\subset q\cap S$. 
On the other hand $q\cap S$ implies only 
$h^2(-\frac{3}{4}f_1^2+2g_1h+3f_0h^2)=0$, i.e. we do not
necessarily come to lie on $h$, there is still another divisor 
$R_{8c_1-2t}$ with $qS=q(2h+R)$ relevant. So one has a disjoint
decomposition $q\cap S=(q\cap h)\cup (q\cap R)$ and we will actually
consider the loci $h\cap q=h\cap S$ and 
$q\cap R=(q\cap S)\backslash (h\cap q)$
where the local forms of the singularities of $D_1$ are respectively
\beqa
h_x\cap q_y&\rightarrow &x^3y^2+(x^2+xy^2+y^4)z+z^2\nonumber\\
q_y\cap R_x&\rightarrow &y^2+z(x+y^2+y^4)+z^2\sim y^2+zx+z^2\sim 
y^2+x^2+z^2
\eeqa
So at $\#\, h\cap q=hq$ resp.
$\#\, ((q\cap S)\backslash (h\cap q))
=\#\, q\cap R=qR=q(S-2h)=4q^2-2hq$ points
one has an singularity of weighted homogeneous 
standard form $x^4+y^8+z^2$ (as the defining
equation is of weights $(2,1,4)$ in $(x,y,z)$)
resp. an ${\bf A_1}$ singularity which lead to corrections $\alpha$
resp. $\beta$ to the Euler number of the {\em singular} surface $D_1$
or in general to a correction $-\mu$ where $\mu$ is the colength (which is
finite as the singularity is isolated) of the Jacobian ideal (being also
the Euler number of the Milnor fibre minus 1). So 
$\alpha=1\cdot 3\cdot 7=-21$ and $\beta =-1$.

Now let us come to the fiber enhancements.
Now one has $C\cap r=h\cap q$ and
\beqa
\# (C\cap r)&=&hq\nonumber\\
e(C\cup h\cup q)&=&e(C)+e(h)+e(q)-hq
\eeqa
giving ($l$ and $k$ parametrize the fiber enhancements at the
codimension three loci)
\beqa
e(X)&=&1(e(D_1)-e(C)-e(h)-e(q)+hq)\nonumber\\
    & &+2(e(C)-hq)\nonumber\\
    & &+7(e(B_2)-e(h)-e(q)+hq)\nonumber\\
    & &+8(e(h)-hq+e(q)-hq-Rq)\nonumber\\
    & &+lhq+kRq\nonumber\\
    &=&e(D_1)+e(C)+7e(B_2)\nonumber\\
    & &+(l-10)hq+(k-8)Rq\nonumber\\
    &=&e(D_1)+e(C)+7e(B_2)\nonumber\\
    & &+(l-2k+6)hq+(k-8)4q^2
\eeqa
With
\beqa
e(D_1)&=&e(D_1)^{ord}+\Delta_{cusp}-2hq+\alpha hq+\beta Rq\nonumber\\
     &=&e(D_1)^{ord}+\Delta_{cusp}+(\alpha -2 -2\beta)hq +\beta 4q^2\nonumber\\
      &=&e(D_1)^{ord}+\Delta_{cusp}-21hq-4q^2
\eeqa
one gets
\beqa
e(X)&=&e(D_1)^{ord}+\Delta_{cusp}+e(C)+7e(B_2)\nonumber\\
    & &+(l-31)hq+(k-9)4q^2\nonumber\\
    &=&288+1476c_1^2+936c_1t+380t^2\nonumber\\
    & &+(l-31)hq+(k-9)4q^2\nonumber\\
    &=&288+1476c_1^2+936c_1t+380t^2-8q^2\nonumber\\
    & &+(l-31)hq+(k-9+2)4q^2\nonumber\\
\eeqa
where in the last equation the first line matches with 
the heterotic expectation
\beqa
24n_5=288+1404c_1^2+984c_1t+372t^2
\eeqa
giving corresponding predictions\footnote{Note that the form of
deviation of the $F$-theory result from the heterotic result is
already a non-trivial check since we have to tune only two
coefficients to match a quadratic expression in $c_1$ and $t$ with
three coefficients.} from the vanishing of the other terms.

\subsubsection{$D_6(I_2^*)$ singularity}
This time one has
\beqa
F_1&=& 4c_1+ 6r+ 4t\Rightarrow F_1r=2h\nonumber\\ 
G_1&=& 6c_1+ 9r+ 6t\Rightarrow G_1r=3h\nonumber\\ 
D_1&=&12c_1+16r+12t\Rightarrow D_1r=12c_1-4t=2h+2P_{4c_1-t}
\eeqa
where $h=2c_1-t$ and $P=4c_1-t$ and the last decomposition holds
not only on the level of cohomology classes but {\em actually 
on the level of divisors}
as seen from the equation $D_1=(h^2P^2+{\cal O}(z)=0)$.

\beqa
C_{old}&=&24(c_1+t)^2+18r(4c_1+t)\nonumber\\
C&=&f_1g_1-3(2c_1-t)r=24(c_1+t)^2+(66c_1+21t)r\nonumber\\
Cr&=&3(2c_1-t)(4c_1-t)=3hP\nonumber\\
\#(C\cap r)&=&hP
\eeqa
The cohomology 
classes of the two codimension 3 terms $Cr$ and $hP$ are again proportional.
Note also that not only $Ch\subset Cr$ but that they are actually 
equal as sets. 
The precise multiplicity is given by $C3h=Cr$ as
the part $3h$ occurs in the $Cr$ not only as cohomology class
but even as divisor.

Now $e(D_1)$ is besides contributions from point singularities
(considered below) computed as 
\beqa
e(D_1)^{ord}+\Delta_{cusp}&=&c_2(B_3)D_1-c_1(B_3)D_1^2+D_1^3
+\Delta_{cusp}\nonumber\\
&=&192+6248c_1^2+4296c_1t+1632t^2+\Delta_{cusp}
\eeqa

Now $F_1G_1=C_{old}=3hr+C$ so 
$2\chi_{ar}(C_{old})=2\chi_{ar}(3h)+e(C)-2(C3h)$ and 
\beqa
2\chi_{ar}(C_{old})=-(960c_1^2+498c_1t+240t^2)
\eeqa
giving with $2\chi_{ar}(3h)=-3h(3h-c_1)=-30c_1^2+33c_1t-9t^2$
\beqa
e(C)=-882c_1^2-567c_1t-225t^2
\eeqa
and 
\beqa
\Delta_{cusp}&=&-7CD_1+6c_1(B_3)C-4e(C)\nonumber\\
      &=&-4020c_1^2-2718c_1t-1038t^2
\eeqa
and altogether
\beqa
e(D_1^{ord})+\Delta_{cusp}=192+2228c_1^2+1578c_1t+594t^2
\eeqa
Let us now come to point singularities of the discriminant of equation
$0=h^2P^2+\frac{1}{4}S_{12c_1-3t}z+{\cal O}(z^2)$ where 
\beqa
S=h^3R-3h^2u^2P+3huP^2+4P^3
\eeqa
where $R=-2u^3+8g_0$.
Now a similar inspection as for $D_5$ shows that one has a disjoint
decomposition $P\cap S=(P\cap h)\cup (P\cap R)$ and we will actually
consider the loci $h\cap P=h\cap S$ and 
$P\cap R=(P\cap S)\backslash (h\cap P)$
where the local forms of the singularities of $D_1$ are respectively
(we neglect here some coefficients)
\beqa
h_x\cap q_y&\rightarrow &-\frac{3}{4}x^2y^2
+\frac{1}{4}(x^3R-3u^2x^2y+3uxy^2+4y^3)z+z^2\nonumber\\
P_y\cap R_x&\rightarrow &y^2+(x+y+y^2+y^3)z+z^2
\eeqa
So at $\#\, h\cap P=hP$ resp.
$\#\, P\cap R=PR=P(S-3h)=3P^2-3hP$ points
one has singularities which lead to corrections $\alpha$
resp. $\beta$ to the Euler number of the {\em singular} surface $D_1$.

So altogether one finds
\beqa
e(D_1)&=&e(D_1^{ord}+\Delta_{cusp})+\alpha hP +\beta (3P^2-3hP)\nonumber\\
      &=&192+2228c_1^2+1578c_1t+594t^2 +(\alpha -3\beta)hP+3\beta P^2
\eeqa

Now for $SO(12)$ one has the matter/enhancement scheme: 
$P-->{\bf 12}--> D_7$ of Euler $9=8+1$ 
and $h-->{\bf 32}-->"E_7"$ of Euler $8$ (as in the $G=E_7$ case here
occurred a collision which is not effectively additive;
cf. the six-dimensional discussion).

So now one has for the contributions from the fiber enhancement
\beqa
e(X)&=&1(e(D_1)-e(C)-e(h)-e(P)+hP+\#(C\cap r))\nonumber\\
    & &+2(e(C)-\#(C\cap r))\nonumber\\
    & &+8(e(B_2)-e(h)-e(P)+hP)\nonumber\\
    & &+8(e(h)-hP-\#(C\cap r))+9(e(P)-hP)\nonumber\\
    & &+l\#(C\cap r)+mhP\nonumber\\
    &=&e(D_1)+e(C)+8e(B_2)-e(h)\nonumber\\
    & &+(l-9)\#(C\cap r)+(m-8)hP\nonumber\\
    &=&e(D_1)+e(C)+8e(B_2)-e(h)\nonumber\\
    & &+(l+m-17)hP
\eeqa
giving altogether
\beqa
e(X)&=&288+1340c_1^2+1008c_1t+370t^2\nonumber\\
    & &+(l+m-17+\alpha -3\beta)hP+3\beta P^2\nonumber\\
    &=&288+1260c_1^2+1044c_1t+366t^2\nonumber\\
    & &+(-2+(l+m-17+\alpha -3\beta))hP+(6+3\beta )P^2
\eeqa
Comparing this with the heterotic value
\beqa
24n_5=288+1260c_1^2+1044c_1t+366t^2
\eeqa
gives the prediction\footnote{For the siginificance of this procedure
compare the last footnote.} $\beta=-2$ and $l+m=13-\alpha$.

\subsubsection{$A_1(I_2)$ singularity}

\beqa
F_1&=& 4c_1+ 8r+ 4t\Rightarrow F_1r=2H\nonumber\\ 
G_1&=& 6c_1+ 12r+ 6t\Rightarrow G_1r=3H\nonumber\\ 
D_1&=&12c_1+22r+12t\Rightarrow D_1r=2H+P
\eeqa
where $H=2c_1-2t$ and $P=8c_1-6t$ and the last decomposition holds
not only on the level of cohomology classes but {\em actually 
on the level of divisors}
as seen from the equation $D_1=(H^2P+{\cal O}(z)=0)$.

Now $F_1G_1=C_{old}=3Hr+C$ (on the level of divisors) so
\beqa
C_{old}&=&24(c_1+t)^2+96c_1r\nonumber\\
C&=&f_1g_1-6(c_1-t)r=24(c_1+t)^2+90c_1r+6rt\nonumber\\
Cr&=&\frac{3}{2}HP\nonumber\\
\#(C\cap r)&=&H\cdot \frac{1}{2}P=(2c_1-2t)(4c_1-3t)
\eeqa
where in the last line we used the fact that the term $3H$ occurs in
$Cr$ as divisor and not just as a cohomology class.

With
\beqa
\Delta_{cusp}&=&-5748c_1^2-588c_1t-1728t^2\nonumber\\
e(C)&=&-1170c_1^2-234c_1t-324t^2
\eeqa
\beqa
e(D_1)&=&c_2(B_3)D_1-c_1(B_3)D_1^2+D_1^3+\Delta_{cusp}\\
&=&264+2942c_1^2+906c_1t+756t^2
\eeqa
one finds
\beqa
e(X_4)&=&1(e(D_1)-[e(C)-\#(C\cap r)+e(H)-HP+e(P)])\nonumber\\
      & &+2(e(C)-\#(C\cap r))\nonumber\\
      & &+2(e(B_2)-[e(H)+e(P)-HP])\nonumber\\
      & &+3(e(H)-HP-\#(C\cap r)+e(P)-HP)\nonumber\\
      & &+l\#(C\cap r)+mHP\nonumber\\
      &=&e(D_1)+e(C)+2e(B_2)\nonumber\\
      & &+(l-4)\#(C\cap r)+(m-3)HP\nonumber\\
      &=&288+1770c_1^2+672c_1t+430t^2\nonumber\\
      & &+(\frac{1}{2}l+m-5)HP\nonumber\\
      &=&288+1866c_1^2+504c_1t+504t^2\nonumber\\
      & &+(\frac{1}{2}l+m-17)HP
\eeqa

Comparing this with the heterotic side 
\beqa
24n_5=288+1866c_1^2+504c_1t+504t^2
\eeqa
leads again to a corresponding prediction.

\subsubsection{$A_2(I_3)$ singularity}

\beqa
F_1&=& 4c_1+ 8r+ 4t\Rightarrow F_1r=4h\nonumber\\ 
G_1&=& 6c_1+ 12r+ 6t\Rightarrow G_1r=6h\nonumber\\ 
D_1&=&12c_1+21r+12t\Rightarrow D_1r=4h+P
\eeqa
where $h=c_1-t$ and $P=8c_1-5t$ and the last decomposition holds
not only on the level of cohomology classes but {\em actually 
on the level of divisors}
as seen from the equation $D_1=(h^4P+{\cal O}(z)=0)$.

Now $F_1G_1=C_{old}=8hr+C$ (on the level of divisors) so
\beqa
C_{old}&=&24(c_1+t)^2+96c_1r\nonumber\\
C&=&f_1g_1-8(c_1-t)r=24(c_1+t)^2+88c_1r+8rt\nonumber\\
Cr&=&8(c_1-t)(3c_1-2t)\nonumber\\
\#(C\cap r)&=&(c_1-t)(3c_1-2t)
\eeqa
where in the last line we used the fact that the term $8h$ occurs in
$Cr$ as divisor and not just as a cohomology class.

With (note that the discriminant (\ref{I3dis}) shows an intrinsic
singularity of $D_1$ at $h\cap Q_{3c_1-2t}$) where
\beqa
\Delta_{cusp}&=&-5656c_1^2-776c_1t-1632t^2\nonumber\\
e(C)&=&-1112c_1^2-328c_1t-288t^2\nonumber\\
e(D_1)&=&c_2(B_3)D_1-c_1(B_3)D_1^2+D_1^3+\Delta_{cusp}+\alpha hQ\nonumber\\
      &=&252+2627c_1^2+1360c_1t+600t^2+\alpha hQ
\eeqa
one finds
\beqa
e(X_4)&=&1(e(D_1)-[e(C)-\#(C\cap r)+e(h)-hP+e(P)])\nonumber\\
      & &+2(e(C)-\#(C\cap r))\nonumber\\
      & &+3(e(B_2)-[e(h)+e(P)-hP])\nonumber\\
      & &+4(e(h)-hP-\#(C\cap r)+e(P)-hP)\nonumber\\
      & &+l\#(C\cap r)+mhP\nonumber\\
      &=&e(D_1)+e(C)+3e(B_2)\nonumber\\
      & &+(l-5)h(3c_1-2t)+(m-4)hP\nonumber\\
      &=&288+1512c_1^2+1032c_1t+312t^2\nonumber\\
      & &+(l-5+\alpha )h(3c_1-2t)+(m-4)hP\nonumber\\
      &=&288+1704c_1^2+720c_1t+432t^2\nonumber\\
      & &+(l-5+\alpha )h(3c_1-2t)+(m-4)hP-24hP\nonumber\\
      &=&288+1704c_1^2+720c_1t+432t^2\nonumber\\
      & &+(l-5+\alpha )h(3c_1-2t)+(m-28)hP
\eeqa
Comparing with the heterotic side 
\beqa
24n_5=288+1704c_1^2+720c_1t+432t^2
\eeqa
leads to a corresponding prediction.

Finally we remark that the heterotic prediction also matches with the 
computation for the case $B_2=F_0$ and $t=0[b]+3[f]$ given in table 
(6.4) of \cite{KLRY}.

\subsubsection{$A_3(I_4)$ singularity}
\beqa
F_1&=& 4c_1+ 8r+ 4t\Rightarrow F_1r=4h\nonumber\\ 
G_1&=& 6c_1+ 12r+ 6t\Rightarrow G_1r=6h\nonumber\\ 
D_1&=&12c_1+20r+12t\Rightarrow D_1r=4h+P
\eeqa
where $h=c_1-t$ and $P=8c_1-4t$ and the last decomposition holds
not only on the level of cohomology classes but {\em actually 
on the level of divisors}
as seen from the equation $D_1=(h^4P+{\cal O}(z)=0)$.

Now $F_1G_1=C_{old}=12hr+C$ (on the level of divisors) so
\beqa
C_{old}&=&24(c_1+t)^2+96c_1r\nonumber\\
C&=&f_1g_1-12(c_1-t)r=24(c_1+t)^2+84c_1r+12rt\nonumber\\
Cr&=&3hP\nonumber\\
\#(C\cap r)&=&hP
\eeqa
where in the last line we used the fact that the term $3h$ occurs in
$Cr$ as divisor and not just as a cohomology class.

Furthermore
\beqa
\Delta_{cusp}&=&-5736c_1^2-624c_1t-1704t^2\nonumber\\
e(C)&=&-972c_1^2-564c_1t-192t^2\nonumber\\
\Delta_{tacn}&=&-2268c_1^2+348c_1t-122t^2\nonumber\\
e(D_1)&=&c_2(B_3)D_1-c_1(B_3)D_1^2+D_1^3+\Delta_{cusp}
              +\Delta_{tacn}-e(h)\nonumber\\
&=&240+2140c_1^2+2084c_1t+328t^2+\Delta_{tacn}-e(h)\nonumber\\
&=&240+1914c_1^2+2431c_1t+207t^2
\eeqa

Now for $SU(4)$ one has the matter/enhancement schemes: 
$P-->{\bf 4}-->SU(5)$ of Euler $5=4+1$ 
and $h-->{\bf 6}-->SO(8)$ of Euler $6=4+1 +1$ so that
we need to add an $e(h)$ in $e(X^4)$. So one gets for the fiber
enhancements
\beqa
e(X_4)&=&1(e(D_1)-[e(C)-\#(C\cap r)+e(h)-hP+e(P)])\nonumber\\
      & &+2(e(C)-\#(C\cap r))\nonumber\\
      & &+4(e(B_2)-[e(h)+e(P)-hP])\nonumber\\
      & &+6(e(h)-hP-\#(C\cap r))+5(e(P)-hP))\nonumber\\
      & &+l\#(C\cap r)+mhP\nonumber\\
      &=&e(D_1)+e(C)+4e(B_2)+e(h)\nonumber\\
      & &+(l-7)\#(C\cap r)+(m-6)hP\nonumber\\
      &=&288+938c_1^2+1868c_1t+14t^2\nonumber\\
      & &+(l+m-13)hP
\eeqa
which matches with the 
codimension one expectation from \cite{KLRY}
for $t=c_1$ (cf. section (7.1))
giving $e(X_4)=288+2820c_1^2$.

\subsubsection{$A_4(I_5)$ singularity}

\beqa
F_1&=& 4c_1+ 8r+ 4t\Rightarrow F_1r=4h\nonumber\\ 
G_1&=& 6c_1+ 12r+ 6t\Rightarrow G_1r=6h\nonumber\\ 
D_1&=&12c_1+19r+12t\Rightarrow D_1r=4h+P
\eeqa
where $h=c_1-t$ and $P=8c_1-3t$ and the last decomposition holds
not only on the level of cohomology classes but {\em actually 
on the level of divisors}
as seen from the equation $D_1=(h^4P+{\cal O}(z)=0)$.

Now $F_1G_1=C_{old}=15hr+C$ (on the level of divisors) so
\beqa
C_{old}&=&24(c_1+t)^2+96c_1r\nonumber\\
C&=&f_1g_1-15(c_1-t)r=24(c_1+t)^2+81c_1r+15rt\nonumber\\
Cr&=&4h\cdot 6h+15ht=3h(8h+5t)=3hP\nonumber\\
\#(C\cap r)&=&hP
\eeqa
by the similar arguments as in the earlier cases.
Note also that not only $Ch\subset Cr$ but that they are actually 
equal as sets. 
The precise multiplicity is given by $C3h=Cr$ as
the part $3h$ occurs in the $Cr$ not only as cohomology class
but even as divisor.

Now (up to a codimension two correction from the higher double point
curve along $h$ and codimension three contributions from corrections 
from point singularities)
\beqa
e(D_1)&=&c_2(B_3)D_1-c_1(B_3)D_1^2+D_1^3+\Delta_{cusp}\nonumber\\
&=&228+7469c_1^2+3210c_1t+1878t^2+\Delta_{cusp}
\eeqa
One has 
$-7CD_1+6c_1(B_3)C=-9222c_1^2-3495c_1t-2259t^2$.
Now $F_1G_1=C_{old}=15hr+C$ so 
$2\chi_{ar}(C_{old})=2\chi_{ar}(15h)+e(C)-2(C15h)$ and 
$2\chi_{ar}(C_{old})=-1296c_1^2-432t^2$
giving with $2\chi_{ar}(15h)=-15h(15h-c_1)=-210c_1^2+435c_1t-225t^2$ 
and $15Ch=5Cr=15hP$ that
\beqa
e(C)=-846c_1^2-765c_1t-117t^2
\eeqa
and 
\beqa
\Delta_{cusp}=-5838c_1^2-435c_1t-1791t^2
\eeqa
and altogether (up to the limitations mentioned above, i.e.
the codim 2 and codim 3 contributions from corrections along
the higher double point curve $h$ and at their intersection $Ch$)
\beqa
e(D_1)&=&228+7469c_1^2+3210c_1t+1878t^2+\Delta_{cusp}\nonumber\\
&=&228+1631c_1^2+2775c_1t+87t^2
\eeqa

Let us now come to the discussion of the codim 3 contributions.
The discriminant equation $\Delta$ for $D$ is given by equ. (\ref{I5discr})
so that one has inside the $r$-plane the equation $h^4P=0$ with
$P=2h^2g_1-3f_1qh-3Hq^2$. Now 
note that near $h$ the equation of $D_1$ is given by (with $e:=f_1H+q^2$)
\beqa
-3Hq^2(h^4-2h^2Hz+H^2z^2)+(-H^2(2g_1H+3f_1^2)+\frac{9}{4}e^2)z^3
+(f_1^3+3g_1e)z^4+g_1^2 z^5 =0\nonumber\\
\label{I5discronplane}
\eeqa
So the complete square structure of the leading terms showed 
that we do not
have a generic tacnode structure, just as for $I_6$ but in contrast 
to the case $I_4$.
With $w:=Hz-h^2$
the terms up to third order became (everything up to coefficients)
$w^2+z^3\rightarrow h^6+3h^4w+w^2+w^3$ near $(h,w)=(0,0)$ which
goes with $w:=v-\frac{3}{2}h^4$ to the normal form $h^6+v^2$.

Now let us compare
with the heterotic side where one has the spectral cover
equation $a_0+a_2x+a_3y+a_4x^2+a_5xy=0$ with $a_i$ in the class
$\eta_1-ic_1=(6-i)c_1-t$, so (up to inessential factors)
\beqa
h_{c_1-t}&=&a_5\nonumber\\
H_{2c_1-t}&=&a_4\nonumber\\
q_{3c_1-t}&=&a_3\nonumber\\
f_1=f_{4c_1-t}&=&a_2\nonumber\\
g_1=g_{6c_1-t}&=&a_0
\eeqa
Further one finds coincidence\footnote{The coincidence of weights
for the $P$'s and $a_i$ was already noticed in \cite{FMW}; here we
give the $F$-theoretic $P$ in $h,H,q,f_1,g_1$.}
of the heterotic expression 
$P\sim a_0a_5^2-a_2a_3a_5+a_3^2a_4$ for this matter curve 
with the $F$-theoretic $P$ (up to inessential factors).

The singularity structure, i.e. the overall higher tacnode structure
along $h$, will change if at special loci coefficient functions vanish
(say $R$ at $R\cap h=(h\cap q)\cup (h\cap H)$)
so that one gets degenerations of the structure equ. (\ref{I5discronplane}).
Now the relevant loci are $h\cap q$ and $h\cap H$ and one has
\beqa
hP&=&h(2q+H)\nonumber\\
hs&=&h(2q+2H)\nonumber\\
hR&=&h(2q+3H)
\eeqa
Both, $h\cap R$ and  $h \cap P$, lie in (and are actually equal to)
$h\cap (q\cup H)$.
In the following we will divide $h \cap P$ and $R\cap h$
into $h\cap q$ and $h\cap H$.

This leads to the consideration of the following 
loci and singularities (up to coefficients):
\beqa
h_x\cap q_y &\Rightarrow &x^4(x^2+xy+y^2)+x^2(x^2+xy+xy^3+y^2)z\nonumber\\
& &+(x^3y+x^2+x^2y^2+xy^3+xy+y^2)z^2+z^3
\eeqa
respectively
\beqa
h_x\cap H_y &\Rightarrow &x^4(x^2+x+y)+x^2(x^2+x^2y+xy+x+y^2)z\nonumber\\
& &+(x^3+x^2+x^2y+x^2y^2+xy+xy^2+y^3)z^2+z^3
\eeqa
So we expect further corrections (denoted below by $\alpha$ and $\beta$) 
to $e(D_1)$ at these $hq$ resp. $hH$ points.

There are '$B_3$-intrinsic' corrections hidden in 
\beqa
e(D_1)&=&c_2(B_3)D_1-c_1(B_3)D_1^2+D_1^3+\Delta_{\rm cusp}
+\Delta_{\rm hightac}-e(h)
+\Delta_{C\cap r}+\Delta_{hq}+\Delta_{hH}\nonumber\\
         &=&e(D_1)^{ord}+\Delta_{\rm cusp}+\Delta_{\rm hightac}-e(h)
             +(\gamma-2)\#(C\cap r)\nonumber\\
         & & +(\alpha -4)\#(h\cap q)+(\beta -4)\#(h\cap H)\nonumber\\
         &=&e(D_1)^{ord}+\Delta_{\rm cusp}+\Delta_{\rm hightac}-e(h)
             +(2\gamma+\alpha -8)hq+(\gamma+\beta -6)hH\nonumber\\
&=:&e(D_1)^{ord}+\Delta_{\rm cusp}+\Delta_h
\eeqa
where this time $\Delta_{C\cap r}$ comprises two effects: first, that
the points of $C$ which lie in $r$ are no longer cusp points, and
secondly that, as the cusp curve $C$ and the tacnode curve $hr$ intersect,
the corrections $\Delta_{\rm cusp_C}$ and $\Delta_{\rm tac_h}$ get
a third term 
$\Delta_{\rm cusp\cap hightac}=\gamma \#(C\cap h)=\gamma \#(C\cap r)$
describing the influence of the intersection locus.
Also at the loci $h\cap q$ and $h\cap H$ we have the point singularities
of $D_1$ measured 
by $\alpha$ and $\beta$, where we then have to subtract
their pointwise tacnode contribution $"+4"$, just as we did for the 
points of $C\cap r$ which were not cusp points where we subtracted their
pointwise cusp contribution $"+2"$.

Now again we now that $Cr=3hP$ as cohomology classes and 
$C\cap r\subset h$. One has
$C\cap r\not\subset h\cap P$, i.e.
that $C\cap r$ lies neither in $h\cap q$ nor in $h\cap H$. Then
one has with the disjoint decomposition 
$h\cap P=(h\cap q)\cup (h\cap H)$ that\footnote{the 
left hand side of the first equation 
is the intersection of $h$ with a divisor of class
$P$ but {\em not with the divisor $P$ itself} $\;$ by assumption;
note that the second equation is 
the intersection of $h$ with {\em the divisor} $P$}
\beqa
\# (C\cap r)&=&hP=2hq+hH\nonumber\\
\#(h\cap P)&=&\#(h\cap q)+\#(h\cap H)=hq+hH
\eeqa

Now for $SU(5)$ one has the matter/enhancement schemes: 
$P-->{\bf 5}-->SU(6)$ of Euler $6=5+1$ 
and $h-->{\bf 10}-->SO(10)$ of Euler $7=5+1 +1$ so that
we need to add an $e(h)$ in $e(X^4)$.
So one finds
\beqa
e(X)&=&1(e(D_1)-e(C)-e(h)-e(P)
               +\#(C\cap r)+\#(h\cap P))\nonumber\\
       & &+2(e(C)-\#(C\cap r))\nonumber\\
       & &+5(e(B_2)-e(h)-e(P)+\#(h\cap P))\nonumber\\
       & &+7(e(h)-\#(h\cap P)-\#(C\cap r))+6(e(P)-\#(h\cap P))\nonumber\\
       & &+l\#(C\cap r)+m\#(h\cap q)+k\#(h\cap H)\nonumber\\
       &=&e(D_1)+e(C)+5e(B_2)+e(h)\nonumber\\
       & &+(l-8)\#(C\cap r)+\#(m-7)(h\cap q)+(k-7)\#(h\cap H)\nonumber\\
       &=&e(D_1)+e(C)+5e(B_2)+e(h)\nonumber\\
       & &+(2l+m-23)hq+(l+k-15)hH\nonumber\\
       &=&288+780c_1^2+2011c_1t-31t^2+\Delta_h\nonumber\\
       & &+(2l+m-23)hq+(l+k-15)hH
\eeqa
Comparing this with the heterotic value
\beqa
24n_5=288+1410c_1^2+975c_1t+375t^2
\eeqa
gives the condition (note that $\Delta_h$ was proportional to $h$ too)
\beqa
\Delta_h+(2l+m-23)hq+(l+k-15)hH&=&h(630c_1-406t)
\eeqa

\subsubsection{$A_5(I_6)$ singularity}

\beqa
F_1&=& 4c_1+ 8r+ 4t\Rightarrow F_1r=4h\nonumber\\ 
G_1&=& 6c_1+ 12r+ 6t\Rightarrow G_1r=6h\nonumber\\ 
D_1&=&12c_1+18r+12t\Rightarrow D_1r=12c_1-6t=4h+P_{8c_1-2t}
\eeqa
where $h=c_1-t$ and $P=8c_1-2t$ and the last decomposition holds
not only on the level of cohomology classes but {\em actually 
on the level of divisors}
as seen from the equation $D_1=(h^4P+{\cal O}(z)=0)$.

Now $F_1G_1=C_{old}=18hr+C$ (on the level of divisors) so
\beqa
C_{old}&=&24(c_1+t)^2+96c_1r\nonumber\\
C&=&f_1g_1-18(c_1-t)r=24(c_1+t)^2+78c_1r+18rt\nonumber\\
Cr&=&4h\cdot 6h+18ht=3h(8h+6t)=3hP
\eeqa
Note again that not only $Ch\subset Cr$ but that they are actually 
equal as sets. 
The precise multiplicity is given by $C3h=Cr$ as
the part $3h$ occurs in the $Cr$ not only as cohomology class
but even as divisor.

Now
\beqa
e(D_1)&=&c_2(B_3)D_1-c_1(B_3)D_1^2+D_1^3+\Delta_{cusp}
            +\Delta_h\nonumber\\
&=&216+7062c_1^2+3642c_1t+1764t^2+\Delta_{cusp}+\Delta_h
\eeqa
One has
$-7CD_1+6c_1(B_3)C=-8820c_1^2-4068c_1t-2088t^2$.
Now $F_1G_1=C_{old}=18hr+C$ so 
$2\chi_{ar}(C_{old})=2\chi_{ar}(18h)+e(C)-2(C18h)$ and 
$2\chi_{ar}(C_{old})=-1296c_1^2-432t^2$
giving with $2\chi_{ar}(18h)=-18h(18h-c_1)=-18(c_1-t)(17c_1-18t)=
-18(17c_1^2-35c_1t+18t^2)=-306c_1^2+630c_1t-324t^2$ 
and $18Ch=6Cr=18hP=18(c_1-t)(8c_1-2t)$ that
\beqa
e(C)&=&-702c_1^2-990c_1t-36t^2\nonumber\\
\Delta_{cusp}&=&-6012c_1^2-108c_1t-1944t^2
\eeqa
and thus
\beqa
e(D_1)=216+1050c_1^2+3534c_1t-180t^2+\Delta_h
\eeqa

Now for $SU(6)$ one has the matter/enhancement schemes: 
$P-->{\bf 6}-->SU(7)$ of Euler $7=6+1$ 
and $h-->{\bf 15}-->SO(12)$ of Euler $8=6+1{\bf +1}$ so that,
like for $SU(5)$ we need to add an $e(h)$ in $e(X^4)$.
So now one has (let us include in $\Delta_h^{\prime}$ as opposed to
$\Delta_h$ also the fiber enhancements at the codimension three loci,
which are also in $h$)
\beqa
e(X^4)^{-->cod2}&=&e(D_1)+e(C)+e(h)+6e(B_2)+\Delta_h^{\prime}\nonumber\\
&=&288+342c_1^2+2545c_1t-217t^2+\Delta_h^{\prime}
\eeqa
Comparing this with the heterotic value
\beqa
24n_5=288+1266c_1^2+1035c_1t+369t^2
\eeqa
gives as prediction
$\Delta_h^{\prime}=h(1150c_1-665t)$.

\subsection{An observation on the 4D Higgs chains}

Let us close with an observation relating the Euler numbers in
neighbouring cases in certain Higgs-chains.

\subsubsection{6D case}

For the chain ${\bf E_7 \rightarrow E_6 \rightarrow D_5 \rightarrow I_5 }$
with the Euler numbers 
\beqa
e(E_7)=-284-56n\nonumber\\
e(E_6)=-300-54n\nonumber\\
e(D_5)=-312-52n\nonumber\\
e(I_5)=-320-50n
\eeqa
one gets
\beqa
e(higher)-2e(enh. locus)=e(lower)
\eeqa
where the enhancement locus is in this chain always the $h$ locus 
of the {\em higher} group, i.e. $f_1=4c_1-t=8-n$, $g_1=3c_1-t=6-n$,
$h=2c_1-t=4-n$ and $h=c_1-t=2-n$ for $E_7, E_6, D_5$ 
and $I_5$ respectively.

The reason for these relations
is of course easy to see. Take for example the case 
of $E_7$ and $E_6$. One has $h^{1,1}(E_7)-h^{1,1}(E_6)=1$ and
from $h^{2,1}=n_H-1=
dim {\cal M}_{12+n}(E_8)+dim {\cal M}_{12-n}(H_k)+h^{1,1}(K3)-1
=112+30n+k(12-n)-(k^2-1)+19$
where $H_k=SU(k)$ is the commutant with $k=2$ and $3$ one gets
that $h^{2,1}(E_7)-h^{2,1}(E_6)=
[k(12-n)-(k^2-1)]-[(k+1)(12-n)-((k+1)^2-1)]=-(12-n)+2k+1$ for $k=2$
and so $e(E_7)-e(E_6)=2(12-n-2k)=2(2(6-k)-n)$ where $2(6-k)-n$
is just the degree of the relevant enhancement curve.

Similarly for the $I_k$ series one gets with
\beqa
e(I_6)=-288-54n\nonumber\\
e(I_5)=-320-50n\nonumber\\
e(I_4)=-352-44n\nonumber\\
e(I_3)=-384-36n\nonumber\\
e(I_2)=-420-24n
\eeqa
that again
\beqa
e(higher)-2e(enh. locus)=e(lower)
\eeqa
where the relevant curve is in this chain always the $P$ locus
of the {\em higher} group, i.e. $8c_1-(8-k)t=16-(8-k)n$ for $I_k$ the
higher group (note that in our set-up we made the switch $n\rightarrow
-n$).

\subsubsection{4D case}

For the chain ${\bf E_7 \rightarrow E_6 \rightarrow D_5 \rightarrow I_5 }$
with the euler numbers 
\beqa
e(E_7)=288+1350c_1^2+1020c_1t+366t^2\nonumber\\
e(E_6)=288+1386c_1^2+ 999c_1t+369t^2\nonumber\\
e(D_5)=288+1404c_1^2+ 984c_1t+372t^2\nonumber\\
e(I_5)=288+1410c_1^2+ 975c_1t+375t^2 
\eeqa
one gets
\beqa
e(higher)-3e(enh. locus)=e(lower)
\eeqa
where the enhancement locus is in this chain always the $h$ locus 
of the {\em higher} group, i.e. $f_1=4c_1-t$, $g_1=3c_1-t$,
$h=2c_1-t$ and $h=c_1-t$ for $E_7, E_6, D_5$ 
and $I_5$ respectively.

Again let us see from a heterotic point of view how this structure 
emerges. From 
\beqa
c_2(SU(n))=\eta \sigma-\frac{n^3-n}{24}c_1^2-\frac{n}{8}\eta(\eta-nc_1)
\eeqa
one sees that
\beqa
c_2(SU(n+1))=c_2(SU(n))-\frac{1}{8}[(n^2+n)c_1^2+\eta(-nc_1+\eta-c_1-nc_1)]
\eeqa
where the correction term is
\beqa
-\frac{1}{8}[(n^2+n)c_1^2+(6c_1-t)((5-2n)c_1-t)]
=\frac{1}{8}e(h_{\eta-nc_1})
\eeqa
with $\eta-nc_1=(6-n)c_1-t$ the class of the $h$ matter curve ($a_n=0$).
This leads in $24n_5$ to the searched for $3e(h)$.

Furthermore inside the $I_k$ series one has the same relation with
\beqa
e(I_6)&=&288+1266c_1^2+1035c_1t+369t^2\nonumber\\
e(I_5)&=&288+1410c_1^2+ 975c_1t+375t^2
\eeqa
where the relevant curve is in this chain the $P$ locus
of the {\em higher} group, i.e. $8c_1-(8-k)t$ for $I_k$ the
higher group.
\newpage
\appendix

{\Large {\bf Appendix}}

In {\em section A} we give some background and notation concerning
four-dimensional $F$-theory models, i.e. 
Calabi-Yau four-folds and recall some facts related to the four-flux.

In {\em section B} the different constructions of the vector bundles
of the dual heterotic models are described and the computation of the
second Chern class of $E_k$ bundles in the parabolic framework is 
derived.

In {\em section C} some known resp. expected connections between the
moduli spaces of compactifications in the two dual pictures of the
heterotic string and $F$-theory are described, especially the analogy
between the gamma class of the bundle description and the four-flux on
the $F$-theory side is recalled.

{\em Section D} has a somewhat different flavour in that it lists some
computational details concerning the explicit discriminant equations
and spectrum computations for the case of Calabi-Yau {\em three}-folds.

\section{$4d$ $F$-theory models}

{\resetcounter}

\subsection{The geometry of the four-fold}

We will consider Calabi-Yau fourfolds $X$ which are elliptically
fibered $\pi: X\rightarrow B_3$ over a complex three-dimensional 
base $B_3=B$, 
let $\sigma$ be a section. $X$ can be described by a Weierstrass equation 
$y^2z = x^3+g_2xz^2+g_3z^3$
which embeds $X$ in a ${\bf P}^2$ bundle $W \rightarrow B$ which 
is the projectivization of a vector bundle  
${\bf P}({\cal L}^2 \oplus {\cal L}^3 \oplus {\cal O}_B)$ with ${\cal L}$ 
being a line bundle 
over $B$. Since the canonical 
bundle of $X$ has to be trivial we get from $K_X=\pi^*(K_B+{\cal L})$ the 
condition 
${\cal L}=-K_B$. Further we can think of $x,y$ and $z$ as homogeneous 
coordinates 
on the ${\bf P}^2$ fibers, i.e. they are sections of ${\cal O}(1)\otimes 
K_B^{-2}$, ${\cal O}(1)\otimes K_B^{-3}$ and ${\cal O}(1)$, whereas
$g_2$ and $g_3$ are sections of $H^0(B, K_B^{-4})$ and $H^0(B, K_B^{-6})$ 
respectively and the Weierstrass 
equation is a section of ${\cal O}(1)^3\otimes K_B^{-6}$.
The section $\sigma$ can be thought of as the point 
at infinity $x=z=0, y=1$. The descriminant of the elliptic fibration 
is given by 
$\Delta=4g_2^3-27g_3^2$ which is a section of $K_B^{-12}$. 
If $\Delta=0$ at a 
point $p\in B$ 
the type of singular fiber is determined by the orders of vanishing 
$\rm{ord}(g_2)=a$,
$\rm{ord}(g_3)=b$ and $\rm{ord}(\Delta)=c$, and given by 
Kodaira's classification of elliptic fiber singularities: 
\begin{center}
\begin{tabular}{c|c|c|c|c}
$a$ & $b$   & $c$ & $\rm{fiber}$  & 
\rm{singularity}  
\\ \hline
  $\ge 0$ & $\ge 0$          & $0$        & $smooth$ & $none$      \\     
  $\ \ \ 0$ & $\ \ \ 0$          & $n$        & $I_n$    & $A_{n-1}$\\    
  $\ge 1$ & $\ \ \ 1$          & $2$        & $II$     & $none$    \\  
  $\ge 1$ & $\ge 2$          & $3$        & $III$    & $A_1$      \\
  $\ge 2$ & $\ \ \ 2$          & $4$        & $IV$     & $A_2$      \\ 
  $\ \ \ 2$ & $\ge 3$          &$n+6$       & $I_n^*$  & $D_{n+4}$  \\
  $\ge 2$ & $\ \ \ 3$          &$n+6$       & $I_n^*$  & $D_{n+4}$   \\ 
  $\ge 3$ & $\ \ \ 4$          & $8$        & $IV^*$   & $E_6$       \\
  $\ \ \ 3$ & $\ge 5$          & $9$        & $III^*$  & $E_7$      \\ 
  $\ge 4$ & $\ \ \ 5$          & $10$       & $II^*$   & $E_8$     
\end{tabular}
\end{center}
Let us denote by $F = -4K_B$, $G = -6K_B$ and $D = -12K_B$ 
the classes of the divisors associated to the vanishing 
of $g_2, g_3$ and $\Delta$ respectively.

We will assume $B_3$ to be a ${\bf P}^1$ bundle 
$\sigma:B_3\rightarrow B_2$ which is the projectivization 
${\bf P}(Y)$ of a vector bundle 
$Y={\cal O}\oplus {\cal T}$, with ${\cal T}$ a line bundle 
over $B_2$ and ${\cal O}(1)$ 
a line bundle on the total space of ${\bf P}(Y)\rightarrow B_2$ 
which restricts on each ${\bf P}^1$ 
fiber to the corresponding line bundle over ${\bf P}^1$. 
Further let $u$, $v$ be 
homogeneous coordinates of the ${\bf P}^1$ bundle and 
think of $a$ and $b$ as sections of ${\cal O}(1)$ and 
${\cal O}(1)\otimes {\cal T}$ over $B_2$. 
If we set $r=c_1({\cal O}(1))$ and $t=c_1({\cal T})$ 
and $c_1({\cal O}\otimes {\cal T})=r+t$ 
then the cohomology ring of $B_3$ is 
generated over the cohomology ring of $B_2$ 
by the element $r$ with the relation $r(r+t)=0$, 
i.e. the divisors $u=0$ resp. $v=0$, which are dual to 
$r$ resp. $r+t$, do not intersect. From adjunction 
$c(B_3)=(1+c_1(B_2)+c_2(B_2))(1+r)(1+r+t)$
one finds for the Chern classes\footnote{Unspecified Chern classes 
refer to $B_2$.}
$c_1(B_3)= c_1+2r+t, \;
c_2(B_3)= c_2+c_1t+2c_1r$.
Note that $B_2$ will be chosen to be rational. For this
recall that the arithmetic genus $p_a$ of $B_3$ 
has to be equal to one 
$1=p_a=\frac{1}{24}\int_{B_3}c_1(B_3)c_2(B_3)
=\frac{1}{12}\int_{B_2}c_1^2+c_2$ (for more details cf. \cite{andis})
in order to satisfy the $SU(4)$ holonomy condition for $X$, 
otherwise there are 
non-constant holomorphic differentials on $B_3$ which 
would pull back to X. 

\subsection{The four-flux}

In order to obtain consistent F-theory compactifications 
to four-dimensions
on Calabi-Yau fourfold $X$, it is necessary to turn on a
number $n_3$ of space-time filling three-branes for tadpole cancellation 
\cite{SVW}. This is related to the fact that compactifications of
the type IIA string on $X$ are destabilized at one loop by 
$\int B\wedge I_8$, where $B$ is the NS-NS two-form which couples 
to the string and $I_8$ a linear combination of the Pontryagin classes 
$p_2$ and $p_1^2$ \cite{VW}.
So, compactifications of the type IIA string to two-dimensions lead to
a tadpole term $\int B$ which is proportional to the 
Euler characteristic of $X$. Similar, in M-theory compactifications to 
three-dimensions on $X$ arises a term $\int C\wedge I_8$ with $C$ being 
the M-theory three-form, and integration over 
$X$ then leads to the tadpole term which is proportional to $\chi(X)$, 
and couples to the 2-brane \cite{DLM},\cite{BB}. Now, as M-theory 
compactified to three-dimensions on $X$ is related to a F-theory 
compactification on $X$ to four-dimensions \cite{V}, one is lead to 
expect a term $\int A$ with $A$ now being the R-R four form potential, 
which couples to the three-brane in F-theory \cite{SVW}. 
Taking into account the proportionality constant \cite{Wphase}, 
one finds $\frac{\chi(X)}{24}=n_3$ three-branes in F-theory 
(or membranes in M-theory resp. strings in type IIA theory) \cite{SVW}.    

Furthermore the tadpole in M-theory will be corrected
by a classical term $C\wedge dC\wedge dC$, which appears if $C$ gets
a background value on $X$ and thus leads to a contribution 
$\int dC\wedge dC$ to the tadpole \cite{BB}. Also one has 
\cite{W4flux} a quantization law for the four-form field strength $G$ 
of $C$ (the four-flux) the modified integrality condition 
$G=\frac{dC}{2\pi}=\frac{c_2}{2}+\alpha$ with $\alpha\in H^4(X,{\bf Z})$ 
where $\alpha$ has to satisfy \cite{CD} the bound
$-120-\frac{\chi(X)}{12}\le \alpha^2+\alpha c_2\le -120$,
in order to keep the wanted amount of supersymmetry in a consistent 
compactification.  
Altogether\footnote{the presence of a non-trivial instanton 
background can also contribute to the anomaly \cite{BJPS}; this adds
a term $\sum_{j} \int_{\Delta_j} c_2(E_j)$ to $n_3$ where 
$\int_{\Delta_j} c_2(E_j)=k_j$ are the instanton numbers of
possible background gauge bundles $E_j$ inside the 7-brane \cite{BJPS} 
and $\Delta_j$ denotes the discriminant components in $B_3$}
one finds \cite{DM}
\beqa
\frac{\chi(X)}{24}=n_3+\frac{1}{2}\int G\wedge G
\eeqa
for consistent $N=1$ F-theory compactifications on $X$ to 
four-dimensions.

{}From the connection of $F$-theory to $M$-theory via 
$S^1$ compactification
one expects some lifting of the four-flux of $M$-theory 
to play a role in
$F$-theory. This is a $(2,2)$ form in integral cohomology 
(essentially; its
precise quantization law leading to half-integrality 
is reviewed below) 
and so according to the Hodge conjecture supported 
on an algebraic surface $S$ in $X_4$.
{}From the primitivity condition (again further reviewed below) 
for the self-dual
four-flux ($g^{b\bar{b}}$ the Kaehler metric) \cite{BB}
\beqa
F_{a\bar{a}b\bar{b}}g^{b\bar{b}}=0
\eeqa
one gets a well-comed 
constraint $\int_XJ\wedge J\wedge F=0$ on the moduli.

\section{$4D$ heterotic models}

\subsection{The spectral cover method for $SU(n)$ bundles}
\label{spectralcovermethod}

Let us recall the idea of the spectral cover description of an 
$SU(n)$ bundle 
$V$: 
one considers the bundle
first over an elliptic fibre and then pastes together these descriptions
using global data in the base $B$. Now over an elliptic fibre $E$ 
an $SU(n)$ bundle $V$ over $Z$ (assumed to be fibrewise semistable)
decomposes as a direct sum of line bundles of degree zero; this is 
described as a set of $n$ points which sums to zero. If you now let 
this vary over the base $B$ this will give you a hypersurface 
$C \subset Z$
which is a ramified $n$-fold cover of $B$. If one denotes the cohomology
class in $Z$ of the base surface $B$ (embedded by the zero-section 
$\sigma$) by 
$\sigma \in H^2(Z)$ one finds that the globalization datum suitable 
to encode 
the information about $V$ is given by a class $\eta \in H^{1,1}(B)$ 
with
\beqa
C=n\sigma + \eta
\eeqa
as $C$ is given as a locus 
$w=a_0+a_2x+a_3y+\dots \, a_nx^{\frac{n}{2}}=0$, 
for $n$ even say and $x,y$ the usual
elliptic Weierstrass coordinates, 
with $w$ a section of 
${\cal O}(\sigma)^n\otimes {\cal N}$ 
with a line bundle ${\cal N}$ of class $\eta$. 
Note that $a_i$ is of class $\eta-ic_1$.

The idea is then to trade in the $SU(n)$ bundle $V$ over $Z$,
which is in a sense essentially a datum over $B$, for a line bundle 
$L$ over
the $n$-fold (ramified) cover $C$ of $B$: one has 
\beqa
V=p_*(p_C^*L\otimes {\cal P})
\eeqa
with $p:Z\times_B C\ra Z$ and $p_C: Z\times_B C\ra C$ the projections
and ${\cal P}$ the global version of the Poincare line bundle over 
$E\times E$ 
(actually one uses a symmetrized version of this), i.e. the 
universal bundle
which realizes the second $E$ in the product as the 
moduli space of degree zero
line bundles over the first factor.

A second parameter in the description of $V$ is given by a
half-integral number $\lambda$ which occurs because one gets 
{}From the condition $c_1(V)=\pi_*(c_1(L)+\frac{c_1(C)-c_1}{2})=0$ 
that with $\gamma \in ker\, \pi_*:H^{1,1}(C)\ra H^{1,1}(B)$ one has
\beqa
c_1(L)=-\frac{1}{2}(c_1(C)-\pi_{*}c_1)+\gamma
\eeqa
where $\gamma$ is being given by 
($\lambda \in \frac{1}{2}{\bf Z}$)\footnote{actually $\lambda$ 
has to be strictly half-integral resp. integral for $n$ odd resp. even}
\beqa
\gamma=\lambda(n\sigma-\eta+nc_1)
\eeqa
as $n\sigma|_C-\eta+nc_1$ is the only generally given class which
projects to zero.

\subsection{$E_k$ bundles and del Pezzo description}

\subsubsection{fibrewise}

The del Pezzo surface $dP_k$ ($k=0,...\, 8$) is given by blowing up
$k$ points in ${\bf P}^2$. So the lattice $L=H^2(dP_k,{\bf Z})$ has the 
signature
$(+1,_1^k)$ in the basis given by the line $H$ from ${\bf P}^2$ and
the exceptional divisors $E_i$ ($i=0,\dots\, 8$); all of these classes are
$(1,1)$. The anti-canonical class, an elliptic curve, is ample and 
given by $-F$ for
$F=3H-\sum_iE_i$. For $k=9$ (and the nine points lying on the 
intersection of two cubics) one gets the (almost del Pezzo) rational 
elliptic surface $dP_9$, 
with $F$ the elliptic fibre class. "Exceptional" or "$(-1)$" curves are
the curves $c$ with $c^2=-1$ and $c\cdot F=+1$. 
The orthogonal complement of $F$
is the $E_k$ lattice. The $A_{k-1}$ lattice occurs too, with the basis
$E_i-E_{i+1}$ ($i=1, \dots \, k-1$); the additional root, 
which leads to $E_k$
and does not lie on the line of the $A_{k-1}$ Dynkin diagram, 
is given by
$H-(E_1+E_2+E_3)$. For $D_{k-1}$ take the representation of $dP_k$
as Hirzebruch surface $F_1$ blown up in $k-1$ points lying on different
fibers, denote the two $P^1$ in each of the $k-1$ special fibers of type
$A_2$ by $L_{\pm i}$ ($i=1,\dots k-1$) of classes $l_{\pm i}$ and by $f$ 
the fibre class ($f=l_i+l_{-i}$); $f^{\bot}$ is the sublattice 
generated by the $l_i$ ($\pm i=1,\dots k-1$) and $(f+K)^{\bot}$ 
is generated 
by the root system $R=\{(l_{\pm i}-l_{\pm j})\},\pm i,\pm j=1, \dots k-1,
i\neq \pm j$ of type $D_{k-1}$.

Now, as every point ${\bf w}$ in the weight lattice $L/F{\bf Z}$ of $G$ 
determines a representation $\rho_{\bf w}$ of the maximal torus and, 
from taking a flat connection $A$ in the representation $\rho_{\bf w}$, 
a line bundle ${\cal L}_{\bf w}$ on $E$ - thus providing a homomorphism 
to $Jac(E)=E$, a semistable $G=E_k$ bundle over an elliptic curve $E$ 
corresponds to a homomorphism from $L$ to $E$ mapping $F$ to zero. 
This first dictionary is further translated via the Torelli theorem
to essentially\footnote{for a subtlety involving a certain $(9-k)^{th}$ 
root of ${\cal L}_F|_F$ cf. \cite{FMW}}
the space of complex structures of a $dP_k$ surface keeping 
a divisor $D$ of class $F$ fixed: namely, keeping $D$ fixed, 
$y\cdot F=0$ for $y\in L^{\bot_F}$ means
that ${\cal L}_y|_D$ is of degree zero, so defines a point in $Jac(E)$.
One may rephrase (cf. \cite{Moore}) the association, 
in the case of $E_8$ say, saying that the flat gauge field on the 
elliptic curve $D$ is mapped to the set of eight points on $D$ 
which represent the intersection of $D$ with divisors
generating the $E_8$ piece in the lattice of the del Pezzo surface. 
One gets the del Pezzo surface writing a second cubic 
(besides the elliptic curve $D$) on
${\bf P}^2$ and forcing nine points to lie on their intersection, 
where the nine points are got in flat coordinates on $D$ 
from the flat $E_8$ gauge field 
represented as $\vec{w}=(w_1,\dots \, , w_8)$ 
in a Cartan basis by $w_i=z_i-z_0$ with $z_0+z_1+\dots +z_8=0$.

\subsubsection{globally}

The root system is describing (cf. for example 
[\ref{MS}],[\ref{DKV}],[\ref{Ma}],
[\ref{Dem}]) a certain part of the $H^{1,1}(dP,{\bf Z})$ 
so that the variation of the fibre of the spectral cover over $B$ 
describes the variation of certain (-1) curves $l$ in their variation 
in a family of surfaces over $B$
(expressing the effective replacement of these lines by points, causing
the (1,1)-shift). This leads also to the necessary  relation between
$G_i^2$ and $l^2 \pi^i_*\gamma ^2=-\pi^i_*\gamma ^2$ (for $i=1,2$).

Actually we will see the spectral cover as parametrization of 
exceptional lines
in a surface fibration over $B$. This occurs by taking into account 
the description of the 'enlarged' root system in surface cohomology. 
Note that as the same moduli space ${\cal W}_G$ parametrizes $G$ bundles 
over an elliptic curve $E$ and del Pezzo surfaces $dP_G$ 
(with $E$ of class $-K$ fixed) one gets
by adiabatic extension over the base $B$ that to the bundle $V$ over $Z$
corresponds a fibration $W^{het}_G \ra B$ of $dP_G$ surfaces via
pulling back the universal object (now the universal surface not the 
universal bundle) along the section $s:B\ra {\cal W}_G={\cal M}_{Z/B}$.

So for $G=E_8$ both data, the  spectral cover and the del Pezzo 
fibration are equivalent. The obvious analogue works for type $E_n$: 
the character lattice $\Lambda$ of  $E_n$ is still 
isomorphic to the primitive cohomology group  $H^2_0(dP_n, {\bf Z})$.
For type $A_n$ or $D_n$ we use the fact that the corresponding character 
lattices can be embedded into the $E_n$ lattice as the 
orthogonal complement of an appropriate fundamental weight 
(corresponding to one of the ends of the Dynkin diagram). 
So one can define a del Pezzo fibration of type 
$A_n$ or $D_n$  to be a 
del Pezzo fibration $\pi:U\ra B$ of type $E_n$ together 
with a section of the family of $E_n$ lattices  $R^2\pi _* {\bf Z}$ 
which, in each fiber, is in the $W$ orbit of that fundamental weight. 
For $A_n$, for example, this additional  data consists, in each fiber, 
of specifying the pullback of a line of the original $P^2$.

\subsubsection{Why spectral covers for $SU(n)$ bundles and del Pezzo 
fibrations for $E_k$ bundles}

When one tries to transfer these results to the 
(D- and especially to the) E-series, one faces the following 
problem. For the E-series one does not describe \cite{FMW} the bundles
via the spectral cover construction but instead via the associated 
del Pezzo fibration, giving not a covering of $B$ but\footnote{This 
admits also a representation-theoretic explanation. 
The Weyl group of $A_n$ admits a small permutation representation
${\bf n+1}$
which decomposes into the sum of {\em two} irreducible representations:
the trivial one and the weights, ${\bf Z}[W/W_0]\cong {\bf 1}\oplus 
\Lambda $. But every permutation representation of $W_{E_n}$ 
contains at least {\em three} irreducible constituents, so the 
cohomology of an associated spectral cover contains
additional pieces. To get an object with the right cohomology one must 
either go up in dimension
or restrict attention to classes which transform 
correctly under some correspondences. }
a fibration over
it by surfaces. This is related to the following 
(cf. [\ref{KMV}]): consider the type IIA string on an
elliptic K3 with ADE singularity times $T^2$; the $N=1$ content of this
4D $N=4$ theory includes three adjoint chiral fields $X$, $Y$, $Z$, 
whose Cartan vevs, parametrizing the Higgs branch,
correspond to blowing up respectively deforming the singularity 
respectively 
giving Wilson lines to the ADE gauge group on $T^2$; the R-symmetry
induces an equivalence of the corresponding moduli spaces. 
This gives the main theorem on the structure of the moduli space 
${\cal M}_G$ of flat $G$-bundles on an elliptic curve.

Concretely let us take as elliptic curve $E=P_{1,2,3}[6]$ of equation
$e:=z^6+x^3+y^2+\mu zxy=0$ leading (with $w$ of sect. 
\ref{spectralcovermethod}) to the deformation {$e+vw$} of the $SU(n)$ 
singularity showing at the same time Looijenga's moduli space  
${(a_0, a_2, a_3,\dots,a_n)\in {\bf P}^{n-1}}$ of flat $SU(n)$ bundles
over $E$ as well as the 0D spectral geometry consisting of n points 
${(e=0)}\cap {(w=0)}$ on $E$. Note that {\it in this case of the $A_n$ 
group} it is possible to effectively replace
\footnote{For the general phenomenon
relating even (0D to 2D, of symmetric intersection form) resp. 
odd (1D to 3D,
of antisymmetric intersection form) cohomology cf. [\ref{L}]; the same 
relation underlies the extraction [\ref{KLMVW}] of the 1D 
Seiberg-Witten curve from the 3D periods of a Calabi-Yau and the relation 
between $K3$ singularities and $ADE$ gauge groups.} 
a 2D geometry of ${\bf P}^1$'s by the zero dimensional 
representatives as $v$ occurs only linear and so in the process of 
period integral evaluation to describe the variation of Hodge structure 
relevant here can be integrated out.

By contrast the same decoupling phenomenon {\it does not} take 
place\footnote{Correspondingly there occurs a situation
involving $E$ groups, where the Coulomb branch of an $N=2$ system
does not reduce to a Riemann surface but is described
in terms of 3-form periods [\ref{KMV}].}
for the the $E_k$ case: there one finds instead for the deformation 
$e+\sum_{i=1}^6 a_iv^iz^{6-i}+b_2v^2x^2+b_3v^3y+b_4v^4x$ of zero locus
$dP_8={\bf P}_{1,1,2,3}[6]$ showing the 2D spectral geometry of the 
del Pezzo surface with $H^{1,1}(dP,{\bf Z})^{\bot _E}=E_8$ 
and moduli space ${\bf P}_{1,2,3,4,5,6,2,3,4}$.

\subsection{The parabolic approach and the characteristic classes
of $E_8, E_7, E_6$}

In order to get a heterotic prediction for the Euler characteristic of 
$X$ with a section of $I_2,I_3$ singularities, we 
have to compute the second Chern class of the corresponding 
$E_7, E_6$ bundles on the 
heterotic side. Also our 'second' bundle will always be an $E_8$ bundle.

In order to do so we can use the parabolic bundles description which
leads one to easily compute their Chern classes 
\cite{FMW},\cite{A}. In this approach one starts with an unstable 
bundle on a single elliptic curve $E$.
For this one fixes a point $p$ on $E$ with the associated 
rank $1$ line bundle ${\cal O}(p)=W_1$. 
Rank $k$ line bundles $W_k$ are then
inductively constructed via the unique non-split extension 
$0\rightarrow 
{\cal O}\rightarrow W_{k+1}\rightarrow W_k\rightarrow 0$. If 
one writes the dual of $W_k$ as $W_k^*$ 
then the unique (up to translations 
on $E$) minimal unstable bundle with 
trivial determinant on $E$ is given by 
$V=W_k\oplus W^*_{n-k}$. 
This can be deformed by an element of
$H^1(E,W^*_k\otimes W^*_{n-k})$ to a stable bundle 
$V'$ which fits then into 
the exact sequence $0\rightarrow W^*_{n-k}\rightarrow V'\rightarrow 
W_k\rightarrow 0$.
To get a global version of this construction on replaces the $W_k$ 
by their global versions, i.e. 
replace ${\cal O}(p)$ by ${\cal O}(\sigma_1)$.
The global versions of $W_k$ are inductively constructed 
by an exact sequence 
$0\rightarrow 
{\cal L}^{n-1}\rightarrow W_k\rightarrow W_{k-1}\rightarrow 0$. 

\subsubsection{$E_8$ bundle}

The starting point for building an $E_8$ bundle is 
$G=SU(6)\times SU(3)\times SU(2)$ and
let $X_{6,3,2}$ denote the three factors. Then locally one has a 
description of the $X$'s given by $X_6=(W_5\oplus {\cal O})\otimes 
{\cal O}(p)^{-1/6}$ and $X_3=(W_3\otimes {\cal O}(p)^{-1/3})$ and 
$X_2=W_2\otimes {\cal L}$. The global versions are given by \cite{FMW}
\beqa
X_6&=&(W_5\otimes {\cal S}^{-1}\oplus {\cal S}^5
\otimes {\cal L}^{-1})\otimes 
{\cal O}(\sigma)^{-1/6}\otimes {\cal L}^{-3/2} \nonumber\\
X_3&=&W_3\otimes {\cal O}(\sigma)^{-1/3}\otimes {\cal L}^{-1}\nonumber\\
X_2&=&W_2\otimes {\cal O}(\sigma)^{-1/2}\otimes {\cal L}^{-1/2}
\eeqa
the fundamental class $\lambda(V)$\footnote{which is $c_2(V)/60$ 
for $E_8$ bundle, also note that 
$\lambda(V)=c_2(V)/36$ and $\lambda(V)=c_2(V)/24$ for $E_7$ resp. $E_6$ 
bundles which we use below} of this bundle is 
given by
\beqa
\lambda(V)=\eta\sigma-15\eta^2+135\eta c_1-310c_1^2
\eeqa
with $\eta=4c_1+c_1({\cal S})$ and 
which then leads to the following expression for the $\chi(X)$ 
\beqa
24n_5=288+1200c_1^2+1080c_1t+360t^2
\eeqa

\subsubsection{$E_7$ bundle}

Our starting point for $E_7$ bundle is 
$G=SU(4)\times SU(4)\times SU(2)$ and
let $X_{1,2,3}$ denote again the three factors. Then locally one has a 
description of the $X$'s given by $X_1=(W_3\oplus {\cal O})\otimes 
{\cal O}(p)^{-1/4}$ and $X_2=(W_4\otimes {\cal O}(p)^{-1/4})$ and 
$X_3=W_2
\otimes {\cal O}(p)^{-1/2}$. As global versions we choose 
\beqa
X_1&=&(W_3\otimes {\cal S}^{-1}\oplus {\cal S}^3
\otimes {\cal L}^{-1})\otimes 
{\cal O}(\sigma)^{-1/4}\otimes {\cal L}^{-1/2} \nonumber\\
X_2&=&W_4\otimes {\cal O}(\sigma)^{-1/4}\otimes {\cal L}^{-3/2}\nonumber\\
X_3&=&W_2\otimes {\cal O}(\sigma)^{-1/2}\otimes {\cal L}^{-1/2}
\eeqa
the fundamental class of this bundle is given by
\beqa
\lambda(V)=\eta\sigma-6\eta^2+48\eta c_1-\frac{399}{4}c_1^2
\eeqa
with $\eta=7/2c_1+c_1({\cal S})$ and 
which then leads to the following expression for the $\chi(X)$ 
\beqa
24n_5=288+1866c_1^2+504c_1t+504t^2
\eeqa

\subsubsection{$E_6$ bundle}

In order to get a $E_6$ bundle one chooses as the unstable bundle whose 
structure group reduces to a group locally 
$G=SU(3)\times SU(3)\times SU(3)$ following \cite{FMW}. The fundamental 
characteristic class of an $E_6$ bundle whose structure group reduces to
$G$ is 
\beqa
\lambda(V)=c_2(X_1)+c_2(X_2)+c_2(X_3)
\eeqa
Now, it was shown in 
\cite{FMW} that on a single elliptic curve $X_1$ and $X_2$ are given by
$X_{1,2}=W_3\otimes {\cal O}(p)^{-1/3}$ and for $X_3$ one has 
$X_3=(W_2\oplus
{\cal O})\otimes {\cal O}(p)^{-1/3}$. All we have to do now is to give a 
global description of these bundles and compute their Chern classes. 
Therefore
we want to consider bundles which are isomorphic to the $X_i$'s 
on each fiber and have trivial determinant. Thus we can take for 
$X_{1,2}$ 
\beqa
X_{1,2}=W_3\otimes {\cal O}(\sigma)^{-1/3}\otimes {\cal L}^{-1}
\eeqa
and for $X_3$ the most general possibility to write down 
a global version of 
it is 
\beqa
X_3=(W_2\otimes {\cal S}^{-1}\oplus {\cal S}^2
\otimes {\cal L}^{-1})\otimes 
{\cal O}(\sigma)^{-1/3}
\eeqa
where ${\cal S}$ is an arbitrary line bundle on $B$ 
which is introduced using
the fact that one can twist by additional data coming from the base. 
>From this we can now 
compute Chern classes. We find for the fundamental characteristic class of
the $E_6$ bundle 
\beqa
\lambda(V)=\eta\sigma-3\eta^2+21\eta c_1-39c_1^2
\eeqa
where $\eta=c_1({\cal S})+3c_1({\cal L})$
and with the anomaly formula we derive simply the heterotic expectation 
for the Euler characteristic of $X$ with $I_3$ singularity which is
\beqa
24n_5=288+1704c_1^2+720c_1t+432t^2
\eeqa

\section{Comparison of the moduli spaces}

{\em In view of the application indicated in the title of the paper
the most important insight of an association of data between the 
heterotic and the $F$-theory side to be expected will be the following}:
the gamma class $\gamma$ is an element of $H^{1,1}(C)$ where the
spectral cover $C$ is an $n$-fold ramified cover of $B$; think, in a
naive picture, of this as a curve in $C$ or as a curve (null-cohomologous) 
in $B$ each of whose points is covered by some preimages in $C$; now, if one
switches from the spectral cover representation of the bundle (where
over each base point $b\in B$ sits a finite point set) to the
representation by a fibration over $B$ of del Pezzo surfaces (of type
$ADE$ in general or of type $E$ in the more widely known case), the points
on the elliptic fibre over $b$ are traded in for $(1,1)$ cohomology classes
on the del Pezzo surface (sitting now over $b$) which correspond to
divisors intersecting the former elliptic curve (which re-occurs here as
anti-canonical divisor) in the formerly given points; then the
situation is considered to be embedded in an $E_8$ del Pezzo surface
with the new (formerly missing) classes/divisors shrunken to zero;
this is embedded in a $dP_9$ set-up which re-occurs (with a structure
representing precisely the corresponding heterotic bundle) in the
stable degeneration of the Calabi-Yau four-fold on the dual $F$-theory
side; as the points were thickened to $P^1$'s the $(1,1)$ class
becomes a $(2,2)$ class which is the candidate for the four-flux class.

\subsection{General comparison of moduli space and spectra}

The moduli in a 4D N=1 heterotic compactification on an elliptic CY, 
as well as in the dual F-theoretic compactification, break into "base" 
parameters which are even (under the natural involution of the elliptic 
curves), and "fiber" or twisting parameters; the latter include a 
continuous part which is odd, as well as a discrete part. 
In \cite{CD} all 
the heterotic moduli were interpreted
in terms of cohomology groups of the spectral covers, 
and identified with the corresponding F-theoretic moduli in a certain 
stable degeneration. For this one uses the close connection of 
the spectral cover and the ADE del Pezzo fibrations. 
For the continuous part of the twisting moduli, this amounts 
to an isomorphism between certain abelian varieties: the connected 
component of the heterotic Prym variety (a modified Jacobian) and the 
F-theoretic intermediate Jacobian. The comparison of the discrete part 
generalizes the matching of  heterotic 5-branes/$F$-theoretic 3-branes.

By working with elliptically fibered $Z$
one can adiabatically extend the known results about moduli spaces of 
$G$-bundles over an elliptic curve $E=T^2$, of course taking into account
that such a fiberwise description of the isomorphism class of a bundle 
leaves definitely room for {\it twisting along the base $B$}. The latter
possibility actually involves a two-fold complication: there is a 
continuous as well as a discrete part of these data. It is quite easy to
see this  for  $G=SU(n)$:  in this case $V$ can
be constructed via push-forward of the 
Poincare bundle on the spectral cover $C \times _BZ$, possibly twisted
by a line bundle ${\cal N}$ over the spectral surface $C$ (an $n$-fold
cover of $B$ (via $\pi$) lying in $Z$), whose first Chern class 
(projected to $B$)
is known from the condition $c_1(V)=0$. So ${\cal N}$ itself is known
up to the following two remaining degrees of freedom: 
first a class in $H^{1,1}(C)$ which projects 
to zero in $B$ (the discrete part), and second an element of 
$Jac(C):=Pic_0(C)$ (the continuous part; the moduli odd under the 
elliptic involution).

The continuous part is expected
\cite{FMW} to correspond on the $F$-theory side to the odd moduli,
related there to the intermediate Jacobian $J^3(X^4)$ of dimension 
$h^{2,1}$, so that the following picture emerges. The moduli space 
${\cal M}$ of the bundles is fibered ${\cal M}\rightarrow {\cal Y}$, with
fibre $Jac(C)$. There is a corresponding  picture on the 
$F$-theory side: ignoring the Kahler classes (on both sides), the moduli 
space there is again fibered. The base is the moduli space of those 
complex deformations which fix a certain complex structure of $Z$; 
the fibre is the intermediate Jacobian 
$J^3(X^4)=H^3(X,{\bf R})/H^3(X,{\bf Z})$
In total\footnote{Unspecified Hodge numbers refer 
to $X^4$},  $h^{2,1}(Z)+h^1(Z,adV)+1=h^{3,1}+h^{2,1}$.

One expects a general scheme of a duality dictionary 
beyond the previously considered cases of relating 
$h^{2,0}(C)$ and $h^{3,1}(X^4)$ respectively elements of $H^{1,0}(C)$
and $H^{2,1}(X^4)$ 
(cf. \cite{BJPS}, \cite{FMW}, section 1 
and the introduction).
Together with the proposed identification
of the discrete moduli one gets a dictionary of elements related by 
a $(1,1)$ Hodge shift 

\begin{center}
\begin{tabular}{c|c} 
$C$ & $X^4$ \\ \hline
$H^{2,0}$ & $H^{3,1}$ \\
$H^{1,0}$ & $H^{2,1}$ \\
$H^{1,1}$ & $H^{2,2}$ \\
\end{tabular}
\end{center}
where 
in the first line the deformations of $X^4$ preserving the given type 
of singularity (corresponding with the unbroken gauge group; actually
we will consider the parts in the $W_i$) are understood, 
in the second line 
a part of the relative jacobian (see below) is understood, 
and in the last line 
the subspaces $ker\, \pi_*$ respectively $ker\, (J\wedge \cdot)$. 

One can also give \cite{CD} an interpretation
of all the bundle moduli $H^1(Z,ad V)$,
even or odd under the involution, in terms of even respectively 
odd cohomology of the spectral surface , including an 
interpretation of the 
${\bf Z}_2$ equivariant index of \cite{FMW} as giving essentially the 
holomorphic Euler characteristic of the spectral surface.

Now let us recall that the ${\bf Z}_2$ equivariant index $I=n_e-n_o$ of 
\cite{FMW}, 
counting the bundle moduli which are even respectively odd 
under the $\tau$-involution, can be interpreted as giving 
essentially the holomorphic Euler 
characteristic of the spectral surface \cite{CD} which is 
\beqa
1+h^{2,0}(C)-h^{1,0}(C)&=
&\frac{c_2(C)+c_1^2(C)}{12}|_C=\frac{c_2(Z)C+2C^3}{12}\nonumber\\
&=&n+\frac{n^3-n}{6}c_1^2+\frac{n}{2}\eta(\eta-nc_1)+\eta c_1
\label{cohomspectrsurf}
\eeqa
Now one identifies the number of local complex 
deformations $h^{2,0}(C)$ of $C$ with $n_e$
respectively the dimension $h^{1,0}(C)$ of $Jac(C):= Pic_0(C)$
with $n_o$.

In this way one gets from a spectrum comparison 
the following relations \cite{AC}, [\ref{ACL}] 
in a pure gauge case (the case referred to as seperation resp. 
codimension one (what concerns the discriminant) case in the main body
of the paper)
\beqa
h^{1,1}&=&h^{1,1}(Z)+1+r\nonumber\\
h^{2,1}&=&n_o             \nonumber\\
h^{3,1}&=&h^{2,1}(Z)+I+n_o+1
\label{AC-Hodgemap}
\eeqa

Now one has to realize explicitely the map providing the $(1,1)$ shift
in Hodge classes. In the end this goes of course back to the 
additional ${\bf P}^1$
one has in $F$-theory over the heterotic side, as visible already 
in eight dimensions.
More precisely we will reinterpret the spectral cover of $B$ which 
describes the heterotic $SU(n)$ bundle in terms of a fibration of 
del Pezzo surfaces over $B$, where what were $n$ points of $C$ 
lying over $b\in B$ are then 'exceptional' curves\footnote{i.e.
rational curves $l$ of self-intersection $-1$ which have intersection 
$+1$ with the ample anti-canonical class; for the (almost del Pezzo) 
case of $dP_9$ these are just sections of the elliptic fibration}
in the del Pezzo surface over $b$.

Note that the the effective replacing of the ${\bf P}^1$ classes by
points accounts for the missing dimensions causing the mentioned (1,1)
shift in cohomology when comparing the dual results.
The description in the $E_k$ case is already well adapted to the
F-theory picture of having a fibration $W\rightarrow B$ (for each
bundle) of del Pezzo surfaces over $B$.

\subsection{Brane-impurities}
\label{secBrane-impurities}
  
{}From the relations (\ref{AC-Hodgemap}) one finds that $n_3=n_5$ 
as follows 
\cite{AC}.
First one expresses, from the heterotic identification, 
the Hodge numbers of $X^4$ purely in data of the common base $B_2$
(using Noether $12=c_1^2+c_2$ and $\chi(Z)=-60c_1^2$)
\beqa
h^{1,1}(X^4)=12-c_1^2+r\nonumber\\
h^{3,1}(X^4)=12+29c_1^2+I+n_o
\eeqa
and one next inserts
the expression for the index $I$ resulting from (\ref{cohomspectrsurf})
\beqa
I=n-1+\frac{n^3-n}{6}c_1^2+\frac{n}{2}\eta(\eta-nc_1)+\eta c_1
\eeqa
Then one re-expresses $I$ with $c_2(V_i)$
\beqa
I=rk_i-4(c_2(V_i)-\eta_i \sigma +\frac{1}{2}\pi_* \gamma_i^2)+\eta_i c_1
\eeqa
which gives with $rk_1+rk_2=rk=16-r$ and $\eta_1+\eta_2=12c_1$ 
for the total 
index
\beqa
I&=&rk+48c_1\sigma+12c_1^2 -4(c_2(V_1)+c_2(V_2))-4(\frac{1}{2}\pi_*
 \gamma_1^2+
\frac{1}{2}\pi_* \gamma_2^2)\nonumber\\
 &=&rk-48-28c_1^2+4n_5-4(\frac{1}{2}\pi_* \gamma_1^2+\frac{1}{2}\pi_* 
\gamma_2^2)
\eeqa
giving finally
\beqa
n_3+\frac{1}{2}G^2&=&\frac{\chi(X^4)}{24}=
2+\frac{1}{4}(h^{1,1}(X^4)-h^{2,1}(X^4)+h^{3,1}(X^4))\nonumber\\
 &=&n_5-(\frac{1}{2}\pi_* \gamma_1^2+\frac{1}{2}\pi_* \gamma_2^2)
\eeqa

\subsection{Stable degeneration and the map from the heterotic theory 
to $F$-theory}

We consider the stable degeneration \cite{FMW},[\ref{AM}], \cite{AspHyp}
$X^4\rightarrow X^4_{deg}=W_1 \cup_Z W_2$
where the $W_i$ are fibered by del Pezzo surfaces over $B$.
The 8D picture involves a $K3$ degenerating 
into the union of two rational elliptic surfaces 
($dP_9$, almost del Pezzo). The base of the
fibration is the union of two projective lines 
intersecting in a point $Q$ over 
which a common elliptic curve $E$ is fibered; roughly speaking
the two $E_8$ contributions in the $K3$ are separated. Recall that
the chosen $K3$ had Picard number two with section and fiber as the
two algebraic cycles 
(still allowing $18$ deformations which match the heterotic side)
and the transcendental lattice $E_8\oplus E_8 \oplus H$, 
with $H$ the 2-dimensional hyperbolic plane, which leads to the
18-dimensional space $S:=SO(2,18)/SO(2)\times SO(18)$ divided by 
the appropriate
discrete group; one specializes then to two $E_8$ singularities
at positions $z=0,\infty$ in the $P^1$ base, which after the 'separation'
in two surfaces are again re-smoothed; imagine to take 
(for the $dP_9$'s to arise) the two $f_4,g_6$
parts at $z=0,\infty$ of the original Weierstrass data $f_8,g_{12}$ 
of the $K3$.

This corresponds on the heterotic side to the large area degeneration 
of a $T^2$ 
of the same complex structure parameter as $E$ \cite{FMW}, [\ref{AM}]. 
Imagine that the $H$ and its counterpart in $S$ above 
correspond to the degrees of freedom represented by 
the complex structure modulus $\tau$ and the area (+ $B$-field) 
modulus $\rho$ of $E$; then in the $\rho \ra i\infty$ limit one finds
in the corresponding boundary component of the quotient 
(discrete$\backslash S$) the two spaces 
($W_{E_8} \backslash (E_i\otimes \Lambda _c)$), 'glued' together by 
$\tau (E_1)=\tau (E_2)$, describing the moduli of the two $dP_9$'s
($\Lambda _c$ the coroot lattice of $E_8$). 
The heterotic invariant $n_5=c_2Z-c_2V_1-c_2V_2$ is then mirrored on the 
F-theory side by
$n_3=-\frac{\chi(Z)}{24}+\frac{\chi(W_1)}{24}+\frac{\chi(W_2)}{24}$.

Note that the (even) deformations of $V_i$ correspond to those
deformations of $W_i$ which preserve fiberwise the elliptic curve $E$
common with the heterotic side, so preserving in total the Calabi-Yau $Z$ 
common to the $W_i$: their number is given by the dimension of 
$H^1(W_i,T_{W_i}\otimes {\cal O}(-Z))\cong H^{3,1}(W_i)$. These are the 
deformations in the stable degeneration of
$X^4$ which are relevant to the respective bundle. Further
under the stable degeneration $J^3(X)$ splits off the abelian
varieties $J^3(W_i)$, which contain the pieces relevant for the 
comparison. So essentially this construction interprets
those elements
of $H^{2,2}_{prim}(X^4,{\bf Z})$ that are 'captured by' the corresponding 
parts in the $W_i$ cohomology\footnote{for the relation of the 
primitiveness condition to the $W,Z$ geometry cf. the discussion 
in the section on the four-flux}.
concerning complex structure deformations note that 
the distribution into deformations of $Z$ respectively those deformations
$H^1(W_i,T_{W_i}\otimes {\cal O}(-Z))\cong H^{3,1}(W_i)$ 
of $W_i$, which preserve $Z$, reflects the well known distribution of the
deforming monomials of the defining $F$-theory equation for $X^4$ 
into "middle-polynomials" and the rest.

In the representation of the bundle via the del Pezzo construction 
respectively in the stable degeneration on the $F$-theory side the data 
are already in a form appropriate for comparison. For example in the 
case of $E_8$ bundles one has just to blow down the section of the 
$dP_9$ fibre on the $F$-theory side to get the $dP_8$ fibre of the 
heterotic side showing the relation of
the cohomologies and the intermediate jacobians (cf. \cite{CD}).
For a bundle of group $H\neq E_8$ the section 
$\theta :B\rightarrow X^4$ 
of $G$-singularities in the $F$-theory setup 
corresponds (assuming $G$ to be simply-laced) 
to having a bundle of unbroken gauge group
$G$, i.e. an $H$ bundle where $H$ is the commutant of $G$ in $E_8$, 
over the heterotic Calabi-Yau $Z$ respectively having\footnote{at least 
locally over the dense open subset of $B$ where fibres 
correspond to semistable bundles} a section 
$s :B\ra {\cal W}_H={\cal M}_{Z/B}$ 
or, as the fibre of ${\cal W}_H$ 
over $b\in B$ parametrizes the corresponding del Pezzo surfaces, 
a bundle $W_H^{het}\ra B$ of del Pezzo surfaces $dP_H$ fibered 
over $B$. So, if\footnote{this can be done as
we have an $ADE$ system of rational (-2) curves lying
in $H^{1,1}(K3,{\bf Z})$ as well as in $H^{1,1}(dP_8,{\bf Z})^{\bot_F}$
(in the case of the E-series, say; 
$F$ the elliptic curve representing the ample anticanonical divisor); 
note that the complex structures for $dP_H$ are given by 
homomorphisms $H^{1,1}(dP_H,{\bf Z})^{\bot_F} \ra F$
and the complex structures for $dP_8$ keeping the $G$ singularity
are similarly given by the corresponding homomorphisms 
for $dP_8$ mapping the $G$ system of rational (-2) curves to zero 
(i.e. they essentially describe a mapping for the $H$-part)} 
one considers heterotically actually a $dP_8$ 
fibration with $G$ singularity instead of the $dP_H$ fibration,
then locally at $\theta $, i.e. at the singularity along $B$
(local in the $dP$ fibre and global along B), the picture in 
the $K3$ fibre of $X^4\ra B$ respectively the $dP$ fibre 
on the heterotic side is the same.

\subsection{Comparison of the discrete data}

Of course the ultimate goal will be to make the complex two-cycle 
supporting the four-flux explicit and
check the choice with a dual heterotic situation.
Note that, comparing the heterotic contribution of $\gamma^2$ in eq. (1.5)
\beqa
n_{5,\gamma}=n_{5,\gamma=0}+\frac{1}{2}\pi_*(\gamma^2)
\eeqa
with the formula \cite{DM}
\beqa
n_3=\frac{\chi(X^4)}{24}-\frac{1}{2}G^2\; ,
\eeqa
we are led to expect an association letting $\gamma_i$ correspond 
with $G_i$ giving
\beqa \label {-1}
\pi^i_*(\gamma_i^2)=-G_i^2.
\eeqa

Note that two other facts fit with this association of 
$\gamma$ and $G$. 
{\em First}
the shifted integrality (to half-integrality):
the analogy in the data concerned with the discrete part of the 
twisting degrees of freedom (cf. below) is represented in the 
following juxtaposition: on the heterotic side one has 
(cf. sect. (\ref{spectralcovermethod}))
\beqa
\gamma=\frac{c_1(C)-\pi ^* c_1(B)}{2}+c_1({\cal L}),
\eeqa
where the last term is an element of integral cohomology whereas the 
square root $(K_C^{-1}\otimes K_B)^{\frac{1}{2}}$ does not necessarily 
exist as a line bundle. Similarly one has 
on the F-theory side \cite{W4flux}
\beqa
G=\frac{c_2}{2}+\alpha 
\eeqa
where $\alpha \in H^4(X,{\bf Z})$, but $c_2$ is not necessarily even.
Strictly speaking one should consider here the individual
$G_i$ ($i=1,2$) corresponding to the $\gamma_i$ by means of the
association provided by the stable degeneration (cf. the introduction
to this section).

{\em Secondly} the restriction to the subspace $ker\; \pi$ respectively 
primitiveness: the $G$ admissible in an $N=1$ supersymmetric 
compactification are in $ker(J\wedge \cdot)$ \cite{BB}. 
The last condition comes down for the relevant projected classes in 
$H^{2,2}(W)$ to the following:
on the heterotic side the actual spectral cover construction will
in the $E_8$ case involve the corresponding fibration of $dP_8$ surfaces 
over $B$ (the section of $dP_9$ blown down); now, the embeddings of 
the 8D heterotic elliptic curves in the 8D del Pezzos patch together 
to an embedding of $Z$ in the $W_i$, giving a map
$H^{p,q}(W)\rightarrow H^{p,q}(Z)$; but for the $dP_8$ the anti-canonical 
class given by the elliptic curve $E$ is ample, so actually the 
$ker\, (J\wedge \cdot)$ condition leads to a 
$ker\, \cdot |_Z:\, (H^{2,2}(W)\rightarrow H^{2,2}(Z))$
condition, respectively, if one combines with the integration over the 
fibre, in a $ker(H^{2,2}(W)\rightarrow H^{2,2}(Z)\rightarrow H^{1,1}(B))$ 
condition; one has then to divide 
out the class dual to $S_b$, the del Pezzo fibre of $pr:W\rightarrow B$,
corresponding to a differential form supported on the base,
which is mapped to zero in the integration over the $\pi:Z\rightarrow B$
fibre. So finally the space we are concerned with is the 
$(ker:W\rightarrow B)/S_b {\bf Z}$ part in $H^{2,2}(W)$. 
So the primitiveness condition is the analogue of the condition 
$ker \, \pi:H^{1,1}(C,{\bf Z})\rightarrow H^{1,1}(B,{\bf Z})$ on $\gamma$.

This fits in and actually completes the general scheme of a duality 
dictionary beyond the previously considered cases of relating 
$h^{2,0}(C)$ and $h^{3,1}(X^4)$ respectively elements of $H^{1,0}(C)$
and $H^{2,1}(X^4)$ (cf. \cite{BJPS}, \cite{FMW} and the appendix).

Now one has to realize explicitely the map providing the $(1,1)$ shift
in Hodge classes. A naive\footnote{More precisely the right hand side of 
(\ref{-1}) gets contributions also from distinct lines which intersect 
in the del Pezzo surface. $H^i(C)$ breaks into several isotypic pieces 
(five of them, for $E_8$). The values of $\gamma$ coming from bundles 
all live in one of these pieces, where the cylinder map changes 
the intersection numbers by a factor of $-60$ (for $E_8$); so the correct
association sends $\gamma$ to $\frac{1}{60}$ times its cylinder.}
way to obtain the association of $\gamma_i$ with $G_i$ is via the 
cylinder map [\ref{K1}],[\ref{K2}]. This  replaces each point in $C$ 
lying over $b\in B$ by a complex projective line $L$ lying above in the 
del Pezzo surface over $b$. Indeed, $L^2=-1$, suggesting the desired 
relation (\ref{-1}).

\section{6D computations}

We list some spectrum and Euler number computations in 6D for the $I_n$ series.

\subsection{heterotic spectra}

The Euler numbers (cf. [\ref{CF}]) 
match with the heterotic expectations
for the spectrum
(note that the spectra have to fulfil the gravitational
anomaly condition $244+n_V=n_H$ (here occur the fundamental matter and the 
antisymmetric tensors) and that always $h^{1,1}(Z)=3+(k-1)$,
$h^{2,1}(Z)=n_H-1$ and $n_V=k^2-1$):
\beqa
{\bf I_2}\;\;\;\;\;\;\;\;\;\;\;\;\;\;\;\;\;\;\;\;\;\;\;\;\;\;\;\;\;\;\;\;
(16+6n)(2)-3&=&29+12n\nonumber\\
dim_Q({\cal M}_{inst}^{(n_1;n_2)})+h^{1,1}(K3)&=&166+20=186\nonumber\\
n_H^0 &=& 215+12n\nonumber\\
244+3&=&215+12n+2(16-6n)\nonumber\\
{\bf I_3}\;\;\;\;\;\;\;\;\;\;\;\;\;\;\;\;\;\;\;\;\;\;\;\;\;\;\;\;\;\;\;\; 
(18+6n)(3)-8&=&46+18n\nonumber\\
dim_Q({\cal M}_{inst}^{(n_1;n_2)})+h^{1,1}(K3)&=&132+20=152\nonumber\\
n_H^0 &=& 198+18n\nonumber\\
244+8&=&198+18n+3(2-n)+3(16-5n)\nonumber\\
{\bf I_4}\;\;\;\;\;\;\;\;\;\; (2+n)(6)+(16+4n)(4)-15&=&61+22n\nonumber\\
dim_{\cal Q}({\cal M}_{inst}^{(n_1;n_2)})+h^{1,1}(K3)&=&
102+20=122\nonumber\\
n_H^0 &=& 183+22n\nonumber\\
244+15&=& 183+22n+6(2-n)+4(16-4n)\nonumber\\
{\bf I_5}\;\;\;\;\;\;\; (16+3n)(5)+(2+n)(10)-24&=&76+25n\nonumber\\
dim_Q({\cal M}_{inst}^{(n_1;n_2)})+h^{1,1}(K3)&=&72+20=92\nonumber\\
n_H^0 &=& 168+25n\nonumber\\
244+24&=& 168+25n+10(2-n)+5(16-3n)\nonumber\\
{\bf I_6}\;\;\;\;\;\;\; (16+2n)(6)+(2+n)(15)-35&=&91+27n\nonumber\\
dim_Q({\cal M}_{inst}^{(n_1^{(1)},n_1^{(2)};n_2)})+h^{1,1}(K3)&=&
42+20=62\nonumber\\
n_H^0 &=& 153+27n\nonumber\\
244+35&=& 153+27n+15(2-n)+6(16-2n)\nonumber
\eeqa
(note that one gets $G=A_5=I_6$ from an $SU(2)\times SU(3)$ bundle).

\subsection{discriminant equations}

We consider now in detail the discriminant equations (using the
notation $f_i:= f_{4c_1-it},\; g_i:= g_{6c_1-it}$ and $c:=32\cdot 864$).

{\bf The $I_2$ case}

The ansatz with $H=H_{2c_1-2t}$ 
\beqa
f=\frac{1}{48}(-H^2+f_3z)\nonumber\\
g=\frac{1}{864}(H^3+g_5z+g_4z^2)
\eeqa
gives, because of the $z$-linear term $H^3(2g_5+3f_3H)$ and the thereby
enforced choice $g_5=-\frac{3}{2}f_3H$, that
$D_1r=2H+P_{8c_1-6t}$ with $P=-\frac{3}{4}f_3^2+2g_4H$ as
\beqa
c\cdot \Big(4f^3+27g^2\Big) &=&z^2\Bigg[ \Bigg(H^2(-\frac{3}{4}f_3^2+2g_4H
                             )\Bigg) \nonumber\\
           & &+\Bigg( f_3^3-3f_3g_4H\Bigg) z\nonumber\\
& &+\Bigg( g_4^2)\Bigg) z^2\Bigg]
\eeqa

{\bf The $I_3$ case}

The ansatz 
\beqa
f=\frac{1}{48}(-h^4+f_3z+f_2z^2)\nonumber\\
g=\frac{1}{864}(h^6+g_5z+g_4z^2+g_3z^3)
\eeqa
gives
\beqa
c\cdot \Big( 4f^3+27g^2\Big) &=&h^6(2g_5+3f_3h^2)z\nonumber\\
                 & &+(g_5^2-3f_3^2h^4+2g_4h^6+3f_2h^8)z^2\nonumber\\
                 & &+(f_3^3+2g_5g_4-6f_3f_2h^4+2g_3h^6)z^3+{\cal O}(z^4)
\eeqa
Solving for the $I_3$ condition gives at first $g_5=-\frac{3}{2}f_3h^2$
and then $f_3^2=\frac{4}{3}h^2(2g_4+3f_2h^2)$ which leads us to
introduce $Q_{3c_1-2t}$ with  $3Q^2_{3c_1-2t}=2g_4+3f_2h^2$.
\beqa
g_5&=&-\frac{3}{2}f_3h^2=-3h^3Q\nonumber\\
g_4&=&\frac{3}{2}(Q^2-f_2h^2)\nonumber\\
f_3&=&2hQ
\eeqa
Then one gets
\beqa
c\cdot  \Big(4f^3+27g^2\Big) &=&z^3\Bigg[ \Bigg( h^3(-Q^3-3f_2h^2Q
+2g_3h^3)\Bigg) \nonumber\\
& &+\Bigg( h^2(-\frac{3}{4}f_2^2h^2-6g_3hQ+\frac{15}{2}f_2Q^2)
           +\frac{9}{4} Q^4\Bigg) z \nonumber\\
& &+\Bigg( h(-3f_2g_3h+6f_2^2Q)+3g_3Q^2\Bigg) z^2 \nonumber\\
& &+\Bigg( f_2^3+g_3^2\Bigg) z^3\Bigg]
\label{I3dis}
\eeqa

{\bf The $I_4$ case}

The ansatz 
\beqa
f=\frac{1}{48}(-h^4+f_3z+f_2z^2+f_1z^3)\nonumber\\
g=\frac{1}{864}(h^6+g_5z+g_4z^2+g_3z^3+g_2z^4+g_1z^5)
\eeqa
gives
\beqa
c\cdot \Big( 4f^3+27g^2\Big) &=&h^6(2g_5+3f_3h^2)z\nonumber\\
            & &+(g_5^2-3f_3^2h^4+2g_4h^6+3f_2h^8)z^2\nonumber\\
         & &+(f_3^3+2g_5g_4-6f_3f_2h^4+2g_3h^6+3f_1h^8)z^3\nonumber\\
  & &+(3f_2f_3^2+g_4^2+2g_3g_5+h^4(-3f_2^2-6f_1f_3+2g_2h^2))z^4\nonumber\\
& &+(3f_2^2f_3+3f_1f_3^2+2g_3g_4+2g_2g_5+h^4(-6f_1f_2+2g_1h^2))
z^5\nonumber\\
& &+(f_2^3+6f_1f_2f_3+g_3^2+2g_2g_4+2g_1g_5-3f_1^2h^4)z^6\nonumber\\
& &+(3f_1f_2^2+3f_1^2f_3+2g_2g_3+2g_1g_4)z^7\nonumber\\
& &+(3f_1^2f_2+g_2^2+2g_1g_3)z^8\nonumber\\
& &+(f_1^3+2g_1g_2)z^9\nonumber\\
& &+g_1^2z^{10}
\eeqa
Solving for the $I_4$ condition leads to introduction of $H_{2c_1-t}$ with
\beqa
g_5&=&-3h^4H\nonumber\\
g_4&=&\frac{3}{2}h^2(H^2-f_2)\nonumber\\
g_3&=&\frac{1}{2}H(H^2+3f_2)-\frac{3}{2}f_1h^2\nonumber\\
f_3&=&2h^2H
\eeqa
thus giving
\beqa
c\cdot  \Big( 4f^3+27g^2\Big) &=&z^4\Bigg[ h^4\Bigg( h^2(2g_2-3f_1H)
      -\frac{3}{4}H^4-\frac{3}{2}f_2H^2-\frac{3}{4}f_2^2\Bigg) \nonumber\\
& &+h^2\Bigg( 2g_1h^4+h^2(-\frac{3}{2}f_1f_2-6g_2H+\frac{15}{2}f_1H^2)
           +\frac{3}{2}f_2^2H+3f_2H^3+\frac{3}{2}H^5\Bigg) z\nonumber\\
& &+\Bigg( -h^4(6g_1H+\frac{3}{4}f_1^2)+h^2(3g_2H^2-3f_2g_2
           +\frac{15}{2}f_1f_2H-\frac{3}{2}f_1H^3)\nonumber\\
& &\;\;\;\; +f_2^3+\frac{9}{4}f_2^2H^2+\frac{3}{2}f_2H^4
           +\frac{1}{4}H^6\Bigg) z^2\nonumber\\
& &+\Bigg( h^2(3f_1g_2-3f_2g_1+6f_1^2H+3g_1H^2)+3f_1f_2^2+3f_2g_2H
           +g_2H^3\Bigg) z^3\nonumber\\
& &+\Bigg( -3f_1g_1h^2+3f_1^2f_2+g_2^2+3f_2g_1H+g_1H^3\Bigg)
 z^4\nonumber\\
& &+\Bigg( f_1^3+2g_1g_2\Bigg) z^5\nonumber\\
& &+\;\;\; g_1^2 \; z^6\Bigg]
\eeqa

{\bf $I_5$ case}

So starting from the ansatz 
\beqa
F&=&\frac{1}{48}(-h^4+f_3z+f_2z^2+f_1z^3)\nonumber\\
G&=&\frac{1}{864}(h^6+g_5z+g_4z^2+g_3z^3+g_2z^4+g_1z^5)
\eeqa
one has made sure that the constant ($z$-free) term has already cancelled:
\beqa
c\; \Big( 4F^3+27G^2\Big) &=&h^6(3h^2f_3+2g_5)z\nonumber\\
       & &+(g_5^2+h^4(3h^4f_2+2h^2g_4-3f_3^2)) z^2\nonumber\\
    & &+(f_3^3+2g_5g_4+h^4(3h^4f_1+2h^2g_3-6f_3f_2))z^3\nonumber\\
& &+(3f_3^2f_2+g_4^2+2g_5g_3
+h^4(2h^2g_2-3f_2^2-6f_3f_1))z^4\nonumber\\
& &+(3f_3f_2^2+3f_3^2f_1+2g_4g_3+2g_5g_2 
+2h^4(h^2g_1-3f_2f_1))z^5\nonumber\\
& &+(f_2^3+6f_3f_2f_1+g_3^2+2g_4g_2+2g_5g_1-3h^4f_1^2)z^6\nonumber\\
& &+(3f_2^2f_1+3f_3f_1^2+2g_3g_2+2g_4g_1)z^7\nonumber\\
& &+(3f_2f_1^2+g_2^2+2g_3g_1)z^8\nonumber\\
& &+(f_1^3+2g_2g_1)z^9\nonumber\\
& &+g_1^2z^{10}
\eeqa
To get actually $I_5$ the terms up to $z^4$ have to cancel. For this
one solves for $g_5,g_4,g_3,g_2,f_3,f_2$ in terms of 
$h_{c_1-t}, H_{2c_1-t}, q_{3c_1-t}, f_1=f_{4c_1-t}, g_1=g_{6c_1-t}$
with the terms $H$ and $q$ and
thereby gets succesively the following relations (in the
expression in front of $z^5$ {\em all} terms finally come
with an explicit $h^4$ factor)
\beqa
g_5&=&-3h^4H\nonumber\\
g_4&=&3h^2(H^2-hq)\nonumber\\
g_3&=&\frac{3}{2}h(2Hq-hf_1)-H^3\nonumber\\
g_2&=&\frac{3}{2}(f_1H+q^2)\nonumber\\
f_3&=&2h^2H\nonumber\\
f_2&=&2hq-H^2
\label{I5polynomials}
\eeqa

So the discriminant equation $\Delta$ presents itself now in 
the manifest $I_5$ form
\beqa
\Delta&=&z^5\Bigg[ h^4( 2h^2g_1-3f_1qh-3Hq^2) \nonumber\\
& &+h^2\Bigg( -3h^2(\frac{1}{4}f_1^2+2g_1H)
        +qh(6f_1H-q^2)+6H^2q^2\Bigg) z\nonumber\\
& &+\Bigg( -6qh^3g_1+\frac{3}{2}h^2(Hf_1^2+5q^2f_1+4H^2g_1)
           +3qHh(3q^2-Hf_1)-3q^2H^3\Bigg) z^2 \nonumber\\    
& &+\Bigg( -3f_1g_1h^2+6hq(f_1^2+g_1H)-2g_1H^3-\frac{3}{4}f_1^2H^2
            +\frac{9}{2}f_1q^2H+\frac{9}{4}q^4\Bigg) z^3 \nonumber\\  
& &+\Bigg( f_1^3+3g_1(f_1H+q^2)\Bigg) z^4 \nonumber\\ 
& &+ \;\; \; g_1^2 z^5 \Bigg]
\label{I5discr}
\eeqa
giving
\beqa
P=2h^2g_1-3f_1qh-3Hq^2
\eeqa

{\bf The $I_6$ case}

The ansatz
\beqa
f=\frac{1}{48}(-h^4+f_3z+f_2z^2+f_1z^3)\nonumber\\
g=\frac{1}{864}(h^6+g_5z+g_4z^2+g_3z^3+g_2z^4+g_1z^5)
\eeqa
gives
\beqa
c\cdot \Big( 4f^3+27g^2\Big) &=&h^6(2g_5+3f_3h^2)z\nonumber\\
            & &+(g_5^2+h^4(-3f_3^2+2g_4h^2+3f_2h^4))z^2\nonumber\\
         & &+(f_3^3+2g_5g_4+h^4(-6f_3f_2+2g_3h^2+3f_1h^4))z^3\nonumber\\
  & &+(3f_2f_3^2+g_4^2+2g_3g_5+h^4(-3f_2^2-6f_1f_3+2g_2h^2))z^4\nonumber\\
& &+(3f_2^2f_3+3f_1f_3^2+2g_3g_4+2g_2g_5+h^4(-6f_1f_2+2g_1h^2))z^5\nonumber\\
& &+(f_2^3+6f_1f_2f_3+g_3^2+2g_2g_4+2g_1g_5-3f_1^2h^4)z^6\nonumber\\
& &+(3f_1f_2^2+3f_1^2f_3+2g_2g_3+2g_1g_4)z^7\nonumber\\
& &+(3f_1^2f_2+g_2^2+2g_1g_3)z^8\nonumber\\
& &+(f_1^3+2g_1g_2)z^9\nonumber\\
& &+g_1^2z^{10}
\eeqa
Solving for the $I_6$ condition leads at first to the
identifications (\ref{I5polynomials})
and then with the condition that $P=0$ to the introduction of
${\cal F}={\cal F}_{2c_1}$ whose product with $h$
replaces the old $q_{3c_1-t}$, thereby leading to
\beqa
g_5&=&-3h^4H\nonumber\\
g_4&=&3h^2(H^2-h^2{\cal F})\nonumber\\
g_3&=&\frac{3}{2}h^2(2H{\cal F}-f_1)-H^3\nonumber\\
g_2&=&\frac{3}{2}(f_1H+h^2{\cal F}^2)\nonumber\\
g_1&=&\frac{3}{2}{\cal F}(H{\cal F}+f_1)\nonumber\\
f_3&=&2h^2H\nonumber\\
f_2&=&(2h^2{\cal F}-H^2)
\eeqa
giving
\beqa
\Delta &=&z^6\Bigg[ h^4\Bigg( -{\cal F}^3h^2
       -3(\frac{1}{2}f_1+{\cal F}H)^2\Bigg) \nonumber\\
& &+ 3h^2\Bigg( -\frac{1}{2}h^2{\cal F}^2f_1+2H(\frac{1}{2}f_1
              +{\cal F}H)^2\Bigg) z \nonumber\\
& &+3\Bigg( {\cal F}h^2(\frac{3}{4}{\cal F}^3h^2+\frac{1}{2}f_1^2
               +3{\cal F}f_1H+3{\cal F}^2H^2)
             -H^2(\frac{1}{2}f_1+{\cal F}H)^2\Bigg) z^2 \nonumber\\
& &+\Bigg( \frac{9}{2}{\cal F}\Big( {\cal F}^2h^2(f_1+{\cal F}H)
           +f_1H(f_1+{\cal F}H)\Big) +f_1^3\Bigg) z^3 \nonumber\\
& &+\frac{9}{2}{\cal F}^2 (f_1+{\cal F}H)^2 z^4 \Bigg]
\eeqa

\begingroup\raggedright
\end{document}